\documentclass[%
 aapm,
 mph,%
 amsmath,amssymb,
 reprint,%
]{revtex4-2}
\usepackage[english]{babel}
\usepackage{graphicx} 
\usepackage{epstopdf}
\usepackage{amsmath}
\usepackage{tikz}
\usepackage{hyperref}
\usepackage{subcaption}
\usepackage{float}
\usepackage{dcolumn}
\usepackage{bm}
\usepackage{overpic}

\def\selectlanguage#1{}
\newcommand{\pderiv}[2]{\frac{\partial #1}{\partial #2}}
\newcommand{\ppderiv}[2]{\frac{\partial^2 #1}{\partial #2^2}}
\newcommand{\sech}{\text{sech}}

\begin{document}

\preprint{APS/123-QED}

\title{Fundamental and multihumped multimode solitons in fibers : new families of solitons and their properties}

\author{Julien Dechanxhe}
 \author{Pascal Kockaert}%


\author{Spencer W. Jolly}%
 \affiliation{%
    Service OPERA-Photonique, Université libre de Bruxelles (ULB), Brussels, Belgium
 }%


\date{\today}

\begin{abstract}
We describe multimode solitons made of purely spatial modes using the mathematical formalism developed for the elliptical polarization soliton, and we extend this formalism to solitons made of more than two modes. The main difference with the polarization mode case is the asymmetry in the nonlinear coefficients. Strictly speaking, these objects are solitary waves, whose modal temporal profiles remain invariant owing to the incoherent nonlinear interaction between modes. The approach consists of finding the trajectories of a unit-mass ball in an analogous potential, which correspond to stationary solutions of our set of nonlinear coupled equations. Our analysis reveals a rich variety of multimode solitons with more complex temporal profiles, so-called multihumped multimode solitons. These are new and highly complex nonlinear objects. While fundamental multimode solitons are stable against noise and symmetry breaking, the stability of multihumped multimode solitons depends on their position in the parameter space. The fission of multihumped solitons yields several less complex multihumped solitons, together with fundamental multimode solitons. For the latter, Raman nonlinearity can induce a net energy transfer between modes with different group velocities, whereas self-steepening only modifies the overall velocity of the solitary wave.
\end{abstract}

\maketitle

\section{Introduction}\label{sec:introduction}
Over more than five decades~\citep{hasegawa_transmission_1973,mollenauer_experimental_1980} the study of optical solitons has been a major topic in nonlinear optics. Kerr soliton arises from the balance between Kerr nonlinearity and chromatic dispersion. This balance leads to a particle-like behavior of the pulse that is invariant during propagation, stable and robust. The invariant property can be either in time or in space, leading to temporal~\cite{blanco-redondo_bright_2023} or spatial solitons~\citep{stegeman_optical_1999,chen_optical_2012}. A lot of research has also been devoted to find stable spatiotemporal solitons, called light bullets, that are invariant and stable both in space and time~\citep{kartashov_frontiers_2019,dror_symmetric_2011}. The word soliton is also used in an extended sense to describe solitary waves that are not necessarily robust to collision, or strictly invariant, but still stable against perturbations. In this sense, solitons can be found in a large variety of systems, including mode-locked lasers~\citep{grelu_dissipative_2012,wright_spatiotemporal_2017}, dispersion-managed fibers~\cite{turitsyn_physics_2003}, microresonators~\citep{kippenberg_dissipative_2018,rowley_self-emergence_2022} and fiber ring cavities~\cite{leo_temporal_2010,englebert_parametrically_2021}.
Solitons show interest in a wide variety of fields, including laser technology~\cite{song_recent_2019}, neuromorphic computing~\citep{marcucci_theory_2020,silva_reservoir_2021} and analogue physics~\citep{englebert_bloch_2023,philbin_fiber-optical_2008,robson_topological_2019}.

Even though the existence of multimode solitons was predicted in 1980~\citep{hasegawa_self-confinement_1980}, their interest has grown significantly in the last decade as the search for higher data rates photonic networks has motivated the study of multimode fibers. Beyond the benefits of communication links with increased capacity, multimode fibers have proved to be a particularly rich platform for the study of nonlinear optical phenomena~\citep{krupa_multimode_2019,wright_physics_2022}. Unlike in the linear case, propagation modes can interact with one another via nonlinear effects, which can give rise to collective phenomena or complex structures such as multimode solitons~\cite{sun_multimode_2024}. Solitonic properties of multimode solitons emerge in part from nonlinear interactions between modes. They form when instantaneous Kerr nonlinearities, self-phase modulation and cross-phase modulation, balance intermodal and chromatic dispersion. Such balance was experimentally observed in parabolic graded-index multimode fibers~\cite{renninger_optical_2013}, where intermodal dispersion is minimized, but also in step-index multimode fibers~\citep{zitelli_multimode_2022,wu_highly_2023}.

Propagation of multimode solitons is very rich~\citep{wright_spatiotemporal_2015,zitelli_optical_2024} as they can fission, emitting dispersive waves~\citep{wright_ultrabroadband_2015,eftekhar_general_2021} and inducing supercontinuum generation~\citep{wright_controllable_2015,eftekhar_versatile_2017}. They experience a Raman self-frequency shift, which changes their group velocities and induces inelastic collisions between multimode solitons~\cite{sun_multimode_2022}. Raman nonlinearity is also responsible for a driving of the energy from higher-order modes towards the fundamental mode typically over hundreds of meters~\citep{zitelli_single-mode_2021}.

Among multimode solitons, we can distinguish those formed from polarization modes, which are known as elliptical solitons. They are characterized by a polarization state that varies across the pulse profile. Each point in time within the pulse is associated with a polarization ellipse, whose axes rotate at a constant rate along propagation, dictated by the degree of the polarization ellipticity~\citep{haelterman_elliptically_1994,delque_experimental_2005,delque_polarization_2007}. The interplay between polarization modes in normal dispersion regime can also form solitonic structures such as the vector dark soliton~\cite{kivshar_vector_1993} or domain-wall solitons~\citep{haelterman_vector_1994, pitois_polarization_1998,snee_domain_2024}.

The goal of this paper is to describe the fundamental multimode soliton (with a relatively simple temporal profile) and to identify new kinds of multihumped solitons (with more complicated temporal profiles), composed purely of spatial modes, in multimode fibers. To this end, we employ and extend the mathematical tools originally developed to understand the physics of elliptical solitons. This provides an efficient way to identify new families of stationary solutions. An important difference between polarization multimode solitons and spatial multimode solitons is that the former are limited to two modes with symmetric coupling coefficients. As will be demonstrated in this paper, spatial multimode solitons offer much greater possibilities for generating complex structures, whether stable or unstable, which hold great potential for applications such as optical switching\cite{sparapani_intermodal_nodate} or the generation of random numbers~\cite{quinn_random_2023}, for example.

The paper is organized as follows. In Section~\ref{sec:theory}, we derive a reduced model of which multimode solitons are stationary solutions, and show how we deduce an analogous potential from it. Then, in Section~\ref{sec:hyp}, we extend the validity of our model. In Section~\ref{fiber}, we present the multimode fiber that will be considered for the examples. In Section~\ref{subsec:2modes.fundamental}, we adapt the formalism to the fundamental two-mode soliton and extend the main quantities characterizing the soliton to the multimode case. We also show the link between the temporal profiles of the modes and the analogous potential. In Section~\ref{sec:2modes.second_order}, we use the analogous potential method to find multihumped multimode solitons. In Section~\ref{subsec: fund_3modes}, we extend our analysis to the fundamental three-mode soliton, and in Section~\ref{subsec: multihumped_3modes}, we show the existence of a large variety of three-mode multihumped solitons. Finally, in Section~\ref{sec:prop_fund}, we discuss the propagation of the fundamental multimode solitons, their stability, and the impact of higher-order effects such as Raman nonlinearity and self-steepening. In Section~\ref{sec:prop_mutlihump}, we qualitatively discuss the stability of the multihumped solitons.
\section{Mathematical model}\label{Chap2}
In linear propagation regime, the spatial modes propagate independently of each other. Since they have different group velocities, they separate in time involving a spread of the initial pulse. Here, in the nonlinear anomalous dispersion regime, the modes interact with each other via Kerr nonlinearity counteracting the spreading through the coupling of the modes. This leads to the formation of multimode solitons.

In Section~\ref{sec:theory}, we derive a reduced model that accepts multimode solitons as stationary solutions. We then show how to recast the resulting set of nonlinear ODEs as equations of motion of a unit-mass ball, and derive an associated potential. This potential proves to be an efficient tool to identify new families of multimode solitons, as will be shown later in Section~\ref{sec:2modes}. Finally, we discuss the validity of our model (Section~\ref{sec:hyp}) and define the parameters of the multimode fiber that will be considered for the examples (Section~\ref{fiber}).
\subsection{Equations of motion and potential} \label{sec:theory}
Our approach proceeds in three main steps. We first expand the light beam over the spatial modes of the fiber assuming an invariant transverse profile along the propagation for each mode. We then search for multimode solitons by requiring that each mode also maintain an invariant temporal profile throughout the propagation. Finally, we recast the resulting set of nonlinear ODEs as equations of motion and derive an associated potential. 

This approach has already been applied to the case of two polarization modes~\cite{haelterman_vector_1994} as well as for a two-mode fiber~\cite{pitois_generation_1999}, resulting in potentials with adjustable coefficients. The novelty of our work lies in extending this approach to three or more modes with non-symmetric coupling coefficients between modes. As will be shown later (see Section~\ref{subsec: multihumped_3modes}), this gives rise to a remarkably rich nonlinear dynamics, with many kinds of solitons that we will group in families.

When a light beam is launched into a multimode fiber, its envelope can be decomposed into a superposition of the propagation modes of the fiber:
\begin{equation}\label{defEnvcarrier}
    E(x,y,z,t)=\sum_{p}F_p(x,y)A_p(z,t),
\end{equation}
where $F_p(x,y)$ is the normalized transverse profile, and $A_p(z,t)$ is the envelope of mode $p$. The modal envelope propagation is governed by a set of coupled nonlinear equations, called the Generalized Multimode Nonlinear Schrödinger Equation (GMMNLSE). It describes how the modal envelopes evolve during propagation in a multimode fiber, taking into account both linear and nonlinear effects. The GMMNLSE is given by (for mode $p$)\cite{horak_multimode_2012}:
\begin{widetext}
    \begin{equation}\label{GMMNLSE}
        \pderiv{A_p}{z} = \Hat{D}\left\{A_p\right\}+i\frac{n_2\omega_0}{c}\left(1+\frac{i}{\omega_0}\pderiv{}{t}\right)\sum_{l,m,n}\left\{(1-f_R)S^K_{plmn}A_lA_mA^*_n+f_RS^R_{plmn}A_l\left[h\otimes(A_mA^*_n)\right]\right\},
    \end{equation}
\end{widetext}
where $n_2$ is the nonlinear refractive index, $\omega_0=2\pi c/\lambda_0$ is the carrier angular frequency, $c$ is the speed of light in vacuum, $f_R$ is the fractional contribution of the Raman nonlinearity, and $h$ is the Raman response function. $S^K_{plmn}$ and $S^R_{plmn}$ are the mode overlap factors for Kerr and Raman effects respectively. Although the Raman response function is common to all modes, the strength of the Raman nonlinearity differs from mode to mode, as a result of the Raman overlap factors. $\Hat{D}$ is the linear propagation operator given by:
\begin{equation}\label{D}
    \Hat{D}\left\{A_p\right\} = i\Delta \beta_0^{(p)}A_p-\Delta\beta_1^{(p)}\pderiv{A_p}{t}+i\sum_{k\geq 2}\frac{\beta_k^{(p)}}{k!}\left(i\pderiv{}{t}\right)^kA_p,
\end{equation}
where $\Delta \beta_0^{(p)} = \beta_0^{(p)}-\beta_0^{(1)}$ is the difference between propagation constants of mode $p$ and the fundamental mode (mode 1); $\Delta\beta_1^{(p)} = \beta_1^{(p)}-\beta_1^{(1)}$ is the group velocity mismatch between mode $p$ and the fundamental mode; and $\beta_k^{(p)}$ are the coefficients associated to dispersion of order $k$ of mode $p$. To study solitons we consider a conservative system. Consequently, propagation constants are real and therefore we do not consider propagation losses.

Our goal in this section is to derive a reduced model from the GMMNLSE that accepts stationary multimode solutions. In order to present the different steps of our analysis, we consider first the simplest possible case: two modes ($u$ and $v$), with limited interactions. In this simplest picture, the multimode soliton temporal profile results from a balance between second order chromatic dispersion, self-phase modulation and cross-phase modulation. Therefore, higher-order chromatic dispersion and self-steepening are neglected, as well as Raman nonlinearity ($f_R=0$). Their impact on the propagation of the fundamental multimode soliton will be discussed in Section~\ref{sec:prop_fund}. In addition, we also assume that modes spectra are slightly shifted from the central wavelength $\lambda_0$ such that the group velocity mismatch is negligible (see Section~\ref{sec:hyp}).

Furthermore, we assume that the beat length is much shorter than the dispersion length. Consequently, we can average the GMMNLSE over several beat lengths such that beating terms give zero contribution (refer to discussion in Section~\ref{sec:hyp}). Finally, we assume that chromatic dispersion is essentially due to material dispersion, hence we consider the same $\beta_2$ for all modes ($\beta_2^{(p)}=\beta_2,\;\forall p$). This last assumption is not strictly required but simplifies equations.

After the following normalization : $\tau=t/T_0$, $\zeta=z/L_D$ and $\tilde{a}_p=A_p/\sqrt{P_0}$; Eq.~\ref{GMMNLSE} is reduced to a set of two coupled nonlinear equations:
\begin{subequations}\label{normalized_pde}
    \begin{align}
        \pderiv{\Tilde{a}_u}{\zeta} &= i\kappa_u \Tilde{a}_u+\frac{i}{2}\ppderiv{\Tilde{a}_u}{\tau}+iN_u^2|\Tilde{a}_u|^2\Tilde{a}_u+iC|\Tilde{a}_v|^2\Tilde{a}_u,\\
        \pderiv{\Tilde{a}_v}{\zeta} &= i\kappa_v\Tilde{a}_v+\frac{i}{2}\ppderiv{\Tilde{a}_v}{\tau}+iN_v^2|\Tilde{a}_v|^2\Tilde{a}_v+iC|\Tilde{a}_u|^2\Tilde{a}_v,
    \end{align}
\end{subequations}
where $\kappa_p = L_D\Delta\beta_0^{(p)}$ is the normalized propagation constant of mode $p$, 
\begin{subequations}\label{def_N^2}
    \begin{align}
        N_p^2 &= L_DP_0n_2\omega_0S^K_{pppp}/c\\
        &=L_DP_0\gamma_{pp}
    \end{align}
\end{subequations}
is the normalized self-phase modulation coefficient of mode $p$, and 
\begin{subequations}\label{def_C}
    \begin{align}
        C &= 2L_DP_0n_2\omega_0S^K_{ppmm}/c\\
        &=2L_DP_0\gamma_{pm}\\
        &=2L_DP_0\gamma_{mp}
    \end{align}
\end{subequations}
is the normalized cross-phase modulation coefficient. The factor of 2 in $C$ arises because the cross-phase modulation term appears twice in the large sum of Eq.~\ref{GMMNLSE}, and because $S^K_{ppmm}=S^K_{pmpm}$.

As our goal is to identify solitons, we can reduce this set of coupled nonlinear partial differential equations to a set of coupled nonlinear ordinary differential equations by assuming that modal envelopes are invariant during propagation, meaning that the only dependence in the normalized propagation distance $\zeta$ is in the phase of the temporal profiles. Since we assume that modes propagate with the same group velocity, the phase does not depend on time $\tau$. Therefore, we can write the modal envelopes as:
\begin{subequations}\label{forme_sol}
    \begin{align}
            \Tilde{a}_u(\zeta,\tau) &= a_u(\tau)e^{i\left(\kappa_u\zeta+K_u\zeta\right)}, \\
            \Tilde{a}_v(\zeta,\tau) &= a_v(\tau)e^{i\left(\kappa_v\zeta+K_v\zeta\right)},
    \end{align}
\end{subequations} 
where $a_u$, $a_v\in\mathbb{R}$. $\kappa_u$ and $\kappa_v$ account for the linear part of the phase, while $K_u$ and $K_v$ are the normalized nonlinear phase accumulation rates of modes $u$ and $v$ respectively. By substituting Eq.~\ref{forme_sol} into Eq.~\ref{normalized_pde}, we obtain a set of two coupled nonlinear ordinary differential equations:
\begin{subequations}\label{Syst_ODE}
    \begin{align}
       \ppderiv{a_u}{\tau} &= 2K_u a_u-2N_u^2a_u^3-2Ca_v^2a_u,\\
        \ppderiv{a_v}{\tau} &= 2K_va_v-2N_v^2a_v^3-2Ca_u^2a_v.
    \end{align}
\end{subequations}

Solving this set of equations will provide the temporal shapes of $a_u(\tau)$ and $a_v(\tau)$ associated with solitary waves, \textit{i.e.} solutions with a fixed intensity profile in each mode. Since $K_u$ and $K_v$ are constant and represent nonlinear phase accumulation rates, they can only depend on $\int \left|a_u\right|^2d\tau$ and $\int \left|a_v\right|^2d\tau$. These quantities are proportional to the modal energies and are conserved along propagation.

This system of equations is analogous to the equations of motion for a unit-mass ball moving in a potential $V(x,y)$ within an abstract space. We associate each mode to an orthogonal direction in this abstract space by defining $x = a_u N_u/\sqrt{K_u}$, $y = a_v N_u/\sqrt{K_u}$ and $\tilde{\tau} = \sqrt{K_u}\tau$. Because the system is conservative, forces from Eq.~\ref{Syst_ODE} derive from the following potential:
\begin{equation}\label{pot2modes}
    V(x,y) = -x^2-\frac{K_v}{K_u}y^2+\frac{x^4}{2}+\frac{\gamma_{vv}}{\gamma_{uu}}\frac{y^4}{2}+2\frac{\gamma_{uv}}{\gamma_{uu}}x^2y^2,
\end{equation}
where $K_v/K_u$ is the ratio between the normalized nonlinear phase accumulation rates of the two modes. This ratio depends on the relative energies of the modes, and is a free parameter. The ratio $\gamma_{vv}/\gamma_{uu}$ is the ratio between the self-phase modulation coefficients of the two modes, and $\gamma_{uv}/\gamma_{uu}$ is the ratio between the cross-phase modulation coefficient and the self-phase modulation coefficient of mode $u$. These last ratios depend on the nature of the modes and, consequently, on the fiber parameters. 

\begin{figure}
    \begin{minipage}{0.49\linewidth}
        \begin{overpic}[width=\linewidth]{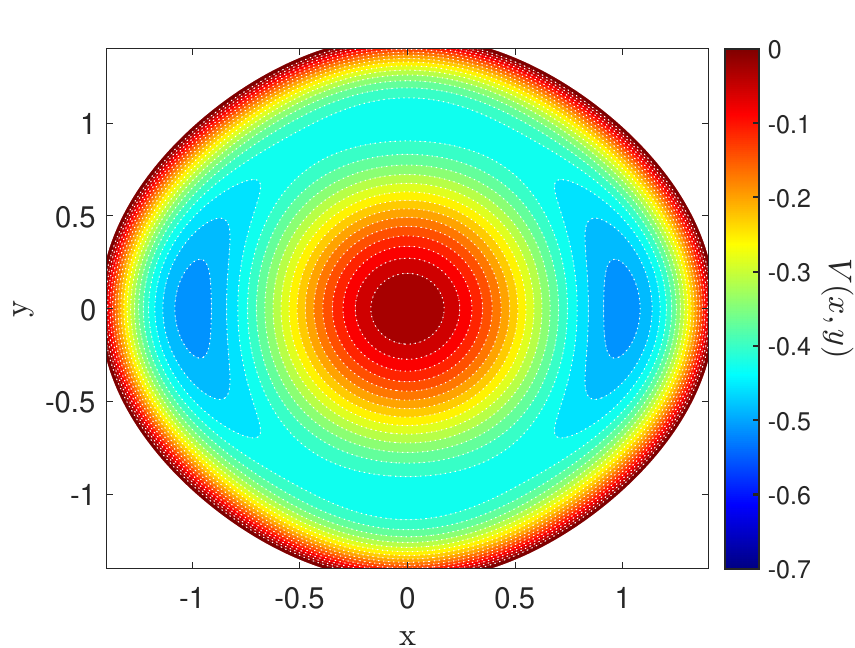}\put(0,65){(a)}
        \end{overpic}
    \end{minipage}
    \begin{minipage}{0.49\linewidth}
        \begin{overpic}[width=\linewidth]{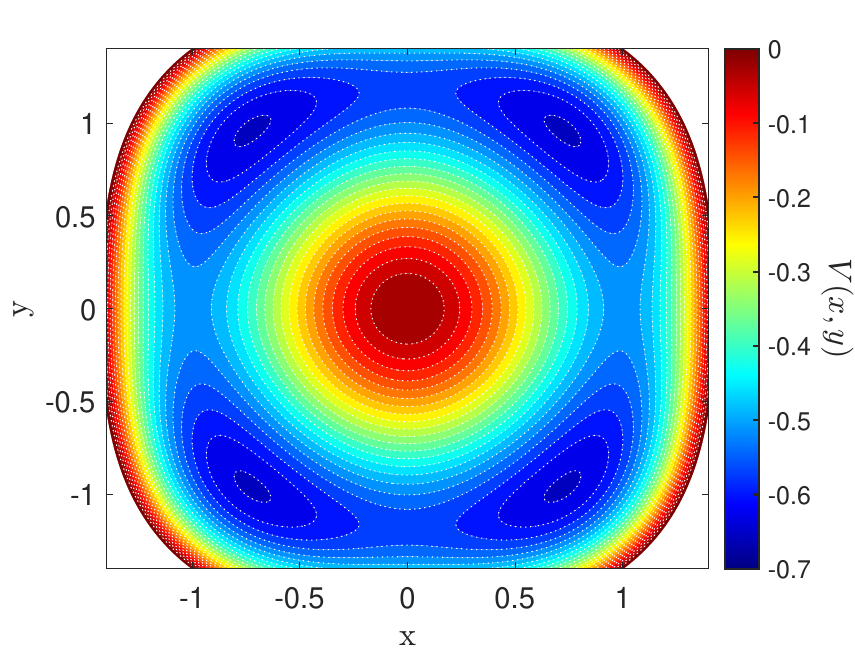}\put(0,65){(b)}
        \end{overpic}
    \end{minipage}
    \caption{a) Analoguous potential when $\gamma_{uu}\gamma_{vv}<4\gamma_{uv}^2$; b) Analoguous potential when $\gamma_{uu}\gamma_{vv}>4\gamma_{uv}^2$. For both cases : $\frac{K_v}{K_u}=0.8$.}
    \label{figure_potentials}
\end{figure}

As shown on Fig.~\ref{figure_potentials}, the potential has a maximum at $(x,y)=(0,0)$, then decreases as one moves away from the origin, and finally increases again towards infinity. There exist two distinct types of potential depending on the values of the nonlinear coefficients. If $\gamma_{uu}\gamma_{vv}<4\gamma_{uv}^2$, the minima are aligned along the cardinal axes of the potential (see Fig.~\ref{figure_potentials}.a). Their locations are fixed, and their deepness depends on the free parameter $K_v/K_u$. The other type corresponds to the case $\gamma_{uu}\gamma_{vv}>4\gamma_{uv}^2$ (see Fig.~\ref{figure_potentials}.b). In this case, the minima are distributed like the vertices of a rectangle. Depending on the value of $K_v/K_u$, the minima can be well separated or very close to each other. The main difference with the previous case is that the location of the minima depends on the free parameter $K_v/K_u$. They can be either close to the cardinal axes or close to the diagonal axes of the potential. The trajectories of the ball that start and end at the center of the potential are associated with solitary waves temporal profiles.

Since modes interact incoherently in pairs, it is straightforward to extend the reduced model to higher numbers of modes. For instance, the three-mode case is described by the following set of three coupled nonlinear ordinary differential equations:
\begin{subequations}\label{Syst_ODE_3modes}
    \begin{align}
       \ppderiv{a_u}{\tau} &= 2K_ua_u-2N_u^2a_u^3-2C_{uv}a_v^2a_u-2C_{uw}a_w^2a_u,\\
        \ppderiv{a_v}{\tau} &= 2K_va_v-2N_v^2a_v^3-2C_{vu}a_u^2a_v-2C_{vw}a_w^2a_v,\\
        \ppderiv{a_w}{\tau} &= 2K_wa_w-2N_w^2a_w^3-2C_{wu}a_u^2a_w-2C_{wv}a_v^2a_w,
    \end{align}
\end{subequations}
where $C_{ij}=C_{ji}$. Adding a mode to our reduced model corresponds to adding a new orthogonal direction in the abstract space of the ball. The potential now depends on three coordinates:
\begin{multline}\label{pot3modes}
    V(x,y,z) = -x^2-\frac{K_v}{K_u}y^2-\frac{K_w}{K_u}z^2+\frac{x^4}{2}+\frac{\gamma_{vv}}{\gamma_{uu}}\frac{y^4}{2}+\frac{\gamma_{ww}}{\gamma_{uu}}\frac{z^4}{2}\\
    +2\frac{\gamma_{uv}}{\gamma_{uu}}x^2y^2+2\frac{\gamma_{uw}}{\gamma_{uu}}x^2z^2+2\frac{\gamma_{vw}}{\gamma_{uu}}y^2z^2.
\end{multline}

Although a direct derivation of the system of equations is easily performed, we will see later that we should define carefully the physical parameters that appear in the extended model and that allow us to describe the observed dynamics.
\subsection{Extending the validity of our model}\label{sec:hyp}
Two main assumptions underlie the mathematical derivations of the previous subsection. The first concerns the null group velocity dispersion mismatch. In the anomalous dispersion regime, cross-phase modulation induces a trapping effect whereby modes tend to lock to each other, cancelling out the group velocity mismatch~\cite{pitois_velocity_2006}. The incoherent interaction between modes shifts their spectra such that their group velocities change until they propagate at the same speed. The spectrum of the fastest mode extends toward lower frequencies, while the opposite occurs for the slowest one, until the modes lock to each other. Naturally, the greater the difference in the group velocity, the stronger this frequency shift becomes. The importance of inter-modal dispersion can be quantified using the walk-off length $L_W=T_0/\overline{\Delta \beta_1}$ where $\overline{\Delta \beta_1}$ is the weighted average of the group velocity mismatch~\cite{zitelli_conditions_2021}. For a small $L_W$, meaning a large group velocity mismatch, more peak power is required to counteract it. This also implies a greater shift in the central wavelength, which can in turn affect all coefficients that depend on it. This trapping process also transforms symmetric spectra into asymmetric ones, which will be important later on (refer to discussion in Section~\ref{sec:prop_fund}). In conclusion, this reduced model remains valid for moderate group velocity mismatch.

The second assumption concerns the intermodal interaction, which is assumed to be purely incoherent. In fact, the reduced model also applies to effective incoherent nonlinear coupling. Indeed, the coherent terms are responsible for a periodic energy exchange between modes. When the associated beat length is much smaller than the other characteristic lengths, it becomes possible to average all quantities over several beat lengths. In such a case, coherent terms are cancelled out, since their contributions average to zero over one period, without significantly affecting the other terms of the equation. The definitions of $K_u$ and $K_v$ as constant quantities proportional to the energies of the modes still hold. For such a system, all the conclusions drawn from the reduced model also hold when energy exchange is taken into account. Keeping the coherent terms, however, makes the mathematics considerably more difficult as the nonlinear phase accumulation rates may vary. This assumption does not hold for degenerate modes. Indeed, for a perfectly circular fiber, the beat length would be infinite. In such a case, there is no periodic energy exchange. Furthermore, the coherent nonlinear coefficient becomes comparable to the nonlinear coefficients associated with the incoherent terms, making the coherent contribution significant. The assumption also does not hold for modes with close propagation constants. Note that this does not necessarily mean that the behavior of the multimode soliton would significantly differ from the ones studied here, but only that our model is not guaranteed to be valid. This will be checked by numerically propagating the solution with all parameters of the GMMNLSE (see Eq.~\ref{GMMNLSE} and Section~\ref{sec:prop_fund}).

\subsection{Parameters of the multimode fiber}\label{fiber}
\begin{figure}
    \begin{minipage}{0.48\linewidth}
        \begin{overpic}[width=\linewidth]{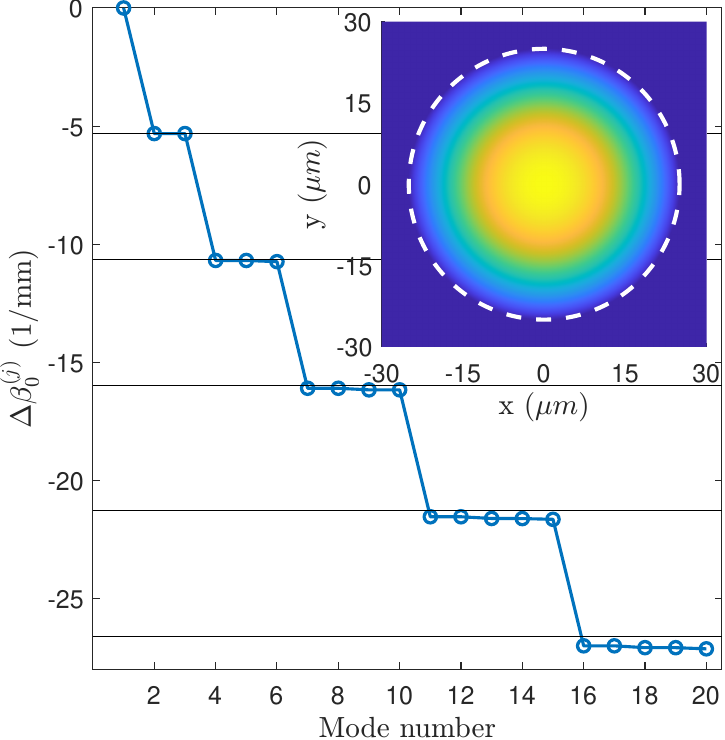}\put(0,101){(a)}
        \end{overpic}
    \end{minipage}
    \begin{minipage}{0.48\linewidth}
        \begin{overpic}[width=\linewidth]{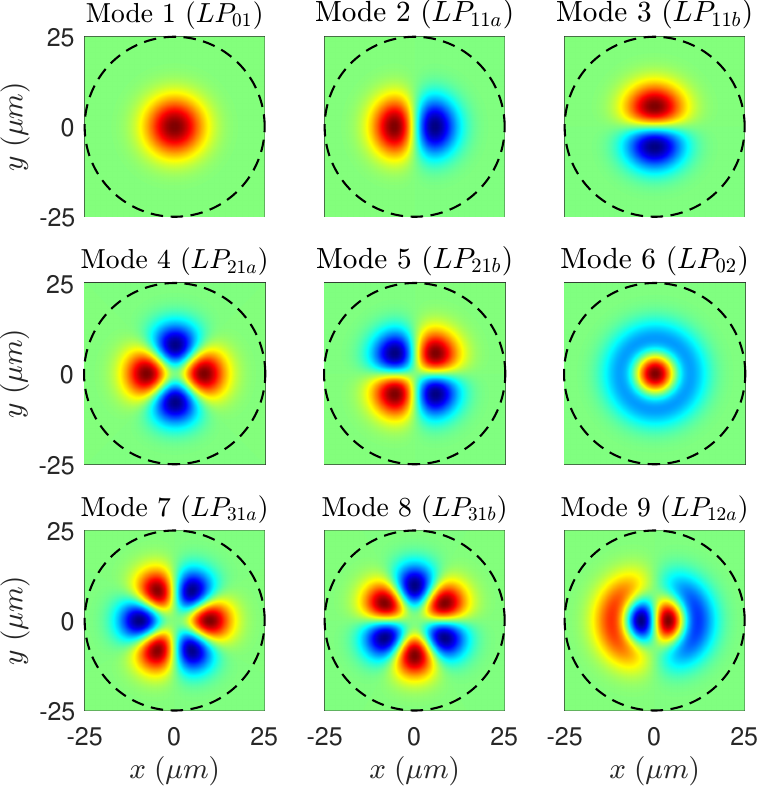}\put(0,100){(b)}
        \end{overpic}
    \end{minipage}
    \caption{a) Propagation constant distribution against mode number and refractive index profile of the multimode fiber; b) Transverse amplitude profile of the spatial propagation modes of the GRIN fiber. The dotted line indicates the core-cladding interface.}
    \label{figure_GRIN}
\end{figure}

To illustrate the general relations discussed in this paper, we need to define the modes and thus impose specific values for the parameters of the fiber. We consider a parabolic graded-index multimode fiber with a $25$ µm core radius and a refractive index difference of $0.0137$ between the core center and the cladding, corresponding to a $0.2$ numerical aperture. 
Graded-index multimode fibers minimize inter-modal dispersion and offer an interesting dispersion profile with a ladder-like structure resulting in some coefficients that are very close and others that are very different (see Fig.~\ref{figure_GRIN}.a). Modes are all supposed to be linearly polarized in the same direction such that the problem can be considered as scalar. At a central wavelength of $\lambda_0=1550$ nm, the fiber supports approximately 100 modes (without taking into account polarization). The fiber has a nonlinear refractive index $n_2=3.2\times 10^{-20} \text{m}^2/\text{W}$.

For this fiber, the shift of wavelength required to counteract the group velocity mismatch between modes $LP_{01}$ (mode 1) and $LP_{11a}$ (mode 2) is given by:
\begin{equation}
    \left|\delta \lambda \right| = \left|\frac{-\lambda_0^2}{2\pi c}\frac{\Delta \beta_1}{\beta_2^{(1)}+\beta_2^{(2)}}\right|=2.7\;\text{nm},
\end{equation}
where $\Delta \beta_1 = \beta_1^{(1)}-\beta_1^{(2)}=0.1194$ fs/mm, $\beta_2^{(1)}=-28.08$ fs$^2$/mm and $\beta_2^{(2)}=-28.22$ fs$^2$/mm are the group velocity mismatch and the second order chromatic dispersion coefficients of modes 1 and 2 respectively. As $\left|\delta \lambda/\lambda_0 \right| \ll 1$, we can neglect the impact of the central wavelength shift on the different coefficients.

Furthermore, the beat length is much shorter than the dispersion length: $L_{beat}=2\pi/\Delta\beta^{(2)}_0=1.1815$ mm $\ll L_D=T_0^2/|\beta_2|=8.9$ m for $T_0=500$ fs. Since $L_{beat} \ll L_D=L_{NL}$, we average the GMMNLSE over several beat lengths such that coherent terms give zero contribution. As already mentioned in the previous section, the beat length is not always negligible compared to the dispersion length for nondegenerate modes with close propagation constants. For instance, the beat length between modes $LP_{21a}$ (mode 4) and $LP_{02}$ (mode 6) is 12.15 cm, and that between modes $LP_{22b}$ (mode 14) and $LP_{03}$ (mode 15) is 20 cm. In such cases, if the soliton formed has a duration on the order of 100 fs, the different characteristic lengths have the same order of magnitude.

The assumption of identical $\beta_2$ coefficients for all modes is valid in our case, as the biggest relative difference is between modes $\text{LP}_{61}$ (mode 23) and $\text{LP}_{04}$ (mode 28): $2\left|\frac{\beta_2^{(23)}-\beta_2^{(28)}}{\beta_2^{(23)}+\beta_2^{(28)}}\right|= 3.78\%$, where $\beta_2^{(23)}=-28.87\;\text{fs}^2/\text{mm}$ and $\beta_2^{(28)}=-27.80\;\text{fs}^2/\text{mm}$.

\section{Two-mode soliton}\label{sec:2modes}
As elliptical solitons are made of two polarization modes, and as they are our reference phenomenon, we begin our study with the case of a soliton composed of two spatial propagation modes. We investigate the impact of different pairs of modes on the soliton dynamics, and hence of the different sets of nonlinear coefficients they entail. This is indeed an important difference between nondegenerate spatial modes and polarization modes, as the latter are symmetrically coupled, which limits the parameter space. To determine the stationary temporal profiles of the modes, we numerically solve Eq.~\ref{Syst_ODE} using the relaxation method~\cite{press_numerical_2007}.

Section~\ref{subsec:2modes.fundamental} presents the fundamental two-mode soliton, its characteristics, and how its temporal profile can be related to the motion of a unit-mass ball in the analogous potential. Section~\ref{sec:2modes.second_order} presents multimode solitons exhibiting nodes in the temporal profile of one of its modes, which we refer to as multihumped solitons.
\subsection{The fundamental two-mode soliton}\label{subsec:2modes.fundamental}
\begin{figure}
    \begin{minipage}{0.49\linewidth}
        \begin{overpic}[width=\linewidth]{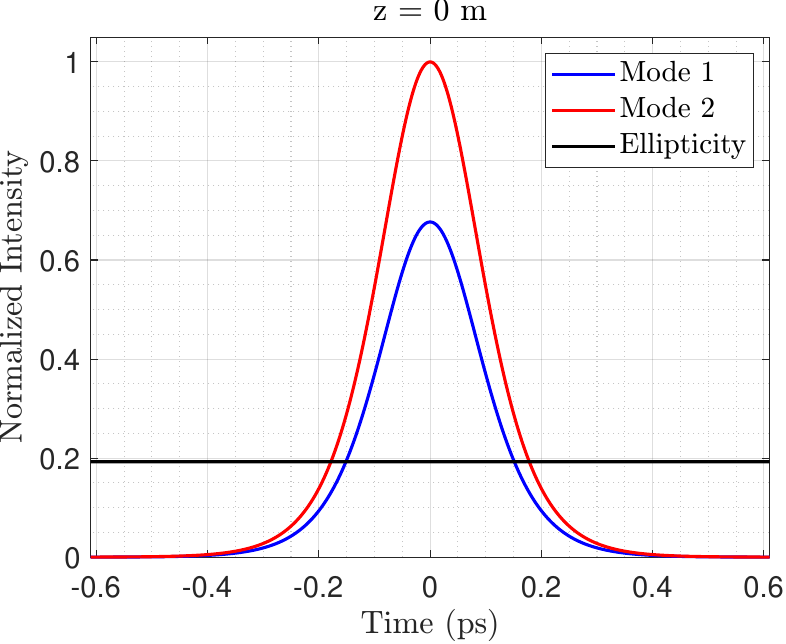}\put(0,80){(a)}
        \end{overpic}
    \end{minipage}
        \begin{minipage}{0.49\linewidth}
        \begin{overpic}[width=\linewidth]{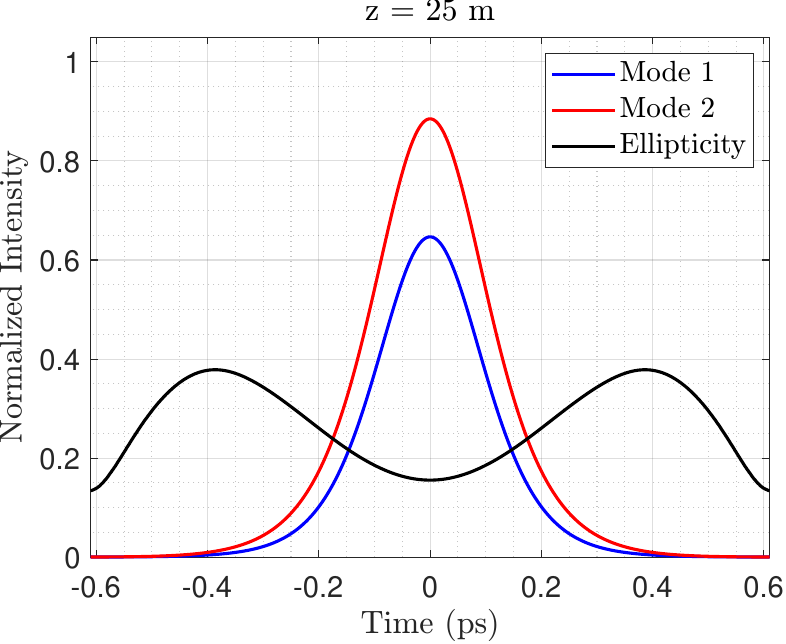}\put(0,80){(b)}
        \end{overpic}
    \end{minipage}
    \caption{a) Temporal profiles of modes 1 ($LP_{01}$) and 2 ($LP_{11a}$) at the beginning of the propagation. Modes are imposed to have the same duration giving a flat ellipticity; b) Temporal profiles of modes 1 and 2 after 25 m of propagation.}
    \label{fig:equ_ellipt}
\end{figure}

The solution for a soliton composed of two spatial modes is shown in Fig.~\ref{fig:equ_ellipt}. In this example, the modes involved are the $LP_{01}$ (mode 1) and $LP_{11a}$ modes (mode 2). Although the two temporal profiles have very similar shapes, they differ in amplitude and temporal width. The same behavior occurs for the elliptical soliton, whose polarization modes also exhibit a difference in temporal evolution, quantified there by a parameter called the ellipticity:
\begin{equation}\label{ellipticity}
    q(\tau) = \frac{|a_v(\tau)|^2-|a_u(\tau)|^2}{|a_u(\tau)|^2+|a_v(\tau)|^2}.
\end{equation}

Note that the ellipticity can also be defined using amplitudes rather than intensities. This quantity is defined in the circular polarization basis. In the case of a time-independent ellipticity, $q=\pm 1$ corresponds to a circular polarization state while a linear polarization state is given by $q=0$. A time-varying ellipticity means that the polarization state of the elliptical soliton varies across the pulse.

The same definition can be used for a soliton composed of spatial propagation modes, although it then loses its meaning as a description of the vectorial evolution of the field across the pulse. Nonetheless, a time-varying ellipticity still indicates that the soliton is intrinsically multimode. Indeed, since the weight of each mode in the modal composition varies from one time point to another across the pulse, it is impossible to find a basis in which the soliton can be described by a single mode at every time point of the pulse.

The ellipticity is also a measure of the equilibrium that defines the multimode soliton. When the temporal profiles of the modes are perturbed, they change their shape until they reach a new equilibrium, as illustrated in Fig.~\ref{fig:equ_ellipt}. Such a soliton has a symmetric, positively curved ellipticity. At the beginning of the propagation, we impose both modes to have the same temporal profile evolution with the only difference being the maximal value. Hence, we impose a constant ellipticity across the pulse (Fig.~\ref{fig:equ_ellipt}.a). During propagation, the temporal profiles of the modes adapt themselves until the ellipticity becomes positively curved again (Fig.~\ref{fig:equ_ellipt}.b). This shows that this multimode structure returns to an equilibrium shape when the temporal profiles are perturbed. We call such a structure a multimode soliton, as the temporal profiles of the modes are invariant during propagation, and stable against noise. Note that it is only the temporal profile of the modes constituting the soliton that is invariant. The total temporal profile of the pulse, as well as its spatial profile, evolve periodically due to beating between the nondegenerate modes.

Crucially, solitonic properties emerge from the interaction between both modes. If one of the modes is cut during propagation, the other will temporally broaden. If it has sufficient peak power, it finds a new equilibrium shape. Otherwise, the pulse becomes a dispersive wave. Since the temporal behavior of one mode is no longer determined solely by that mode, but also by the other mode, the standard definition of the soliton order is no longer valid from the modal point of view. Indeed, when we calculate the order of each mode, we obtain a value lower than one. As an example, for the case presented in Fig.~\ref{fig:equ_ellipt}.b, $\mathcal{N}^2_1=0.41$ and $\mathcal{N}^2_2=0.52$, where the calligraphic N is used to refer to the single-mode definition of the soliton order.

In order to find analytical expressions for relevant quantities characterizing the multimode soliton, as well as their range, we distinguish two cases. The first one, that we call the \textit{equivalence} regime, occurs when both modes have comparable peak powers, such as in Fig.~\ref{fig:equ_ellipt}. In such a case, both modes optically guide each other. Even though their soliton numbers are less than unity, the temporal profiles of the modes can be fitted with sech functions, which suggests that we can take the formalism for single-mode solitons and extend it to the multimode case. Based on this observation, we used Lagrangian formalism with sech functions as ansatz~\citep{anderson_variational_1983,sun_multimode_2024} (see Appendix~\ref{Anx:Lagrange}). This formalism allows us to establish relations between the different quantities characterizing the multimode soliton. The expression for the nonlinear phase accumulation rate is therefore given by:
\begin{equation}\label{K}
    K_u = \frac{1}{2}N_u^2P_u+\frac{1}{4}CP_vG\left(\frac{T_v}{T_u}\right),
\end{equation}
where $P_u$ and $P_v$ are the peak powers of the modes, and $G\left(T_v/T_u\right)$ is a monotonically increasing function resulting from the interaction between two sech functions with different widths in the Lagrangian approach (see Appendix~\ref{Anx:Lagrange} and~\ref{Anx:Barnes}). Note also that $G(1)=2$.

The ansatz describes well temporal profiles of the modes when they have comparable peak powers but loses accuracy for cases with significant peak power differences, as the temporal profile of the weakest mode deviates from a sech function. In order to find the range of the relevant quantities, we consider the limit cases of unilateral optical guiding, i.e. $P_u \ll P_v$ or $P_u\gg P_v$. We refer to these limit cases as the \textit{asymptotic} regime. In this regime, it is possible to find an analytical expression for $K_u$ and $K_v$. Indeed, starting from a single-mode soliton and using perturbation leads to a bifurcation towards a multimode soliton (see Appendix~\ref{Anx:Legendre}). In the following, we will combine the analytical results from the asymptotic and equivalence regimes to characterize the fundamental two-mode soliton.\\

\textbf{Range of the soliton parameters}\\

As explained in Section~\ref{sec:theory}, the ratio between the nonlinear phase accumulation rates is the free parameter characterizing the analogous potential, and thus the multimode soliton. This ratio depends on the energy of both modes. This quantity can be evaluated in the equivalence regime using the Lagrangian method (see Eq.~\ref{K}). Using Eq.~\ref{def_N^2} and Eq.~\ref{def_C} gives:
\begin{equation}\label{frac_K}
    \frac{K_v}{K_u}=\frac{\gamma_{vv}P_v+\gamma_{vu}P_uG\left(T_u/T_v\right)}{\gamma_{uu}P_u+\gamma_{uv}P_vG\left(T_v/T_u\right)}.
\end{equation}

The range of values of the free parameter is given by the limit cases where one mode has a negligible peak power compared to the other. In this asymptotic regime, the free parameter takes the form of Eq.~\ref{def_n_i}:
\begin{equation}\label{bornes_limite}
    \frac{K_j}{K_i} = \frac{1}{4}\left(\sqrt{1+16\frac{\gamma_{ij}}{\gamma_{ii}}}-1\right)^2=n_{ij}^2,
\end{equation}
where $i$ is the guiding mode and $j$ is the guided one ($P_j\ll P_i$). In such a case, while the temporal profile of mode $i$ is given by a $\sech$ function, the one of mode $j$ is given by a $\sech^{n_{ij}}$ function. Hence, our Lagrangian approach loses accuracy when we are close to the asymptotic regime. The free parameter $K_v/K_u$ describing the two-mode soliton is thus bounded between $n_{vu}^{-2}$ ($P_v \gg P_u$) and $n_{uv}^2$ ($P_v \ll P_u$). Note that for polarization modes, $\gamma_{uu}=\gamma_{vv}=\gamma_{uv}$, so $n_{uv}=n_{vu}=n$ and the free parameter characterizing the elliptical soliton is given by $\frac{K_v}{K_u}\in\left[n^{-2},n^2\right]$.

Having an analytical expression for the nonlinear phase accumulation rate, in the equivalence regime, we define the nonlinear lengths associated to mode $u$ as follows :
\begin{subequations}\label{L_NL}
    \begin{align}
        L_{\text{NL}}^{(u)}&=\frac{1}{\gamma_{uu}P_u+\gamma_{uv}P_vG(T_v/T_u)}\\
        &=\left(\frac{1}{L_{\text{SPM}}^{(u)}}+\frac{1}{L_{\text{XPM}}^{(u)}}\right)^{-1}.
    \end{align}
\end{subequations}

Consequently, the order of the multimode soliton can be expressed as:
\begin{subequations}\label{Nouvel_ordre}
    \begin{align}
    \tilde{N}_u^2 &= \frac{L_D}{L_{\text{NL}}}=\frac{T_u^2}{|\beta_2^{(u)}|}\left(\gamma_{uu}P_u+\gamma_{uv}P_vG\left(T_v/T_u\right)\right)\\
    &=\mathcal{N}_u^2\left(1+\frac{\gamma_{uv}}{\gamma_{uu}}\frac{P_v}{P_u}G\left(T_v/T_u\right)\right).
      \end{align}
\end{subequations}

The new definition updates the single-mode soliton order definition by multiplying it with a factor accounting for the ratio between cross-phase modulation coefficient and self-phase modulation coefficient, the peak power ratio between modes, and a function that depends on the ratio between the widths of the temporal profiles of the modes. The peak power ratio implies that the mode with the highest peak power is less impacted by the other mode. The ratio between nonlinear coefficients shows that the mode with the highest self-phase modulation coefficient receives a smaller correction to its soliton order. Using the definition of Eq.~\ref{Nouvel_ordre} for the case presented in Fig.~\ref{fig:equ_ellipt}.b, we get $\tilde{N}_1\approx\tilde{N}_2\approx 1$.\\
\begin{figure}
    \includegraphics[width=\linewidth]{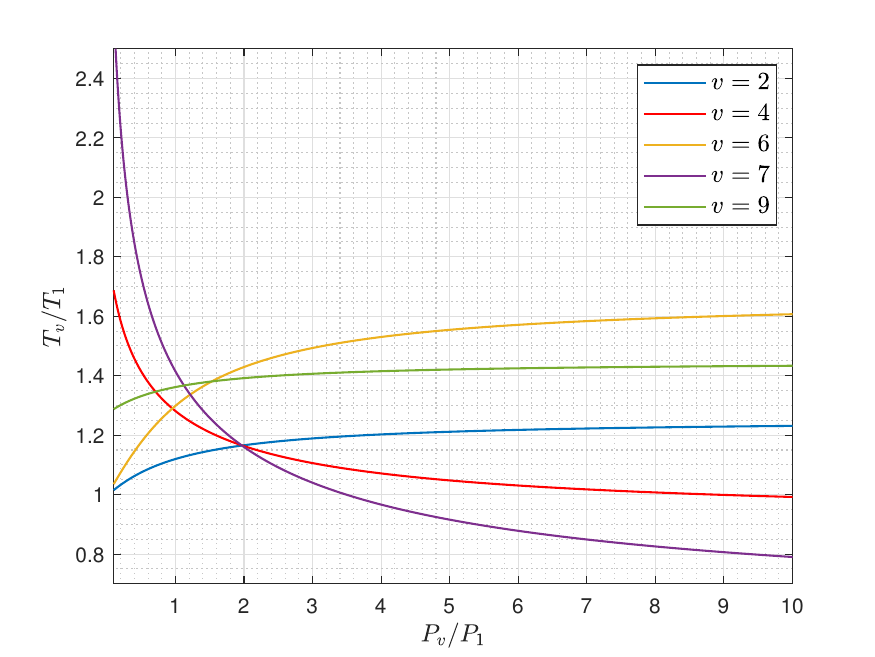}
    \caption{Duration ratio against peak power ratio. Ratios are taken between mode $v$ and the fundamental mode. $\gamma_{11}\gamma_{vv}<4\gamma_{1v}^2$ for $v=2,6,9$, and $\gamma_{11}\gamma_{vv}>4\gamma_{1v}^2$ for $v=4,7$.}
    \label{rapport_temps_vs_rapport_puissance}
\end{figure}

\textbf{Range of the duration ratio}\\

We can exploit the equality between the soliton orders, defined in Eq.~\ref{Nouvel_ordre}, to find the ratio between the durations of the modes. For multimode solitons of unitary order, the duration ratio between modes can be expressed as follows:
\begin{equation}\label{frac_T}
    \frac{T_v}{T_u}=\sqrt{\frac{|\beta_2^{(v)}|}{|\beta_2^{(u)}|}}\cdot\sqrt{\frac{\gamma_{uu}P_u+\gamma_{uv}P_vG\left(T_v/T_u\right)}{\gamma_{vv}P_v+\gamma_{vu}P_uG\left(T_u/T_v\right)}},
\end{equation}
where $\gamma_{uv}=\gamma_{vu} $. We see that, in general, this ratio is not equal to one, implying a time-varying ellipticity, as illustrated in Fig.~\ref{rapport_temps_vs_rapport_puissance}. It shows the duration ratio against the peak power ratio for a multimode soliton composed of the fundamental mode ($u=1$) and a higher-order mode. The figure was obtained by computing $P_v/P_1$ as a function of $T_v/T_1$, and then inverting numerically the function. It is worth noting that, as stated in Section~\ref{sec:theory}, the $\beta_2$ coefficients are approximately the same for all modes. In other words, if material dispersion is dominant over others chromatic dispersion mechanisms, then $\sqrt{\frac{|\beta_2^{(v)}|}{|\beta_2^{(u)}|}}\approx 1$. 

Comparing Eq.~\ref{frac_T} with Eq.~\ref{frac_K}, we see a direct link between these two quantities:
\begin{equation}
    \frac{K_v}{K_u}=\frac{|\beta_2^{(v)}|}{|\beta_2^{(u)}|}\left(\frac{T_v}{T_u}\right)^{-2}.
\end{equation}

Eq.~\ref{frac_T} shows that the duration ratio is bounded between two limits that are given by the asymptotic regime. The first limit is when $P_v\ll P_u$, which gives $\frac{T_v}{T_u}=\sqrt{\frac{|\beta_2^{(v)}|}{|\beta_2^{(u)}|}}\frac{2}{\sqrt{1+16\frac{\gamma_{uv}}{\gamma_{uu}}}-1}$. The second limit is when $P_v\gg P_u$, which gives $\frac{T_v}{T_u}=\frac{1}{2}\sqrt{\frac{|\beta_2^{(v)}|}{|\beta_2^{(u)}|}}\left(\sqrt{1+16\frac{\gamma_{uv}}{\gamma_{vv}}}-1\right)$. The finite range of values for the duration ratio is illustrated in Fig.~\ref{rapport_temps_vs_rapport_puissance}, where the curves saturate for large peak power ratios. Because Fig.~\ref{rapport_temps_vs_rapport_puissance} was made with the expressions from the Lagrangian formalism (equivalence regime), the graph loses in accuracy at the edges (asymptotic regime). The temporal durations of the modes are directly constrained by their spatial overlap, which governs the strength of their mutual interaction. For all pairs of modes that we tested, we observe that the duration ratio decreases with the peak power ratio when $\gamma_{uu}\gamma_{vv}>4\gamma_{uv}^2$. In Fig.~\ref{rapport_temps_vs_rapport_puissance}, this behavior can be observed on curves $v=4$, and $7$. Hence, the duration of a mode decreases when its peak power increases, as for the single-mode soliton case. On the contrary, when $\gamma_{uu}\gamma_{vv}<4\gamma_{uv}^2$, the duration ratio increases with the peak power ratio (see curves $v=2$, 6, and 9). In addition, it is worth noting that the curves cross the line $T_v/T_u=1$ if the following inequality is satisfied: $(2\gamma_{uv}-\gamma_{uu})(2\gamma_{uv}-\gamma_{vv}) \geq 0$. When equality holds, the crossing is trivial: it occurs at $P_v/P_u=0 $ if $2\gamma_{uv}=\gamma_{uu} $, or at $P_v/P_u=\infty$ if $2\gamma_{uv}=\gamma_{vv}$. In Fig.~\ref{rapport_temps_vs_rapport_puissance}, the crossing is approximately trivial for the curves $v=2$ and $v=6$, since $2\gamma_{12}\approx 2\gamma_{16}\approx \gamma_{11}$. In contrast, the curves $v=4$ and $v=7$ cross the line at the nontrivial points $P_4/P_1=9$ and $P_7/P_1=3.5$, respectively.\\

\textbf{Trajectories in the potentials}\\

\begin{figure}
    \begin{minipage}{0.48\linewidth}
        \begin{overpic}[width=\linewidth]{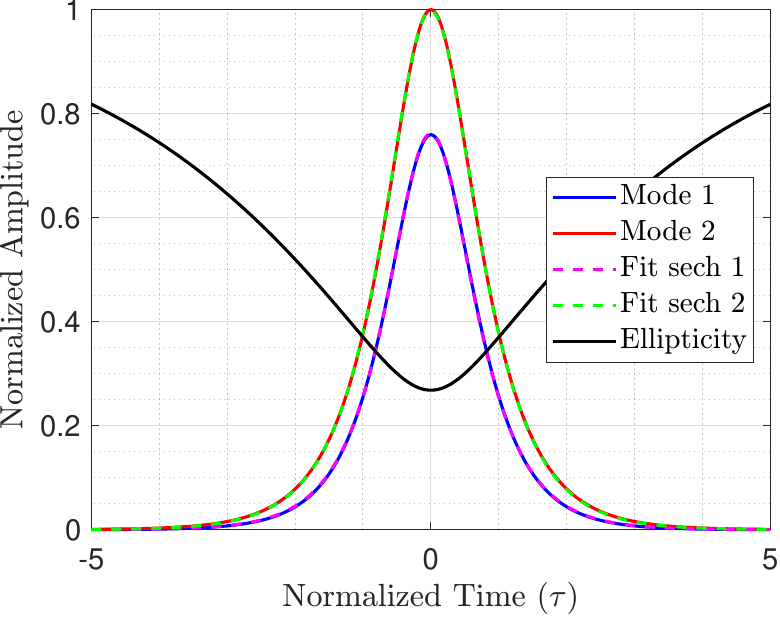}\put(-5,80){(a)}
        \end{overpic}
    \end{minipage}
    \begin{minipage}{0.48\linewidth}
        \begin{overpic}[width=1.1\linewidth]{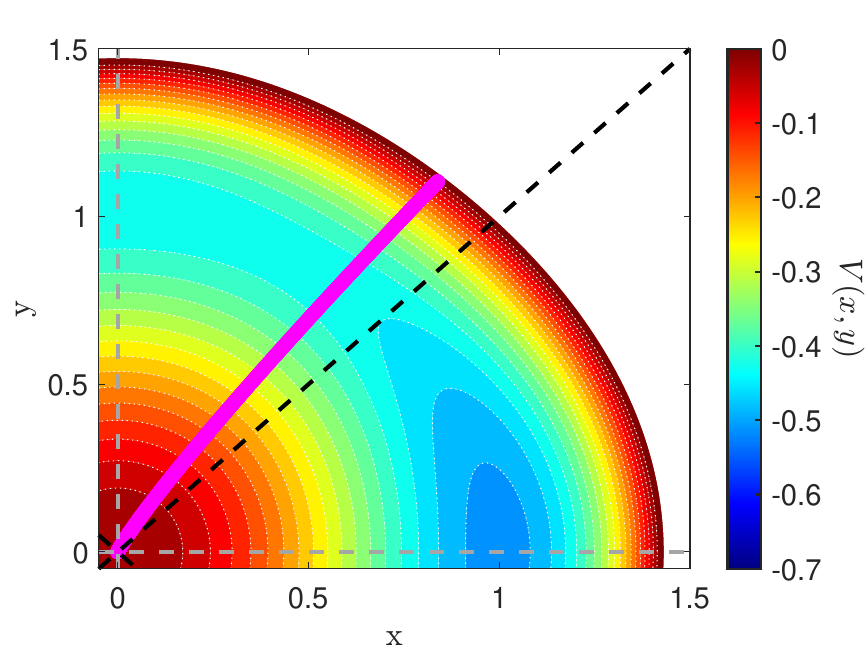}\put(0,74){(b)}
        \end{overpic}
    \end{minipage}
    \begin{minipage}{0.48\linewidth}
        \begin{overpic}[width=\linewidth]{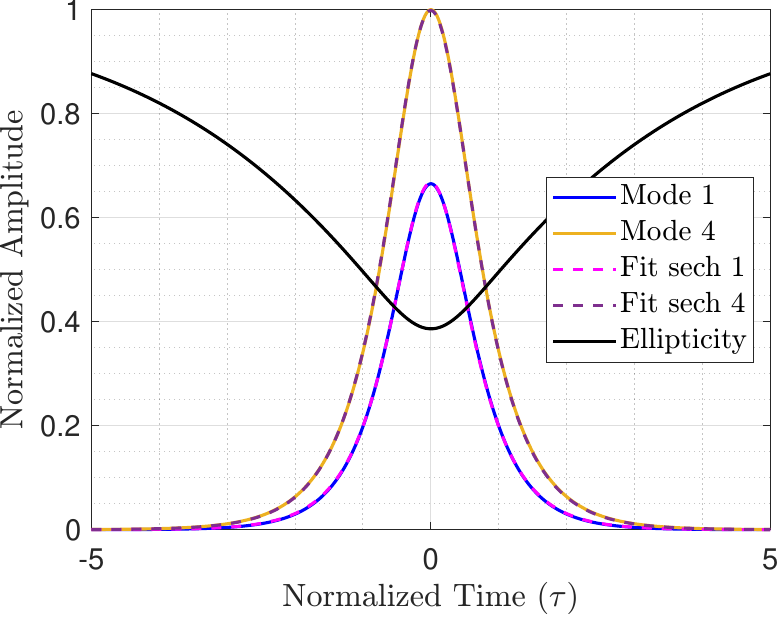}\put(-5,80){(c)}
        \end{overpic}
    \end{minipage}
    \begin{minipage}{0.48\linewidth}
        \begin{overpic}[width=1.1\linewidth]{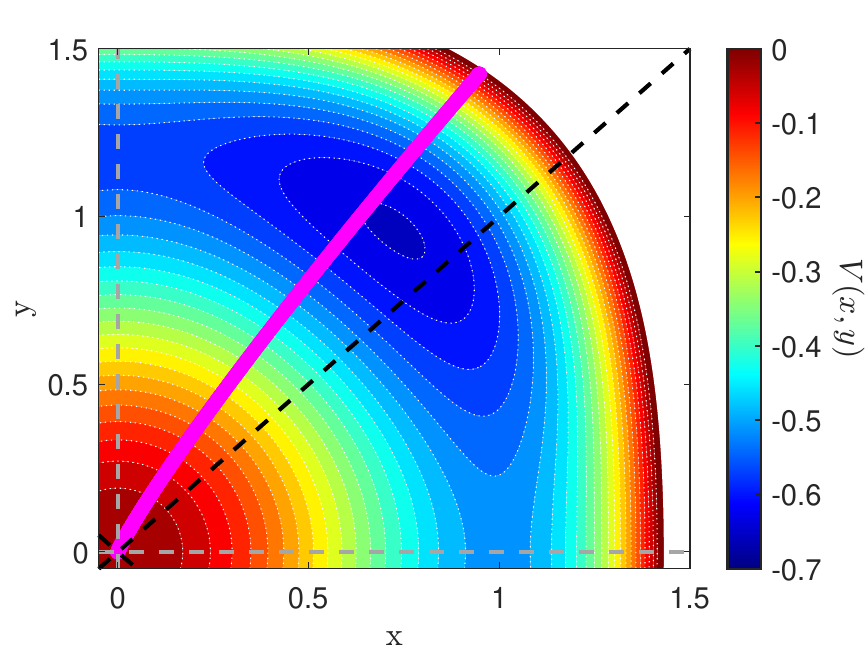}\put(0,74){(d)}
        \end{overpic}
    \end{minipage}
    \caption{a, c) Normalized temporal profiles of the modes forming a multimode soliton and the associated ellipticity. Each profile is fitted with a sech function; b, d) Potential associated to the multimode soliton presented in (a, c): $\frac{K_v}{K_u}=0.8$. Magenta points represent the trajectory of the unit-mass ball in the potential. Gray dotted lines are trajectories associated to a single-mode soliton in that basis. Black dotted line is the trajectory associated to the case where both modes have exactly the same temporal profile.}
    \label{fig:fond_MMS_potential}
\end{figure}

Our interest in this paper is about intrinsically multimode solitons, \textit{i.e.} when the ellipticity varies in time. They appear when the trajectory of the ball in the analogous potential is curved. As we have identified the range of possible values for the ratios of K's, we can analyse the various shapes of the potential in this range and the associated trajectories. Eq.~\ref{K} was validated by checking that the trajectory of the ball in the analogous potential, defined by Eq.~\ref{frac_K}, indeed implies a conservative motion. Fig.~\ref{fig:fond_MMS_potential} shows the link between the stationary solution of Eq.~\ref{Syst_ODE} and the trajectory of the unit-mass ball in the associated analogous potential defined by Eq.~\ref{pot2modes}. The first panel of Fig.~\ref{fig:fond_MMS_potential} corresponds to the case where $\gamma_{uu}\gamma_{vv}<4\gamma_{uv}^2$, while the second one corresponds to $\gamma_{uu}\gamma_{vv}>4\gamma_{uv}^2$.

As already explained in Section~\ref{sec:theory}, the main difference between the two kinds of potentials, shown in Fig.~\ref{fig:fond_MMS_potential}.b and Fig.~\ref{fig:fond_MMS_potential}.d, lies in the location of the minima. The gray dotted lines are trajectories associated with a single-mode soliton solution, while the black dotted lines indicate cases where the modes have the exact same temporal profile. These are the limit cases that can be described with a single-mode equation. For $P_v/P_u\rightarrow 0 $, the trajectory tends toward a straight line along the $x$-axis, which corresponds to a single-mode soliton in mode $u$. For $P_v/P_u\rightarrow \infty $, the trajectory tends toward a straight line along the $y$-axis, which corresponds to a single-mode soliton in mode $v$. In both cases, the ellipticity tends to become constant across the pulse, which is consistent with our previous analysis.

The most interesting solutions are the intermediate cases, where the trajectory is curved, as the one shown in Fig.~\ref{fig:fond_MMS_potential}.a. This multimode soliton is composed of modes 1 and 2, and indeed has a time-varying ellipticity. Fig.~\ref{fig:fond_MMS_potential}.b shows the trajectory of the ball (in magenta) in the associated analogous potential. The $x$-direction is associated with mode 1, and the $y$-direction corresponds to mode 2. The trajectory is homoclinic. The ball starts at the center of the potential ($V(0,0)=0$) and gains speed as it moves down the potential, reaching maximum speed at the minimum along that trajectory, which is near the saddle point of the potential. The ball then climbs up the potential and gradually loses its kinetic energy until it reaches the point where the potential is again equal to zero (the system is conservative, as stated in Section~\ref{sec:theory}). It then moves back in the other direction and finally reaches the center of the potential after an infinite time. Parametrizing $x$- and $y$-coordinates against the normalized time gives the physical invariant temporal profiles constituting the multimode soliton. This single-branch round-trip indicates that the two modes have symmetric temporal profiles. The homoclinic trajectory starting at the center of the potential shows that modal temporal profiles have a finite duration. It is worth noting that the stability of the trajectory of the ball in the analogous potential is uncorrelated to the stability of the physical solution.

When the soliton is only composed of mode 2, the trajectory follows the vertical gray dotted line. In the limit case where mode 1 has energy but a much smaller peak power ($K_2/K_1=0.68$), then the trajectory is still close to a straight line (because of the big difference in peak powers, see Eq.~\ref{ellipticity}) but it is tilted towards the $x$-direction. When $K_2/K_1$ increases, the trajectory becomes more and more tilted, and also becomes curved (an example is shown in Fig.~\ref{fig:fond_MMS_potential}.b). Then for $K_2/K_1$ close to the upper limit, the trajectory becomes a straight line again because mode 2 has not a sufficient peak power to impact the ellipticity. Finally, at the upper limit of $K_2/K_1$, the trajectory is a straight line close to the $x$-direction but slightly tilted towards the $y$-direction. The curvature of the trajectory does not invert while varying $K_2/K_1$ because the only case where modes have the same temporal width is when $K_2/K_1=1.01$, which is its upper limit (when mode 1 is dominant).

The second panel of Fig.~\ref{fig:fond_MMS_potential} corresponds to a multimode soliton composed of mode 1 and mode 4. Fig.~\ref{fig:fond_MMS_potential}.c shows that the dip in ellipticity is less pronounced. The trajectory in the analogous potential is again homoclinic, but less curved, as shown in Fig.~\ref{fig:fond_MMS_potential}.d. Unlike the previous case, the trajectory goes near a minimum of the potential. When the free parameter equals its lower limit ($K_4/K_1=0.4$), the trajectory is a straight line close to the $x$-direction but slightly tilted towards the $y$-direction meaning that the fundamental mode is the dominant one. When the free parameter increases, the trajectory becomes curved (as shown in Fig.~\ref{fig:fond_MMS_potential}.d) and the angle between the $x$-direction and the trajectory increases. When $K_4/K_1=1$, the angle from the $x$-axis is greater than 45°, but the trajectory is perfectly straight as $T_4/T_1=1$. This is different from the elliptical soliton case, where the trajectory is straight at exactly 45° from the horizontal axis. Then by increasing again the free parameter (the upper limit is given by $K_4/K_1=1.17$), the trajectory curves again, but in the other direction. 

The single-branch homoclinic trajectory can lie in any quadrant of the potential. A trajectory in the second or fourth quadrant corresponds to a $\pi$-phase shift of one of the modes, while a trajectory in the third quadrant corresponds to a $\pi$-phase shift of both modes. In all cases, the temporal profiles are symmetric and have a finite duration. In Fig.~\ref{fig:fond_MMS_potential}.b and Fig.~\ref{fig:fond_MMS_potential}.d, the trajectory lies in the upper part of the first quadrant, \textit{i.e.}, above the diagonal. A trajectory in this region means that the temporal profile of the mode associated with the $y$-direction is always higher than that of the other mode. Hence, the temporal profile of the mode associated with the $x$-direction is always contained within the other temporal profile. This corresponds to a positive ellipticity across the entire pulse. If the trajectory lies in the lower part of the first quadrant (below the diagonal), the opposite holds. It is also possible to find a configuration in which the trajectory crosses the diagonal of the potential. In such a case, the temporal profiles of the modes cross each other, and the ellipticity changes sign across the pulse.

One final important point concerns the relative orientation of the spatial profiles of the modes constituting the multimode soliton. For instance, modes 2 and 9 share the same axis of symmetry (see Fig.~\ref{figure_GRIN}.b). Hence, when the modes are aligned, their overlap is maximal, so that $\gamma_{22}\gamma_{99}<4\gamma_{29}^2$, resulting in an analogous potential with the same shape as in Fig.~\ref{figure_potentials}.a. Now if we consider the pair of modes 3 and 9, even though mode 3 has the same transverse profile as mode 2 (see Fig.~\ref{figure_GRIN}.b), they have a minimum overlap since mode 3 is rotated by 90° with respect to mode 9. In this case, we have $\gamma_{33}\gamma_{99}>4\gamma_{39}^2$ and, consequently, an analogous potential with the same shape as in Fig.~\ref{figure_potentials}.b. As these examples illustrate, both the shape of the spatial profile and its relative orientation are crucial in determining the properties of the multimode soliton. In addition, the discrete rotational symmetry of the spatial amplitude profiles is not the same as that of the spatial intensity profiles. Consequently, the transverse spatial beating does not share the same symmetry as the incoherent modal interaction, so two multimode solitons that are identical in time can differ in space. For instance, mode 4 exhibits a 90° rotational symmetry in intensity, but only a 180° rotational symmetry in amplitude (see Fig.~\ref{figure_GRIN}.b). As a result, it interacts with mode 2 in the same way as mode 3 does, even though the transverse spatial beating differs between the two cases.

\subsection{Multihumped two-mode soliton}\label{sec:2modes.second_order}
A main advantage of working with the analogous potential is that it allows to identify different kinds of stationary solutions without solving explicitly the complex set of nonlinear equations such as Eq.~\ref{Syst_ODE}, but by looking for specific trajectories of the ball in that potential. For instance, this method allowed to identify domain-wall solitons in normal dispersion regime~\cite{haelterman_polarization_1994}.
\begin{figure}
    \begin{minipage}{0.48\linewidth}
        \begin{overpic}[width=\linewidth]{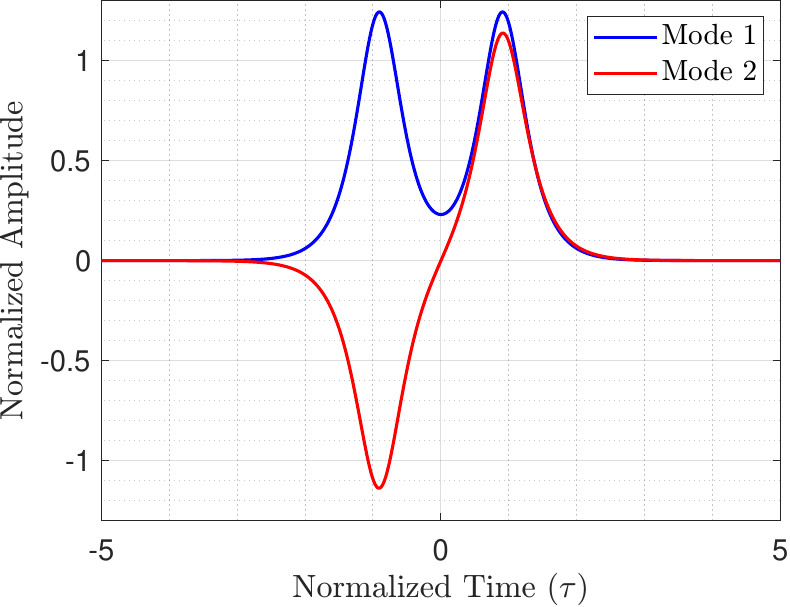}\put(0,77.5){(a)}
        \end{overpic}
    \end{minipage}
    \begin{minipage}{0.48\linewidth}
        \begin{overpic}[width=1.05\linewidth]{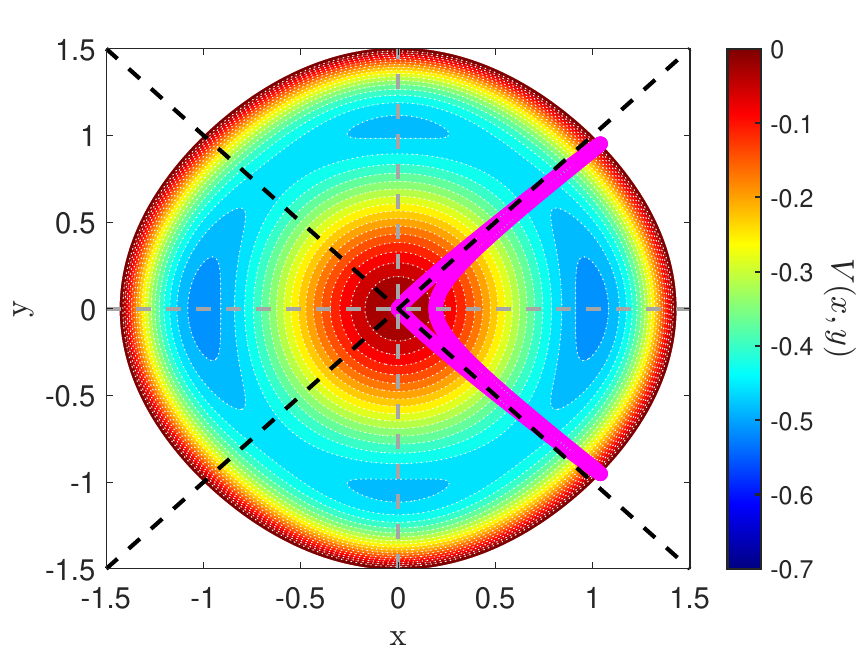}\put(0,75){(b)}
        \end{overpic}
    \end{minipage}
    \begin{minipage}{0.48\linewidth}
        \begin{overpic}[width=\linewidth]{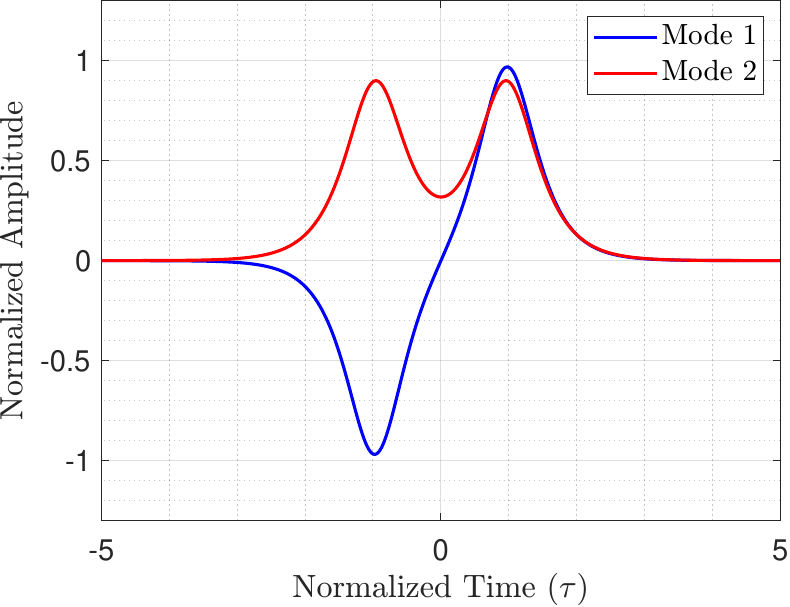}\put(0,77.5){(c)}
        \end{overpic}
    \end{minipage}
    \begin{minipage}{0.48\linewidth}
        \begin{overpic}[width=1.05\linewidth]{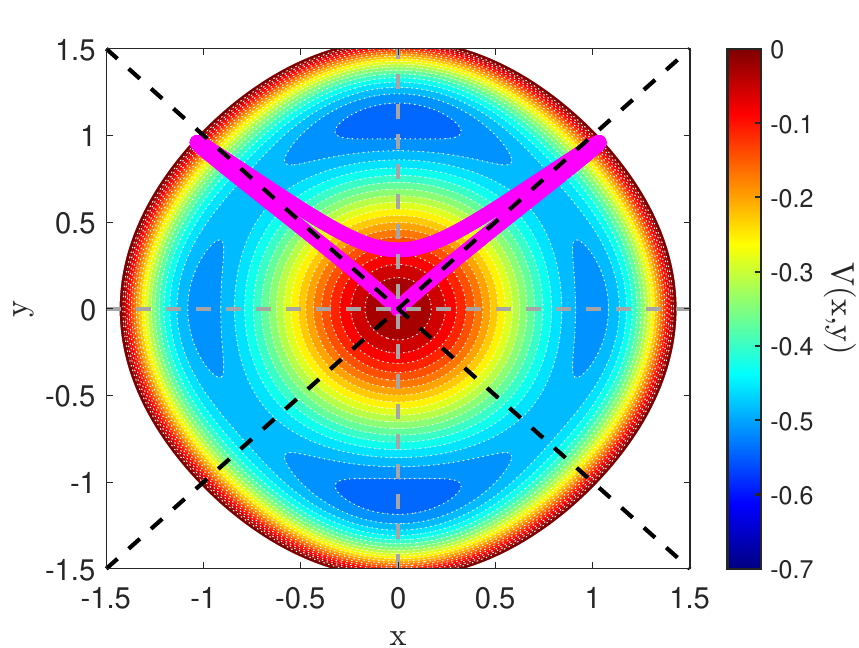}\put(0,75){(d)}
        \end{overpic}
    \end{minipage}
    \begin{minipage}{0.48\linewidth}
        \begin{overpic}[width=\linewidth]{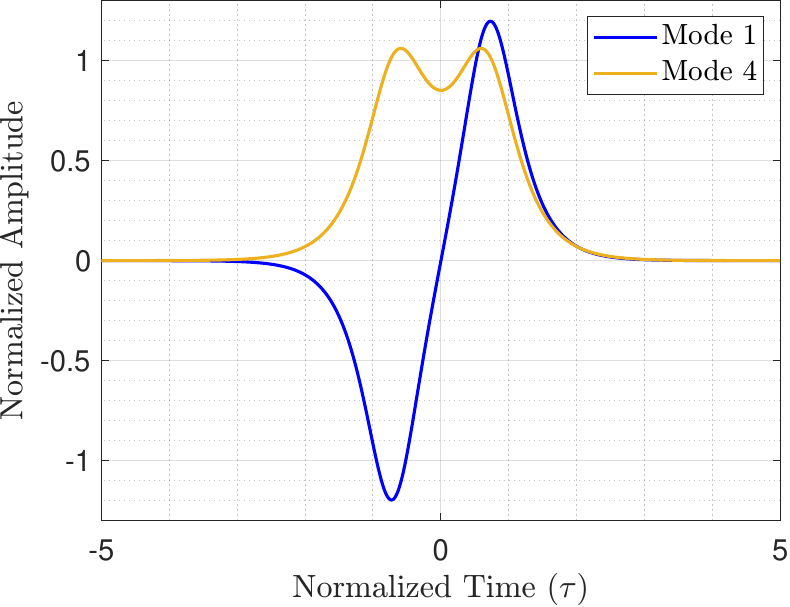}\put(0,77.5){(e)}
        \end{overpic}
    \end{minipage}
    \begin{minipage}{0.48\linewidth}
        \begin{overpic}[width=1.05\linewidth]{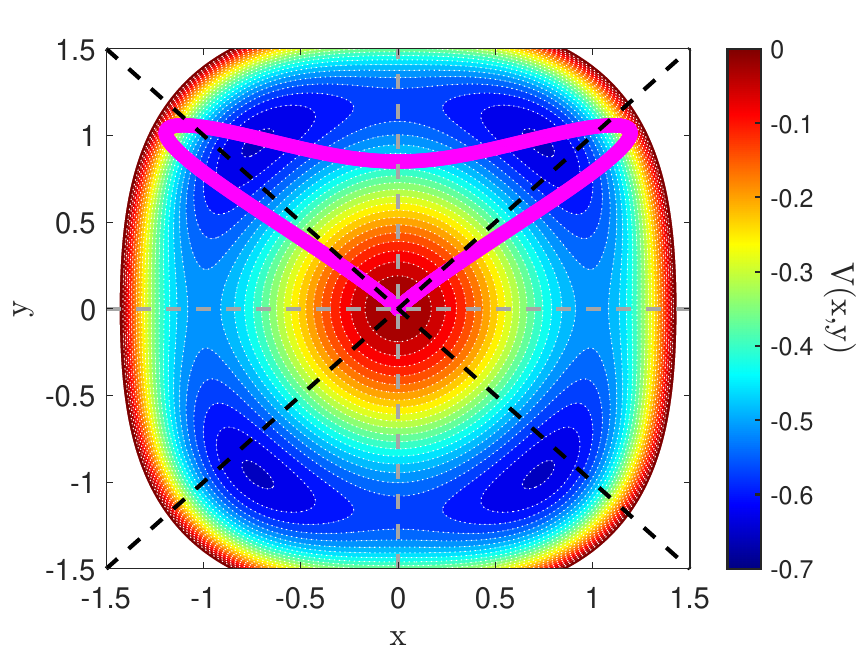}\put(0,75){(f)}
        \end{overpic}
    \end{minipage}
    \begin{minipage}{0.48\linewidth}
        \begin{overpic}[width=\linewidth]{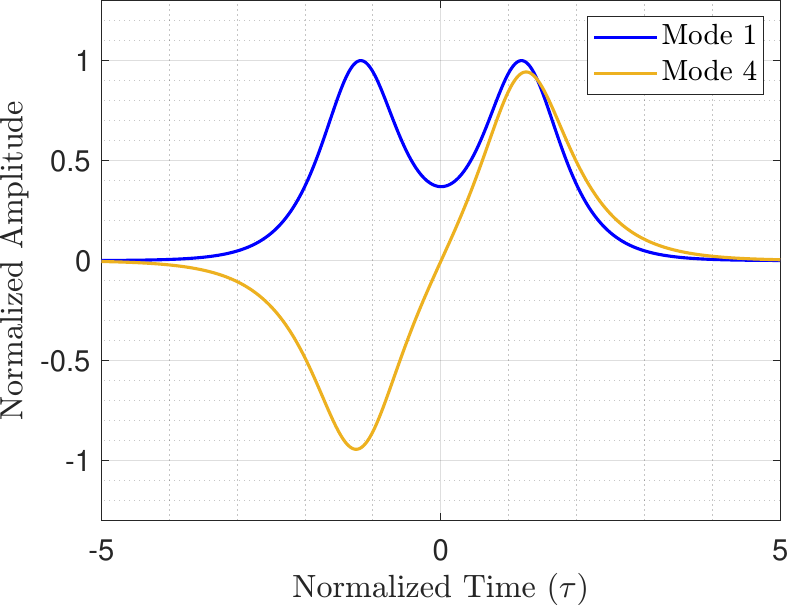}\put(0,77.5){(g)}
        \end{overpic}
    \end{minipage}
    \begin{minipage}{0.48\linewidth}
        \begin{overpic}[width=1.05\linewidth]{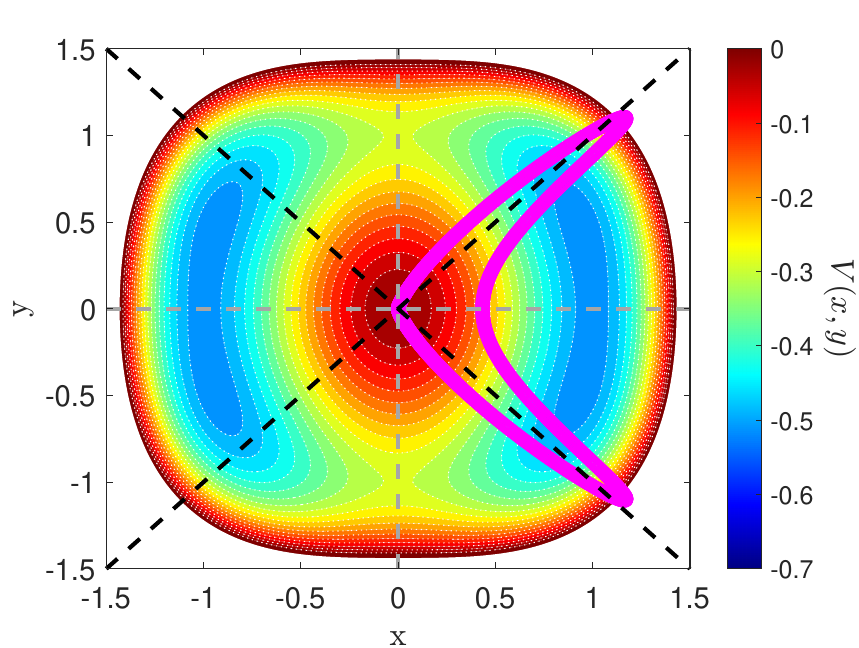}\put(0,75){(h)}
        \end{overpic}
    \end{minipage}
    \caption{Multimode solitons with a node in their temporal profile and the associated analogous potentials. a-b) $K_2/K_1=0.84$; c-d) $K_2/K_1=0.9$; e-f) $K_4/K_1=0.8$; g-h) $K_4/K_1=0.57$.}
    \label{second_order_TwoModes_MMS}
\end{figure}
\begin{figure*}
    \centering
        \begin{minipage}{0.18\linewidth}
            \includegraphics[width=\linewidth]{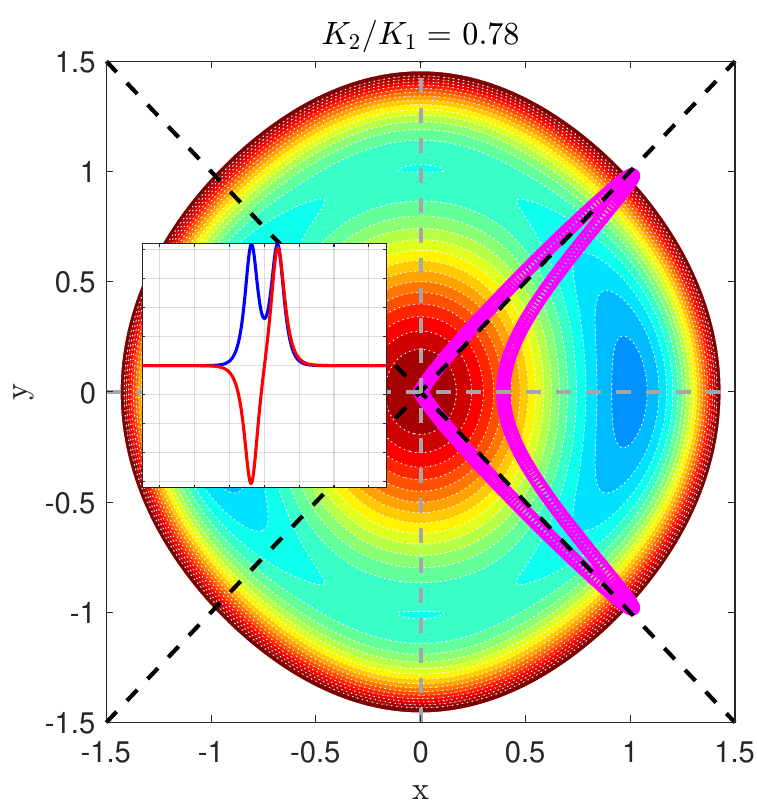}
        \end{minipage}
        \begin{minipage}{0.18\linewidth}
            \includegraphics[width=\linewidth]{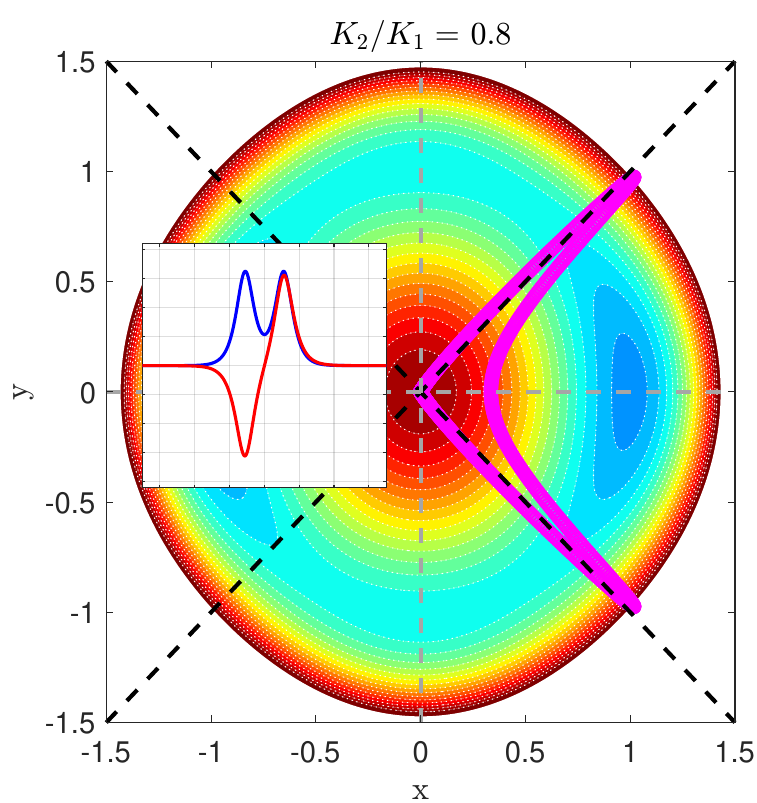}
        \end{minipage}
        \begin{minipage}{0.18\linewidth}
            \includegraphics[width=\linewidth]{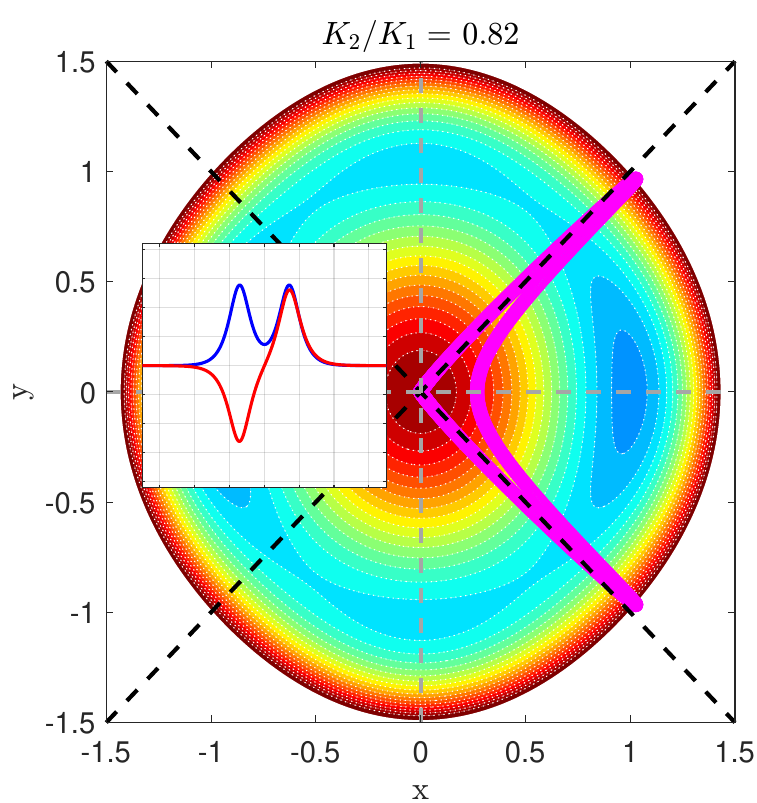}
        \end{minipage}
        \begin{minipage}{0.18\linewidth}
            \includegraphics[width=\linewidth]{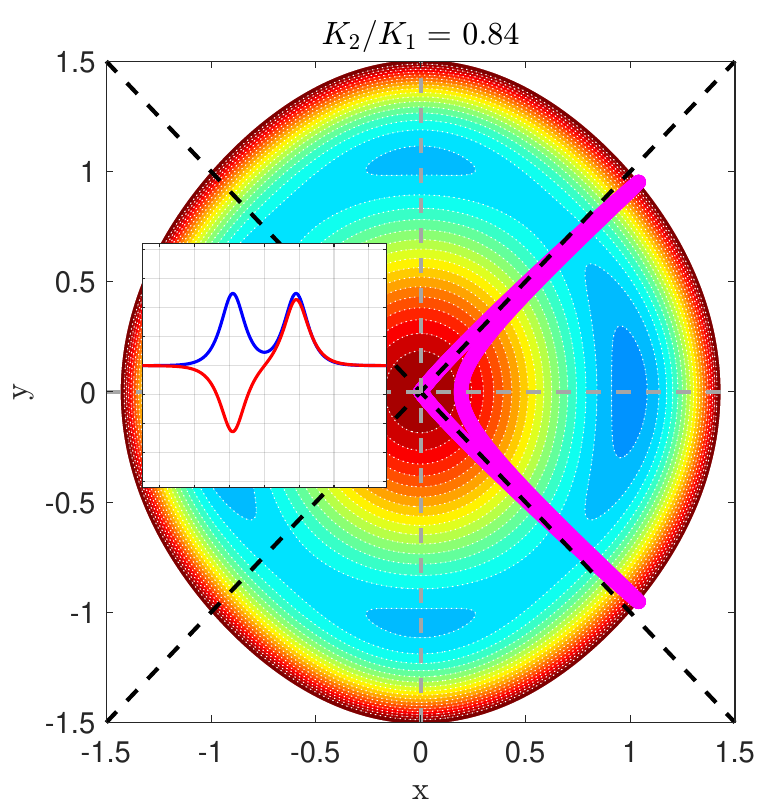}
        \end{minipage}
        \begin{minipage}{0.18\linewidth}
            \includegraphics[width=\linewidth]{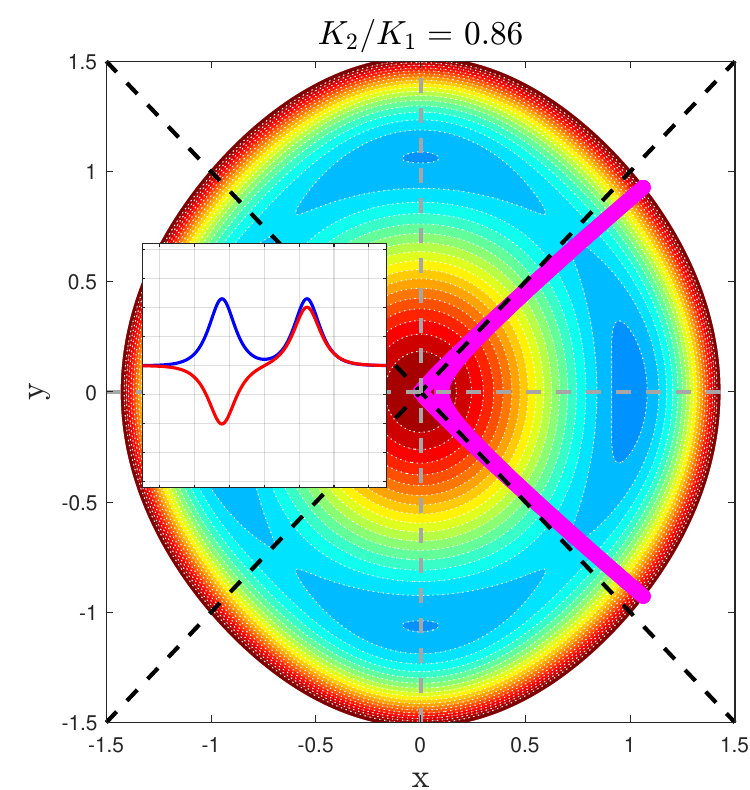}
        \end{minipage}
        \begin{minipage}{0.18\linewidth}
            \includegraphics[width=\linewidth]{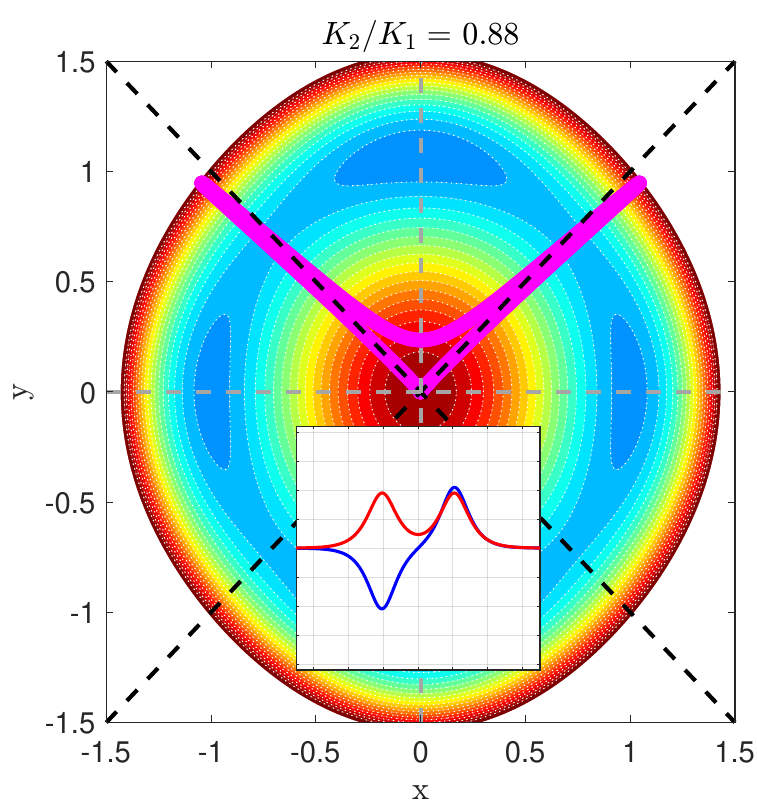}
        \end{minipage}
        \begin{minipage}{0.18\linewidth}
            \includegraphics[width=\linewidth]{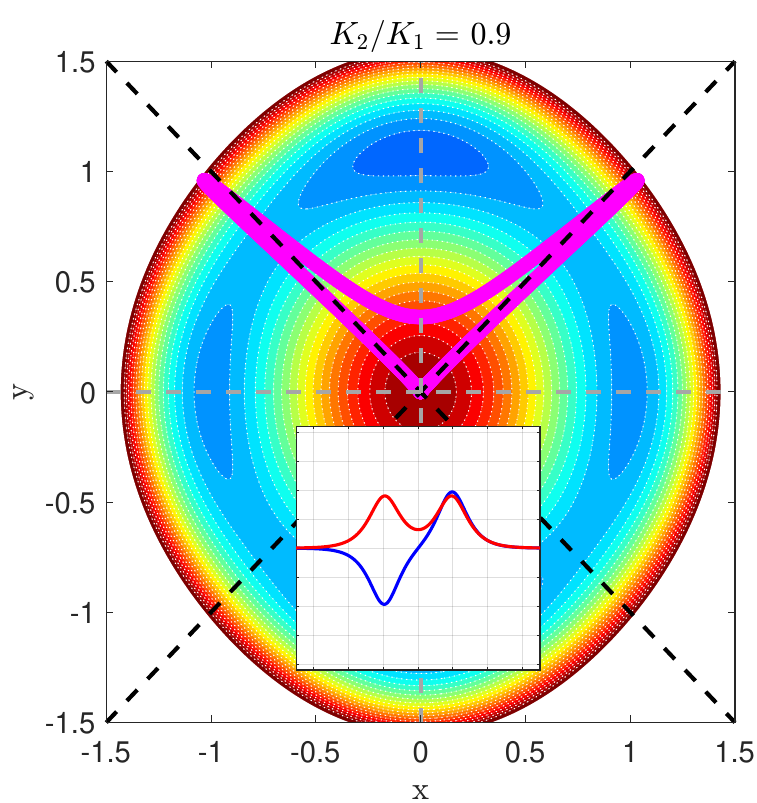}
        \end{minipage}
        \begin{minipage}{0.18\linewidth}
            \includegraphics[width=\linewidth]{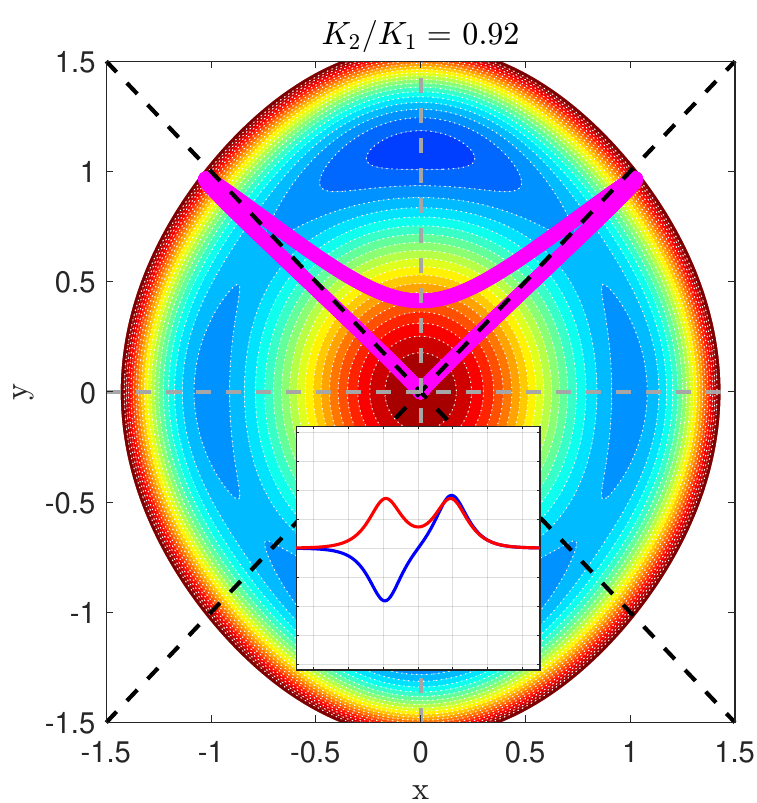}
        \end{minipage}
        \begin{minipage}{0.18\linewidth}
            \includegraphics[width=\linewidth]{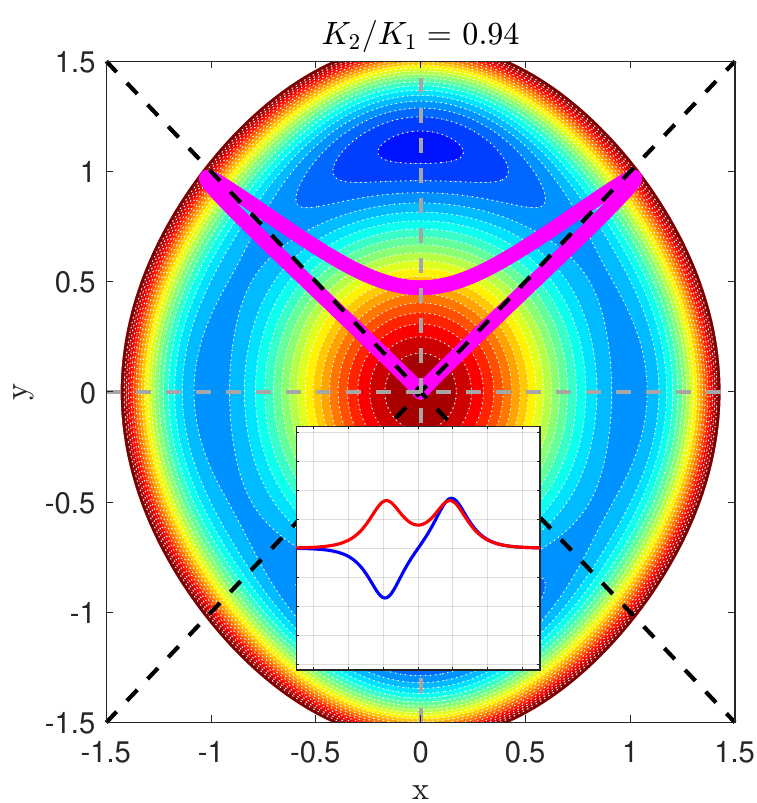}
        \end{minipage}
        \begin{minipage}{0.18\linewidth}
            \includegraphics[width=\linewidth]{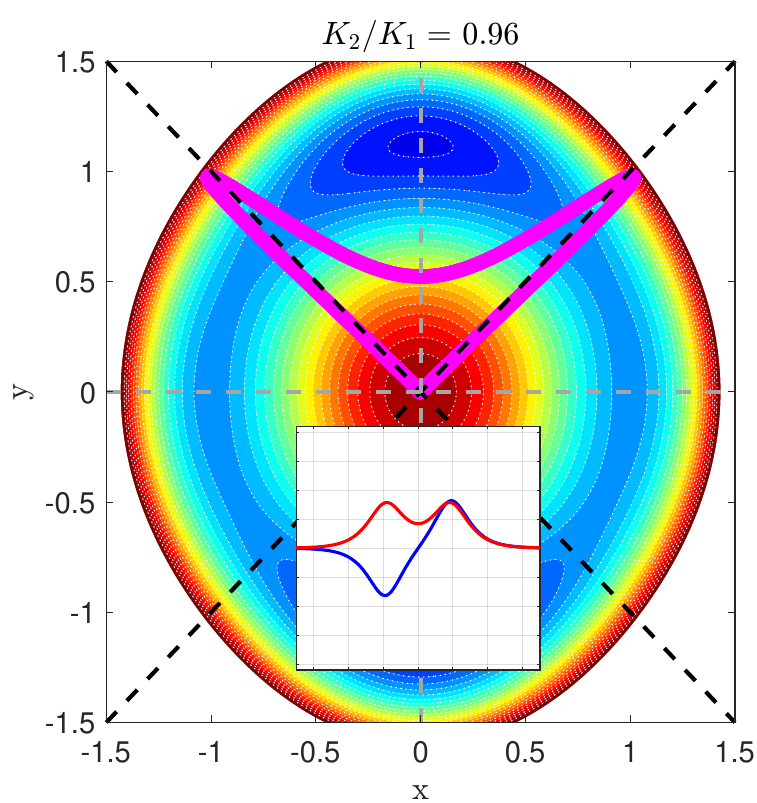}
        \end{minipage}
        \caption{Trajectories for various values of $K_2/K_1$ associated with a multihumped multimode soliton composed of one symmetric temporal profile and one antisymmetric temporal profile (shown in the small inset). $x$-direction corresponds to mode 1 (blue curve) and $y$-direction corresponds to mode 2 (red curve). Color scale is the same as in Fig.~\ref{second_order_TwoModes_MMS}. The amplitudes shown in the insets range from -2 to 2 over a normalized time window from -3.5 to 3.5.}
        \label{fig:evolution_traj_second_ordre}
\end{figure*}

Fig.~\ref{fig:fond_MMS_potential} shows the trajectory of the unit-mass ball that corresponds to a fundamental multimode soliton. This trajectory consists on a round-trip along one single branch. By looking at the potential, we now search for other types of trajectories. Since the soliton should vanish at infinity, the new trajectories must begin and end at the center of the potential like the one for the fundamental soliton. Nonetheless, the shape of the potential gives us the intuition that the trajectory can be constituted of more than one branch. This is demonstrated in Fig.~\ref{second_order_TwoModes_MMS} that shows various multimode solitons of the same kind, and their associated trajectories, that is two-hump structures that are temporally-symmetric in one mode and antisymmetric on the other one. 

In order to clarify the terminology used here, the fundamental multimode soliton solutions are those where the different modes have the same qualitative temporal profile, although we have seen that they will intrinsically have slightly different temporal widths. These are characterized by the quantities defined in Section~\ref{subsec:2modes.fundamental}. However, in the general physical picture and resolvable with the same formalism, there are multimode solitons where the different modes have different qualitative temporal profiles and one or all of them have multiple maxima. We refer to these as multihumped solitons~\cite{mitchell_observation_1998}, shown in the left column of Fig.~\ref{second_order_TwoModes_MMS}. For them, quantities defined in Section~\ref{subsec:2modes.fundamental}, except the ellipticity, are not strictly valid since they are defined for fundamental solitons.

The right column of Fig.~\ref{second_order_TwoModes_MMS} shows that the trajectory is constrained in a region defined by the black dotted line. In the mechanical analogy, first, the ball leaves the center of the potential and moves along a straight line approximately following a half diagonal of the figure. Then the ball goes back towards the center of the potential but does not come at the right angle to climb up the central hill. Consequently, the ball is reflected towards the other half-diagonal and finally reaches the center of the potential by approximately following the half-diagonal. Depending on the part of the graph where the ball moves, it is either one mode or the other that is antisymmetric in time. It is worth mentioning that it is impossible that both modes are antisymmetric. Indeed, the energy is minimized when the light gathers in regions with the highest refractive index. A node in the total temporal profile acts therefore as a repulsive point between both lobes that will then separate to form two identical fundamental multimode solitons.

Importantly, as already mentioned, the definition of $K_u$ given in Eq.~\ref{K} is rigorously valid only when modes have a sech temporal profile but can be used as an approximation. In the case of multimode solitons with the shape presented in Fig.~\ref{second_order_TwoModes_MMS}, the trajectory climbs the edge of the potential slightly too far, reaching $V = 0.05$ instead of exactly zero. This reflects a small inaccuracy in the potentials that are shown. The error is small enough to perform a qualitative study. In principle, refining the parameters would lead to a perfect potential. Note that the physical meaning of $K_u$ and $K_v$ does not depend on the temporal profile of the modes. As in the fundamental soliton case, they represent nonlinear phase accumulation rates, which remain constant along propagation since the intermodal interaction is purely incoherent.

 Fig.~\ref{fig:evolution_traj_second_ordre} shows the evolution of the trajectory of the ball for different values of the free parameter $K_v/K_u$. The corresponding multimode soliton is composed of mode 1 ($x$-direction) and mode 2 ($y$-direction). The range of values for $K_2/K_1$ that we consider for the example presented in Fig.~\ref{fig:evolution_traj_second_ordre} is given by $[0.78,0.96]$. This range was approximately centered at the transition value where modes qualitatively exchange their temporal profiles. Note that other solutions exist for values of the free parameter beyond this range; they will be discussed later.
 
 For $K_2/K_1<0.87$, the trajectory is confined to the right half-plane of the potential and is symmetric against the horizontal line, as shown in the upper panel of Fig.~\ref{fig:evolution_traj_second_ordre}. This trajectory corresponds to a multimode soliton where the temporal profile of mode 1 is symmetric and the one of mode 2 is antisymmetric, as shown in Fig.~\ref{second_order_TwoModes_MMS}.a. When $K_2/K_1$ increases, the minima aligned along the vertical axis are becoming increasingly pronounced, and the ball is being reflected closer and closer to the center of the potential. The sharper the trajectory, the larger the multimode soliton with well separated lobes. Since the modal interaction is incoherent, the trajectory could also be confined in the left half-plane as it corresponds to a $\pi$-phase shift of the fundamental mode.

 For $K_2/K_1>0.87$, the trajectory is rotated by 90° and lies in the upper half of the graph, as shown on the bottom panel of Fig.~\ref{fig:evolution_traj_second_ordre}. In this configuration, it is the fundamental mode that is antisymmetric in time (see Fig.~\ref{second_order_TwoModes_MMS}.b). Again, the trajectory could also lie in the lower half. When $K_2/K_1$ increases, minima on the horizontal axis become less pronounced and the trajectory of the ball is less sharp giving a shorter multimode soliton with closer lobes.

Such complex temporal profiles were already reported for elliptical solitons~\cite{haelterman_bound-vector_1993}. Note that the analogous potential for polarization modes is of the type $\gamma_{uu}\gamma_{vv}<4\gamma_{uv}^2$, because $\gamma_{uu}=\gamma_{vv}=\gamma_{uv}$. However, major differences emerge when considering the asymptotic regime. We consider first a soliton made of  polarization modes. When $P_v\ll P_u$, mode $v$ is optically guided by mode $u$. The optical waveguide induced by the intensity profile of mode $u$ accepts a fundamental mode that has the temporal shape of a sech function, but also accepts a two-lobe temporal profile for mode $v$. This temporal profile is fully contained in the one of mode $u$. When $K_v/K_u$ tends to unity, the temporal profiles of the modes progressively reshape themselves from a two-lobe profile confined in a bell-shaped distribution to a two-lobe temporal profile for both modes, similar the temporal profiles presented in Fig.~\ref{second_order_TwoModes_MMS}. At the limit $K_v/K_u=1$, the temporal profiles of the modes consist of two well-separated lobes, with a constant ellipticity. In other words, the trajectory corresponds to two infinitely separated linearly polarized solitons out of phase by $\pi/2$. This corresponds to the degenerate case, where the four minima of the analogous potential have the same depth and the saddle points are located on the diagonals. The trajectory then takes the form of two straight branches following the diagonals.  Interestingly, for the nondegenerate spatial mode case, the branches do not follow the diagonal exactly because they are still curved. As we can see on Fig.~\ref{fig:evolution_traj_second_ordre}, the potential is never symmetric in $x$- and $y$-direction because of the asymmetry in the nonlinear coefficients (see Eq.~\ref{pot2modes}). As a result, we do not obtain two well-separated multimode solitons with constant ellipticities, unlike in the elliptical soliton case. Moreover, the case of two well separated lobes do not occur when the free parameter is unitary. For instance, for a soliton made of modes 1 and 2, we see on Fig.~\ref{fig:evolution_traj_second_ordre} that temporal lobes are well separated for $K_v/K_u\approx 0.87$. In principle, at the limit value of $K_v/K_u$ where modes qualitatively exchange their temporal profiles, the lobes should be infinitely far apart in time.

Furthermore, for the elliptical soliton, the evolution of the temporal profiles of the modes induced by the continuous change of $K_v/K_u$ is symmetric by exchanging the role of modes $u$ and $v$. On the contrary, for nondegenerate spatial modes, this symmetry does not hold due to the asymmetry in the nonlinear coefficients. Indeed, in the limit of $P_v\ll P_u$, solutions that decay at infinity are given by $P^{\lambda-n_{uv}}_{n_{uv}}\left(\tanh(\sqrt{2}\tau)\right)$ and exist only if $\lambda < n_{uv}$ (see Appendix~\ref{Anx:Legendre}). Since we are looking for a two-lobe temporal profile, we have that $\lambda = 1$. Hence, the solution exists only if $1 < n_{uv}$. For a soliton made of modes 1 and 2, such solution exists for $P_2 \gg P_1$ because $n_{21}=1.21$, but is not possible for $P_2\ll P_1$ because $n_{12}=1$. Consequently, we lose the symmetry in the evolution of the temporal profiles against $K_v/K_u$ by exchanging the role of both modes. When we decrease the value of $K_2/K_1$ below 0.78 (value corresponding to the first graph of Fig.~\ref{fig:evolution_traj_second_ordre}), the lobes of the temporal profile of the fundamental mode get closer to each other but never fusion. Hence, the trajectory in the potential keeps the same qualitative shape.

On the contrary, we can increase $K_2/K_1$ beyond 0.96 (value corresponding to the last graph of Fig.~\ref{fig:evolution_traj_second_ordre}) giving solutions for which the temporal profile of mode 2 is bell-shaped while the one of the fundamental mode has a central node. Concerning the associated trajectory in the potential, the dip becomes less and less pronounced and eventually disappears, resulting in a loop-shaped trajectory. In such a case, the soliton approaches something close to a space-time optical vortex (STOV)~\cite{bekshaev_spatiotemporal_2024}, which we could not see in recent work~\cite{dechanxhe_space-time_2026}. The main difference lies in the temporal profile, which involves a Gaussian function for a true STOV, but a hyperbolic secant in our case along with a different ratio of energies between the two modes. Note that we can see an asymmetry in the evolution of the soliton in Fig.~\ref{fig:evolution_traj_second_ordre}. In the upper row, we observe a more significant evolution of the temporal profiles against the value of $K_2/K_1$ compared to the lower row.

For a soliton composed of modes 1 and 4, there is no solution with a node for $P_1 \gg P_4$ and $P_1 \ll P_4$ (see Fig.~\ref{second_order_TwoModes_MMS}.e and~.g). This also illustrate that the equivalence regime is not simply a continuation of the asymptotic regime, as the latter may not exist.

\begin{figure}
    \begin{minipage}{0.48\linewidth}
        \begin{overpic}[width=\linewidth]{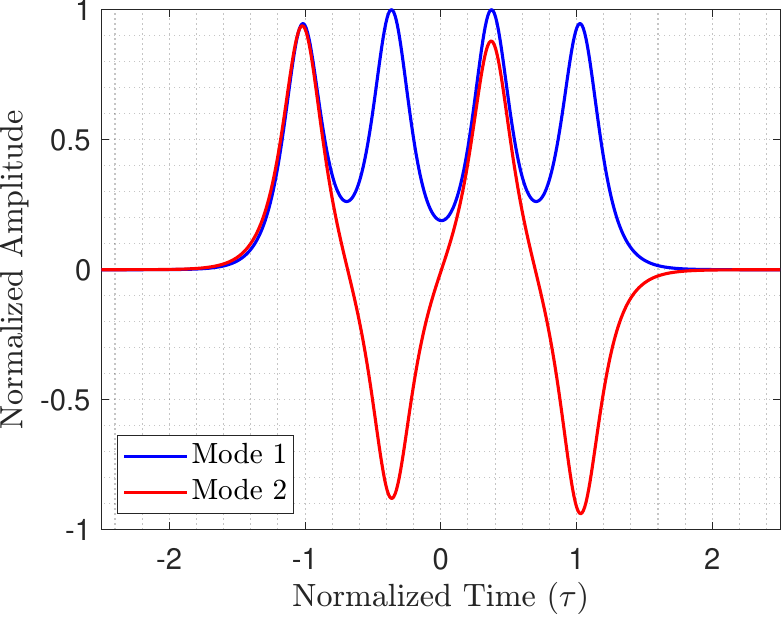}\put(-2,78){(a)}
        \end{overpic}
    \end{minipage}
        \begin{minipage}{0.49\linewidth}
        \begin{overpic}[width=1.05\linewidth]{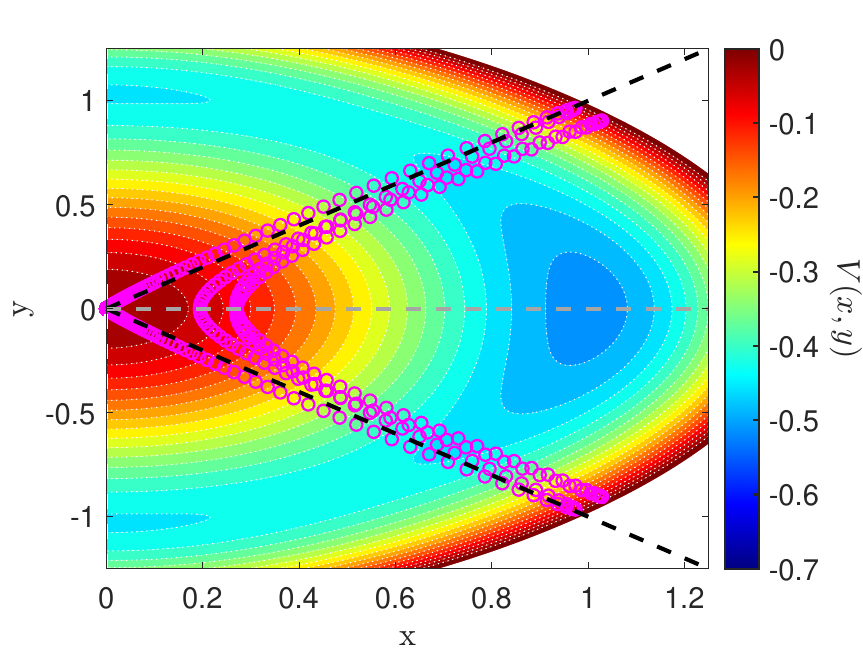}\put(0,73){(b)}
        \end{overpic}
    \end{minipage}
            \caption{Example of a multimode soliton with four nodes in the temporal profiles of the modes (a) and the associated trajectory in the analogous potential (b), for $K_2/K_1=0.81$.}
        \label{fig:double_mol_2modes}
\end{figure}

Finally, more complex temporal structures have also been reported for elliptical solitons~\cite{haelterman_bifurcation_1994}, and we identified the spatial-mode equivalent of one of them, shown in Fig.~\ref{fig:double_mol_2modes}. The trajectory is similar to those shown in Fig.~\ref{second_order_TwoModes_MMS}, except that the ball is reflected twice off the central hill, resulting in a four-branch trajectory rather than a two-branch one. The ball crosses four times the $x$-axis meaning that mode 2 has four nodes in its temporal profile. There exists no limit case ($P_1\ll P_2$ or $P_1\gg P_2$) where one mode has a two-node temporal profile confined under the bell-shaped temporal profile of the other mode. Hence, such a structure only exists when modes have comparable peak powers.

In conclusion, the nondegenerate spatial modes composing a multimode soliton can have temporal profiles as complex as the one of the polarization modes composing an elliptical soliton. Modes can exhibit several nodes in their temporal profile but because these act as repulsive points, the other mode always must have a nonzero field at the same time point to provide the invariance property. These complex temporal profiles may not always exist when one of the mode has a significantly weaker peak power than the other. In contrast to the elliptical soliton, these complex temporal profiles are not symmetric under an exchange of the role of the modes. These major differences between the polarization-mode and the nondegenerate spatial-mode cases come from the fact that for the former the nonlinear coefficients are symmetric while for the last they are not. Another important difference between both cases is that the elliptical soliton is only made of two modes while, in the spatial-mode case, the soliton can be made of more modes, as discussed in Section~\ref{sec:3modes}.
\section{Three-mode soliton}\label{sec:3modes}
A major novelty of this work is that, thanks to the fact that spatial modes interact incoherently by pairs, we can extend the previous framework to a multimode soliton made of more than two modes, which has no counterpart in the elliptical soliton case. For instance, such solitons made of three modes or more can be generated by launching a space-time optical vortex into a multimode fiber~\cite{dechanxhe_space-time_2026}.

In this section, we restrain ourselves to the case of three modes without loss of generality. In Section~\ref{subsec: fund_3modes}, we extend the development presented in Section~\ref{subsec:2modes.fundamental} to the three-mode case. In Section~\ref{subsec: multihumped_3modes} we show that there exists a large variety of multihumped three-mode solitons.
\subsection{The fundamental three-mode soliton}\label{subsec: fund_3modes}
\begin{figure}
     \begin{minipage}{0.48\linewidth}
        \begin{overpic}[width=\linewidth]{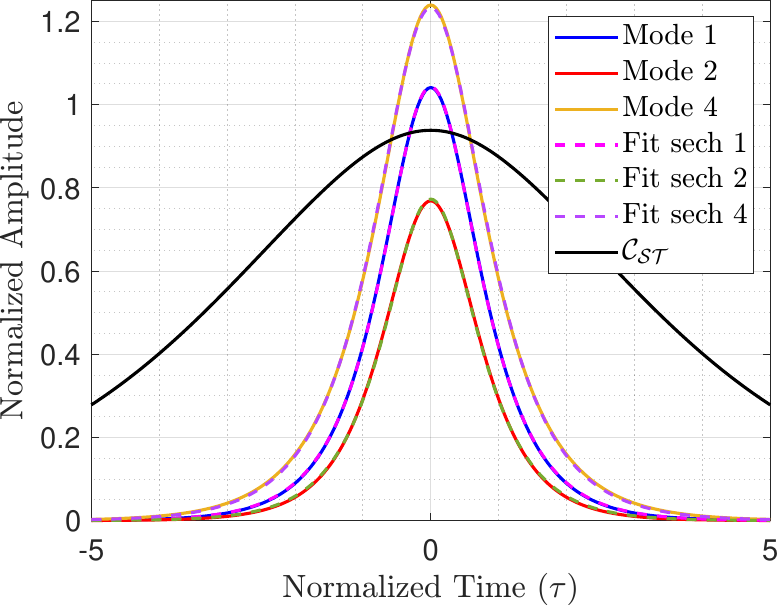}\put(-5,77.5){(a)}
        \end{overpic}
    \end{minipage}
    \begin{minipage}{0.48\linewidth}
        \begin{overpic}[width=\linewidth]{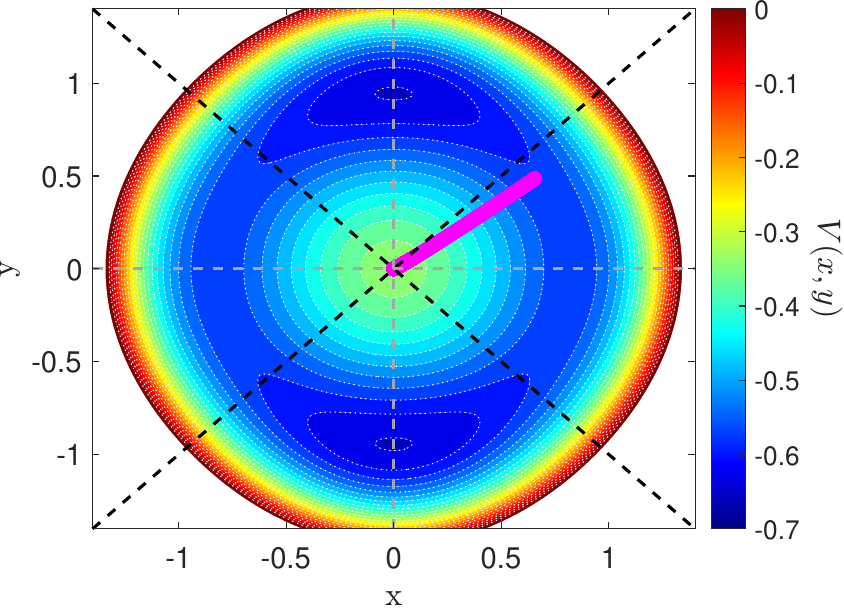}\put(0,75){(b)}
        \end{overpic}
    \end{minipage}
    \begin{minipage}{0.48\linewidth}
        \begin{overpic}[width=\linewidth]{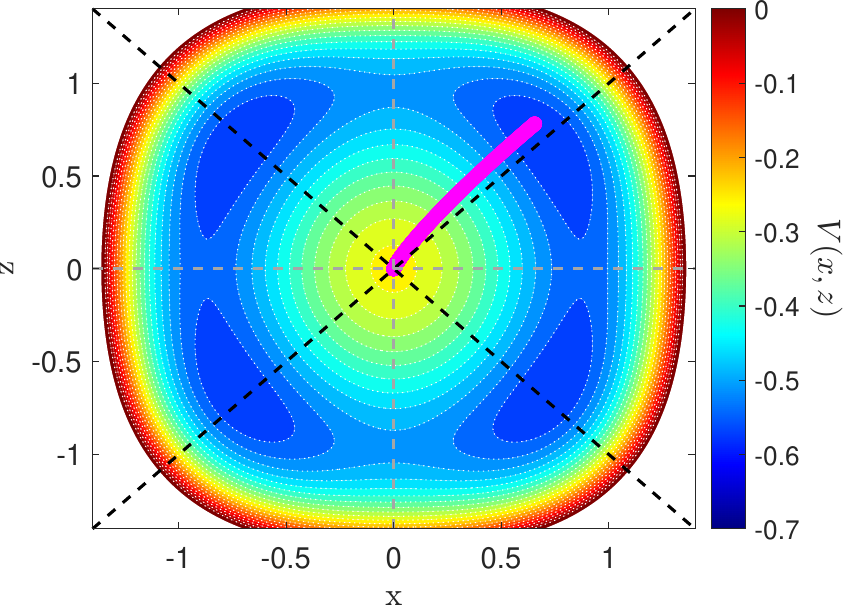}\put(-5,75){(c)}
        \end{overpic}
    \end{minipage}
    \begin{minipage}{0.48\linewidth}
        \begin{overpic}[width=\linewidth]{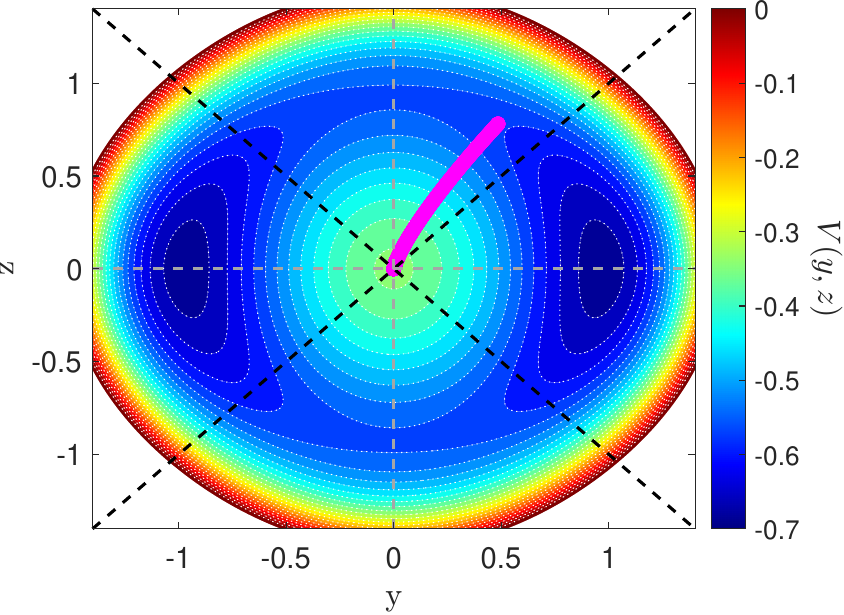}\put(0,75){(d)}
        \end{overpic}
    \end{minipage}
    \caption{Fundamental multimode soliton composed of three spatial propagation modes: 1 (x), 2 (y), and 4 (z); with $K_2/K_1=1.15$ and $K_4/K_1=0.71$. a) Temporal profiles of the modes and the space-time complexity; Analoguous $4D$ potential in the plane: b)  $z=z_{\text{max}}$; c) $y=y_{\text{max}}$; d) $x=x_{\text{max}}$.}
    \label{fig:fund_3modes_MMS}
\end{figure}

A fundamental multimode soliton can be composed of three spatial propagation modes (see Fig.~\ref{fig:fund_3modes_MMS}.a). Such a structure is an invariant solution of Eq.~\ref{Syst_ODE_3modes}. The balance between nonlinearity and chromatic dispersion, that gives the invariance of the temporal profiles, is exactly due to the three modes. If one of them is cut, the balance is broken and the multimode pulse will undergo a complex transient before splitting into solitons and/or dispersive waves. As shown in Fig.~\ref{fig:fund_3modes_MMS}.a, the temporal profiles of the modes are again close to sech functions. We can thus extend Eq.~\ref{K}, Eq.~\ref{L_NL}, and Eq.~\ref{Nouvel_ordre} to the three-mode case:
\begin{align}
        K_u &= \frac{1}{2}\left(N_u^2P_u+\frac{C_{uv}}{2}P_vG(\frac{T_v}{T_u})+\frac{C_{uw}}{2}P_wG(\frac{T_w}{T_u})\right)\label{K_3modes},\\
    L_{\text{NL}}^{(u)}&=\left(\gamma_{uu}P_u+\gamma_{uv}P_vG(\frac{T_v}{T_u})+\gamma_{uw}P_wG(\frac{T_w}{T_u})\right)^{-1},\label{L_NL_3modes}\\
    \tilde{N}^2_u &=\frac{T_u^2}{|\beta_2^{(u)}|}\left(\gamma_{uu}P_u+\gamma_{uv}P_vG(\frac{T_v}{T_u})+\gamma_{uw}P_wG(\frac{T_w}{T_u})\right).\label{Nouvel_ordre_3modes}
\end{align}

For the fundamental multimode soliton shown in Fig.~\ref{fig:fund_3modes_MMS}.a., the soliton order defined in the single-mode case gives $\mathcal{N}_1=0.67$, $\mathcal{N}_2=0.41$, and $\mathcal{N}_4=0.67$. Using Eq.~\ref{Nouvel_ordre_3modes}, we get $\tilde{N}_1=1.02$, $\tilde{N}_2=1.05$, and $\tilde{N}_4=0.96$.

A quantity that cannot be so easily extended to more than the two-mode case is the ellipticity (see Eq.~\ref{ellipticity}) as it is defined in a two-mode basis. A first generalization attempt is to consider pairs of modes and compute the ellipticity for each pair, then summing all contributions:
 \begin{equation}\label{Q}
    Q(\tau)=\sqrt{\sum^{M-1}_{i=1}\sum^{M}_{j>i}q_{ij}^2(\tau)},
 \end{equation}
 where $M$ is the total number of modes. We sum the square of the ellipticity to avoid any cancelling between different ellipticities. This quantity can play the role of the ellipticity for multimode solitons made of more than two modes. While the $q$ parameter could be associated to a polarization state, the physical meaning of $Q$ is less obvious.

A new quantity can be defined by reformulating the role of the ellipticity. This will allow relating $q$ to some entropy function. Starting from Eq.~\ref{ellipticity}, and considering a two-mode case, we can rewrite the ellipticity as follows: $q(\tau)=p_2(\tau)-p_1(\tau)=1-2p_1(\tau)$, where $p_k(\tau)=\left|a_k(\tau)\right|^2\left(\sum_{j=1}^{M}\left|a_j(\tau)\right|^2\right)^{-1}$ is the normalized contribution of mode $k$ to the intensity profile of the multimode soliton at time point $\tau$. For a time point where $q=\pm 1$, it means that only one mode contributes to the multimode soliton. For a time point where $q=0$, it means that all modes contribute equally. The ellipticity can thus be seen as a measure of the variety of the modal contribution at each time point in the multimode soliton. In other words, the ellipticity measures the modal diversity of the multimode soliton at each time point.

 Based on this observation and inspired by the concept of entropy in information theory~\cite{shannon_mathematical_1948}, we define the \textit{space-time complexity}:
 \begin{equation}\label{ST_complexity}
    \mathcal{C_{ST}}(\tau)=\frac{-1}{\ln\left(M\right)}\sum_{k=1}^Mp_k(\tau)\ln\left(p_k(\tau)\right),
\end{equation}
 where $M$ is the total number of modes considered. Like the ellipticity, the space-time complexity is a measure of the modal diversity at each time point in the multimode soliton, but is easily generalizable to an arbitrary number of modes. The word complexity refers to the measure of the diversity/contribution of the modes. If it is zero at a time point $\tau_0$, it means that the soliton is single-mode at that time, so the complexity is minimal. On the contrary, if it is equal to unity at a time point $\tau_0$, the complexity is maximal since it means that all modes contribute equally. The word space-time refers to the fact that we look at the contribution of spatial modes at each time point of the pulse. Moreover, Eq.~\ref{ST_complexity} is a kind of measure of the constraint imposed by the transverse profiles of the modes, and their relative orientation, on their temporal profiles, as already underlined in Section~\ref{subsec:2modes.fundamental}. Of course, as for the ellipticity, a time-independent space-time complexity means that modes have the same temporal evolution. A time-varying complexity implies that there exists no basis in which the soliton can be described with only one mode that would be the same at all times.

Coming back to our potential picture, the unit-mass ball now moves in a 3D space as the potential is given by Eq.~\ref{pot3modes}. Since a new dimension is added in the phase space, the visualisation of the potential to find trajectories corresponding to physical invariant temporal profiles becomes more challenging. Fig.~\ref{fig:fund_3modes_MMS} shows three cross-sections of the potential. The projection of the trajectory of the ball in that plane is shown in magenta. Cross-sections are taken at each maximum coordinate along the trajectory (Fig.~\ref{fig:fund_3modes_MMS}.b: $x=x_{\text{max}}$, Fig.~\ref{fig:fund_3modes_MMS}.c: $y=y_{\text{max}}$, and Fig.~\ref{fig:fund_3modes_MMS}.d: $z=z_{\text{max}}$). As for the fundamental two-mode soliton, the trajectory is homoclinic and curved. In the 3D space, the trajectory of the ball is curved in such a way that it cannot be fully contained in one plane, a confirmation that the soliton cannot be described with only two modes.

\subsection{Multihumped three-mode solitons}\label{subsec: multihumped_3modes}
As in Section~\ref{sec:2modes.second_order}, we now consider trajectories in the analogous potential made of more than one branch. Due to the new dimensionality, we have access to a large variety of trajectories, leading to a rich diversity in invariant temporal profiles for the multimode soliton. Note that all the trajectories start at the center of the potential $\left(V(0,0,0)=0\right)$ and are homoclinic since the soliton has a finite duration. Like for the fundamental soliton, trajectories are twisted in such a way that they cannot be fully contained into one plane. The goal of this subsection is to showcase this diversity without analyzing all possible trajectories in detail. The stability of these solutions will be tested to a certain extent in Section~\ref{sec:prop_mutlihump}.

Stationary solutions were found using the relaxation method. In order to get a nontrivial solution, it was necessary to impose some symmetry on the temporal profiles. They are thus either symmetric or antisymmetric, except for the case where all modes are antisymmetric as already explained in Section~\ref{sec:2modes.second_order}. 
\begin{figure}
    \centering
    \includegraphics[width=\linewidth]{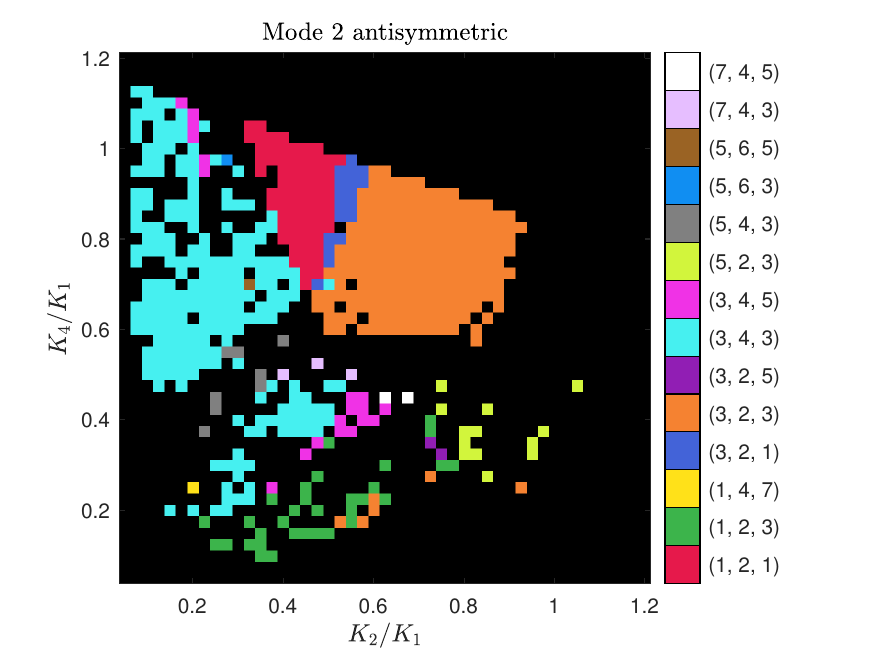}
    \caption{Mapping of multihumped three-mode solitons, with an antisymmetric temporal profile for mode 2, in the free parameter space. Modes 1 and 4 have symmetric temporal profiles. Black pixels indicate cases where the algorithm does not converge, or gives solitons made of less than three modes. Other colors indicate distinct three-mode solutions. The triplet of integers $(m_1,m_2,m_3)$ gives the numbers of extrema in the temporal profiles of modes 1 ($\mathrm{LP}_{01}$), 2 ($\mathrm{LP}_{11}$), and 4 ($\mathrm{LP}_{21}$), respectively.}
    \label{M2_antisym}
\end{figure}

\begin{figure}
        \begin{minipage}{0.48\linewidth}
        \begin{overpic}[width=\linewidth]{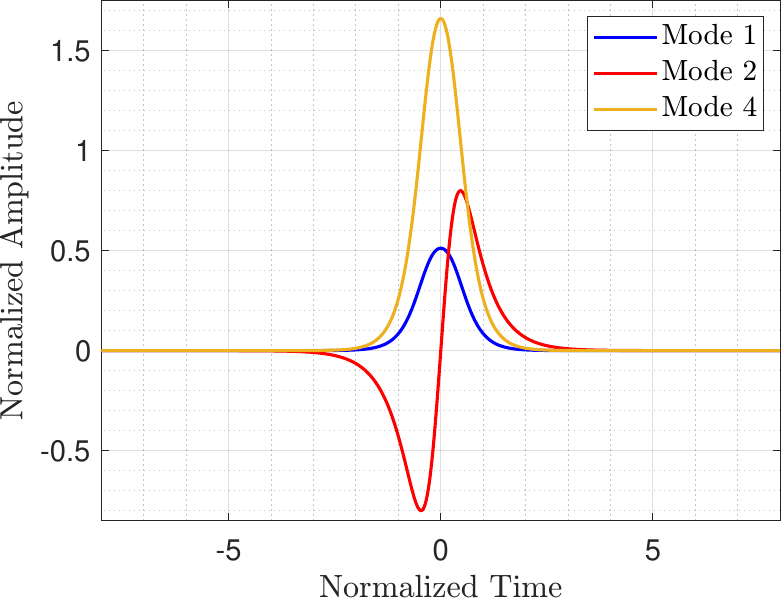}\put(15,65){(a)}
        \end{overpic}
    \end{minipage}
    \begin{minipage}{0.48\linewidth}
        \begin{overpic}[width=\linewidth]{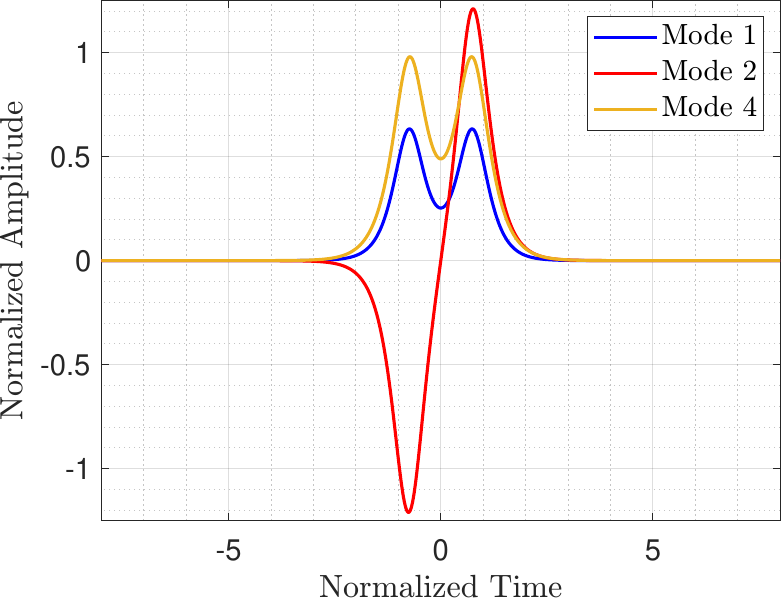}\put(15,65){(b)}
        \end{overpic}
    \end{minipage}
    \begin{minipage}{0.48\linewidth}
        \begin{overpic}[width=\linewidth]{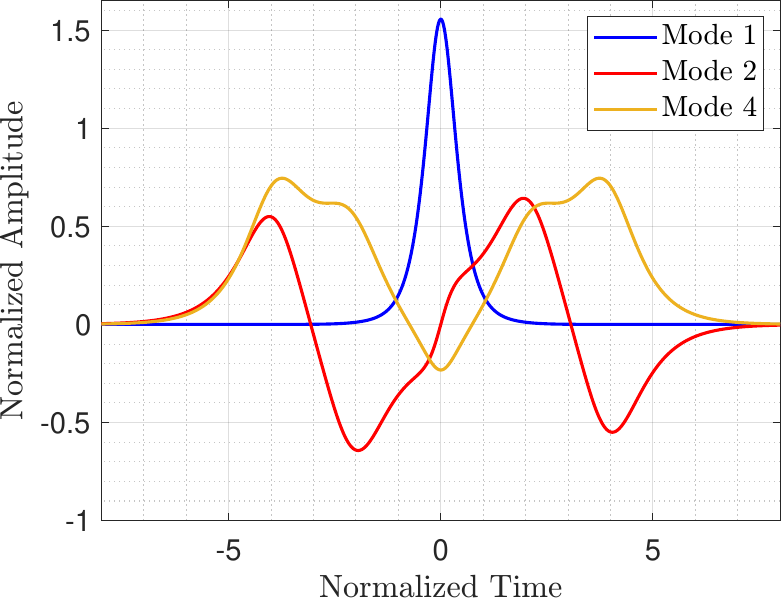}\put(15,65){(c)}
        \end{overpic}
    \end{minipage}
    \begin{minipage}{0.48\linewidth}
        \begin{overpic}[width=\linewidth]{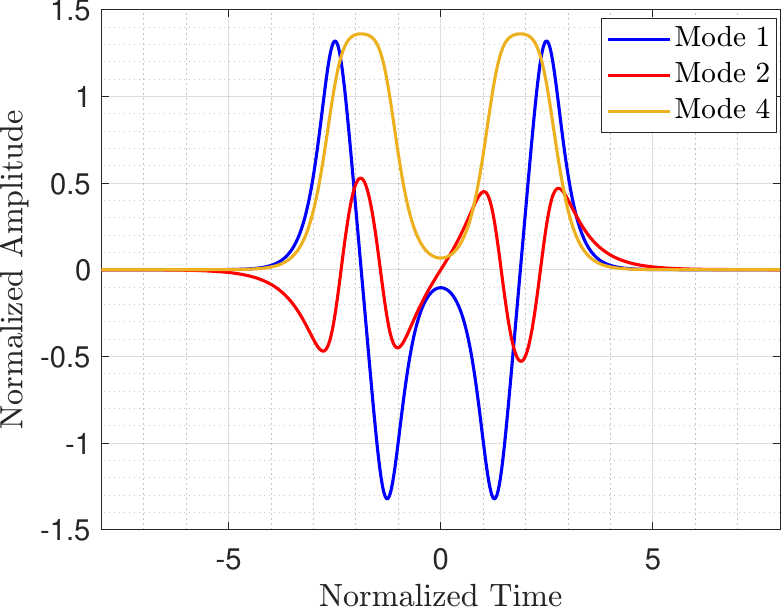}\put(15,65){(d)}
        \end{overpic}
    \end{minipage}
    \begin{minipage}{0.48\linewidth}
        \begin{overpic}[width=\linewidth]{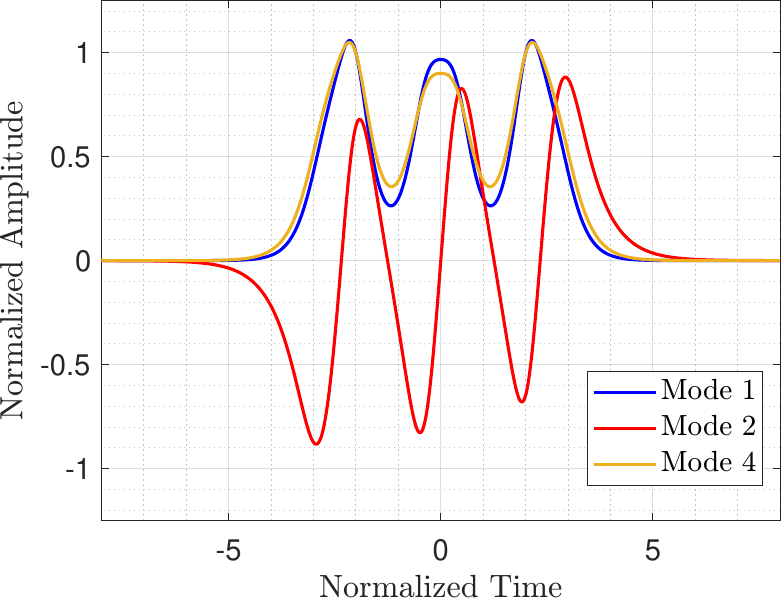}\put(15,65){(e)}
        \end{overpic}
    \end{minipage}
    \begin{minipage}{0.48\linewidth}
        \begin{overpic}[width=\linewidth]{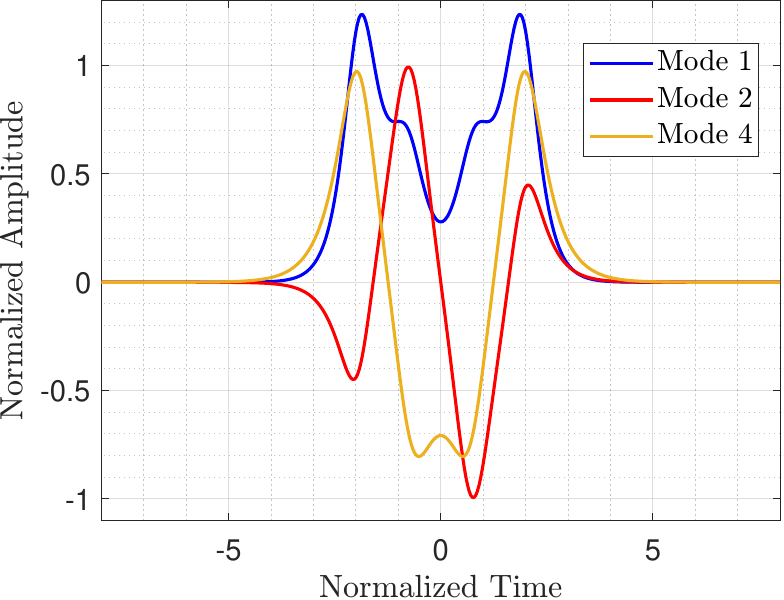}\put(15,65){(f)}
        \end{overpic}
    \end{minipage}
    \caption{Multihumped three-mode solitons where the temporal profile of mode 2 is antisymmetric. a) (1,2,1); b) (3,2,3); c) (1,4,7); d) (5,6,3); e) (5,6,5); f) (7,4,5).}
    \label{fig:exemples_3modesMMS_SecondOrdre}
\end{figure}

In order to deal with the plethora of invariant solutions, we categorize them with a triplet of integers. They correspond to the number of extrema in the temporal amplitude profile of each mode. Note that an antisymmetric temporal profile exhibits an even number of extrema, while a symmetric one has an odd number of extrema. For instance, the triplet (7, 4, 5) means that the temporal profile of mode $u$ exhibits 7 extrema, the one of mode $v$ exhibits 4 extrema, the one of mode $w$ has 5 extrema (see Fig.~\ref{fig:exemples_3modesMMS_SecondOrdre}.d). In the different examples presented in this subsection, we always consider a multimode soliton composed of modes 1, 2, and 4.

Solutions are presented in the form of a mapping in the space of the free parameters $K_v/K_u$ and $K_w/K_u$. Eq.~\ref{K_3modes} is not strictly valid for multihumped solitons since it is defined for sech profiles, but can be used as an approximation to guide our qualitative understanding, like in Section~\ref{sec:2modes.second_order}. Fig.~\ref{M2_antisym} shows the map in the free parameter space for multihumped three-mode solitons with modes 1 and 4 exhibiting a symmetric temporal profile while the one of mode 2 is antisymmetric. Each color of the map corresponds to a unique triplet in that map. Black pixels indicate cases for which the relaxation method does not converge with our standard initial guess, but also cases where the algorithm gives a solution made of at most two modes. The multihumped two-mode solitons among these were already discussed in Section~\ref{sec:2modes.second_order}, so they are disregarded here and represented by a single color. In addition, their location in the map does not make sense as such solutions are described by only one free parameter. The presence of these peculiar pixels is due to the nature of the problem that is very sensitive to the initial guess given at the relaxation method. It also implies that the list of solutions given in this section is not exhaustive. By changing the initial guess, it is possible to find new triplets.

Fig.\ref{fig:exemples_3modesMMS_SecondOrdre} shows examples of stationary solutions made of three spatial propagation modes, all with an antisymmetric temporal profile for mode 2. Hence, these multimode solitons are already presented in Fig.~\ref{M2_antisym}. Temporal profiles are listed in order of complexity. They can be very complex but always maintain a balance between chromatic dispersion and nonlinearity. It should be noted that, regardless of the complexity of the temporal profiles, each node of a mode must correspond to a non-zero field point of another mode. Indeed, the energy is minimized when the electric field is located in the region with the highest refractive index. As the nonlinearity increases the refractive index, region with light acts as an attractor for the other parts of the profile. On the contrary, a node acts as a repulsive point. Hence, to keep an invariant shape, each repulsive point should be counteracted by the appropriate amount of electric field to counteract the repulsion. As for all other cases, if one of the modes is cut, then the temporal structure collapses during propagation.

Furthermore, note that Fig.~\ref{fig:exemples_3modesMMS_SecondOrdre}.a shows a two-lobe temporal profile for mode 2 associated to a bell-shaped temporal profile for the fundamental mode. As already discussed in Section~\ref{sec:2modes.second_order}, this configuration is impossible for a two-mode soliton made of modes 1 and 2. Nonetheless, we see that adding a third mode (here mode 4) enables the existence of such invariant structure. In the analogous potential picture, adding a new dimension allows the ball to go around previously impassable obstacles.
\begin{figure*}
    \centering
        \begin{minipage}{0.32\linewidth}
            \includegraphics[width=1.1\linewidth]{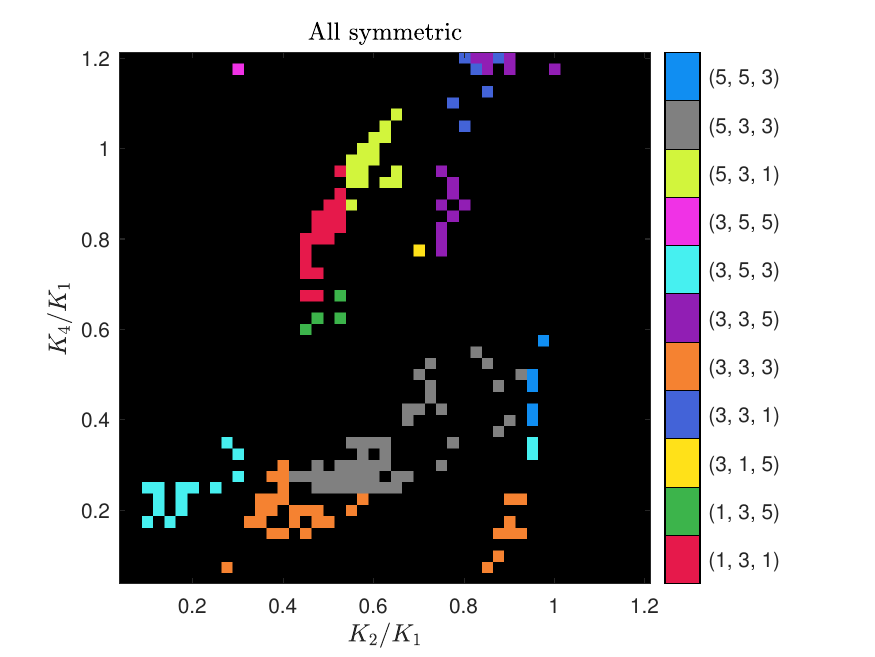}
        \end{minipage}
        \hfill
        \begin{minipage}{0.32\linewidth}
            \includegraphics[width=1.1\linewidth]{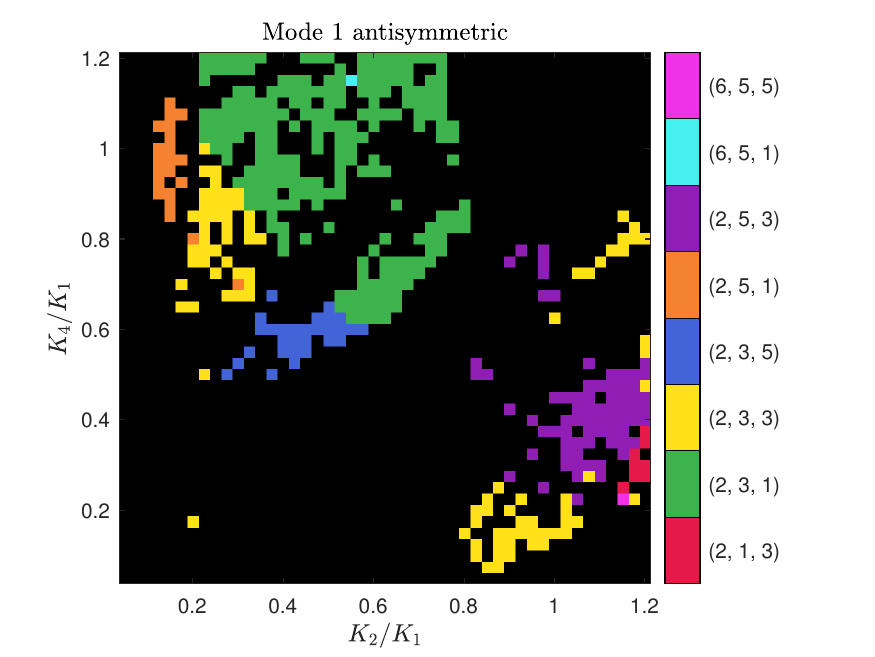}
        \end{minipage}
        \hfill
        \begin{minipage}{0.32\linewidth}
            \includegraphics[width=1.1\linewidth]{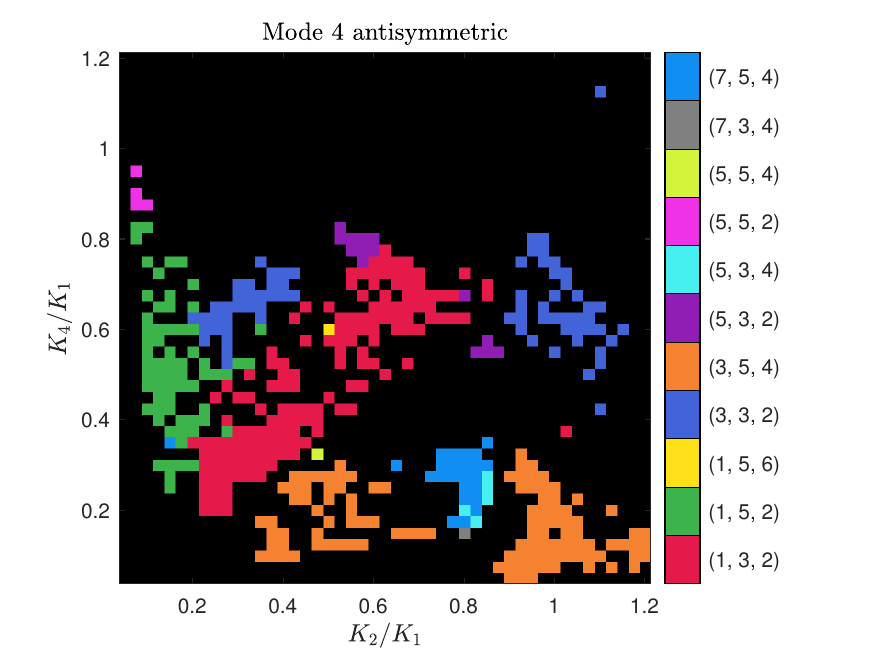}
        \end{minipage}
        \hfill
        \begin{minipage}{0.32\linewidth}
            \includegraphics[width=1.1\linewidth]{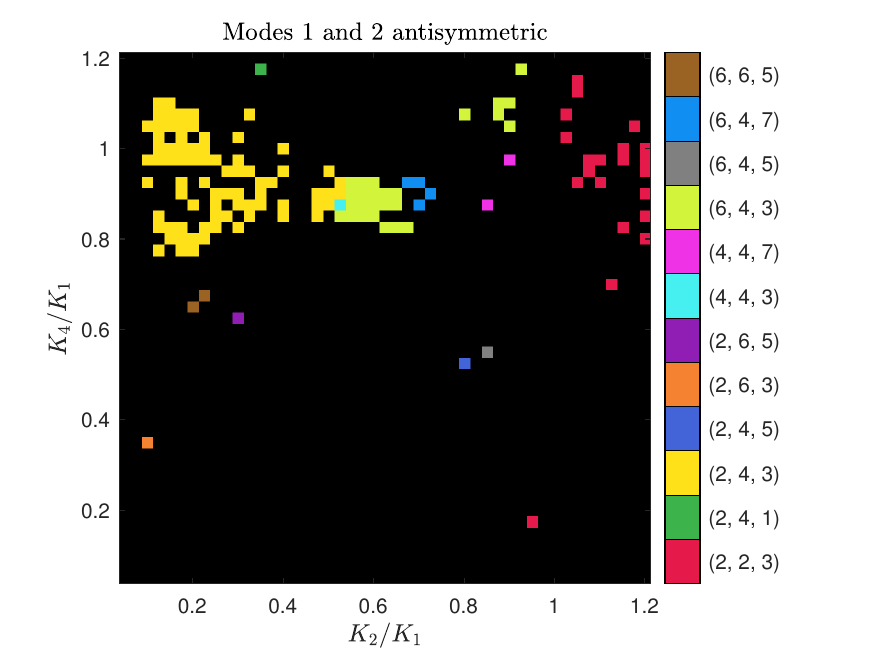}
        \end{minipage}
        \hfill
        \begin{minipage}{0.32\linewidth}
            \includegraphics[width=1.1\linewidth]{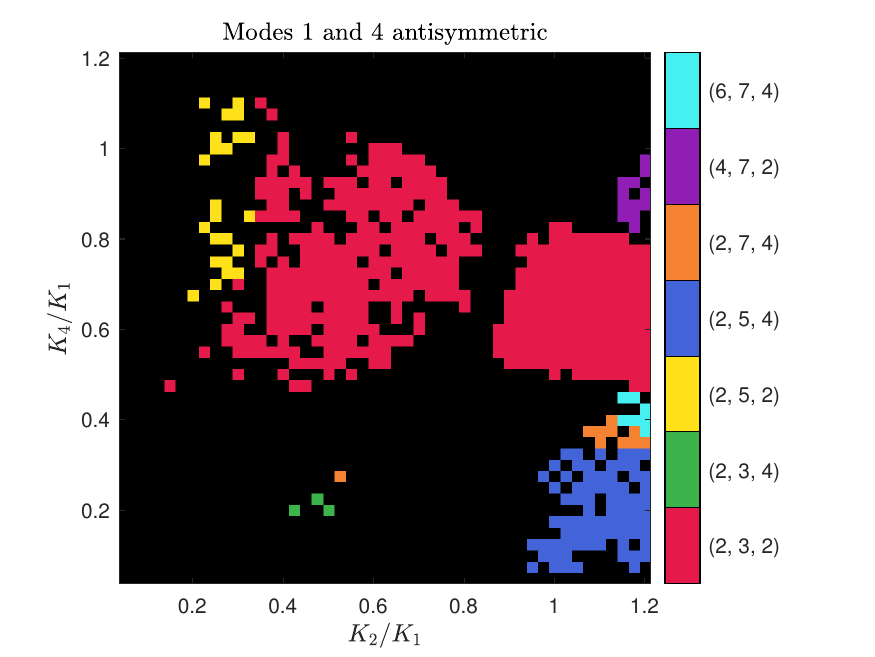}
        \end{minipage}
        \hfill
        \begin{minipage}{0.32\linewidth}
            \includegraphics[width=1.1\linewidth]{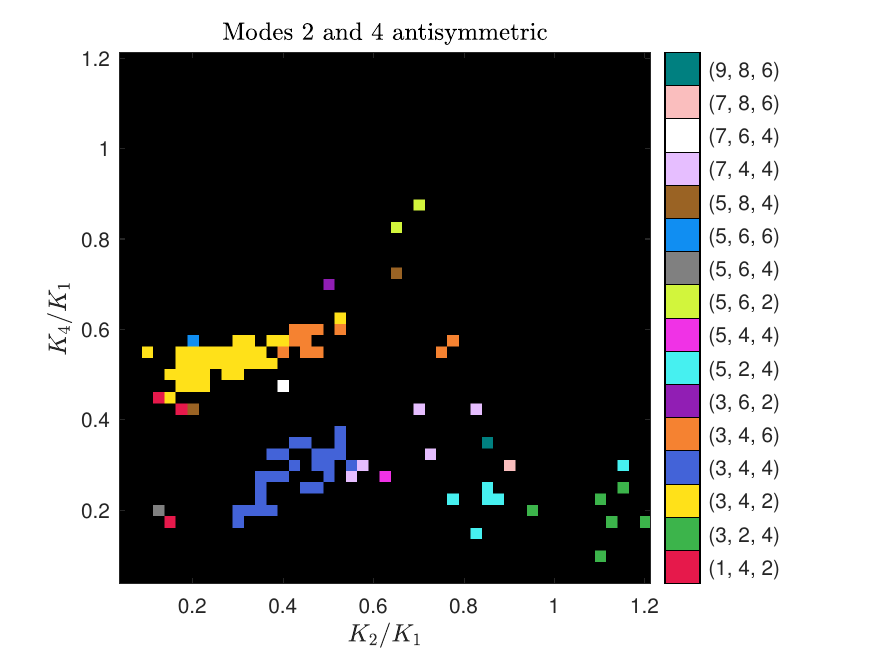}
        \end{minipage}
        \caption{Mappings in the free parameters space $\left(\frac{K_2}{K_1},\frac{K_4}{K_1}\right)$ for different symmetries of the temporal profiles of the modes. Black pixels indicate cases where the algorithm does not converge, or gives solitons made of less than three modes. Other colors indicate unique three-mode solutions. The triplet of integers $(m_1,m_2,m_3)$ gives the numbers of extrema in the temporal profiles of modes 1 ($\mathrm{LP}_{01}$), 2 ($\mathrm{LP}_{11}$), and 4 ($\mathrm{LP}_{21}$), respectively.}
        \label{fig:mappings}
\end{figure*}

Fig.~\ref{fig:mappings} shows the mappings for different imposed symmetries. We combined the results obtained for different initial guesses that differ in the number of nodes in the temporal profiles of the modes. In total, we counted more than seventy stationary solutions with our classification system. \textit{A priori}, the number of possible solutions is infinite. Note that the large variety of solution also implies a large variety of temporal widths that impact the quality of the convergence of our relaxation method. Indeed, as the evolution of the temporal widths against the free parameters is far from trivial, it is difficult to find an appropriate temporal window for each solution. Hence, the no convergence points can be in part explained as cases where the window is too short or too long compared to the width of the solution which prevents the algorithm from converging.

For each map, it appears that a specific solution occupies a region of the free parameters space. However, there are no clear frontiers between the domains. Therefore, our results are a summary of the existence of solutions, but not the rigorous exposition of every solution. Using continuation method, it would be possible to eliminate black pixels in the middle of colored regions and to extend the presented frontier. Furthermore, there is a clear overlap between different regions of solutions. For instance, Fig.~\ref{M2_antisym} shows a brown pixel corresponding to a multimode soliton of type (5,6,5) in the middle of a region occupied by (3,4,3). Starting the relaxation with a guess (5,6,5), it is possible to identify the soliton (5,6,5) in the whole region where we identify solitons (3,4,3) with our standard starting guess. Importantly, it indicates that there is a coexistence of different multimode soliton solutions for the same values of the free parameters. This coexistence is not limited to solutions with the same symmetry. For instance, the triplet (3,4,3) (mode 2 antisymmetric) also shares a part of its region with (2,3,3), (2,5,3), (1,5,2), (3,3,2), (5,5,2), (2,4,3), (2,3,2), (2,5,2), (3,4,2), and (3,4,4). This list, that is not exhaustive, contains triplets from solutions with at least one mode that is antisymmetric. Some solutions with mode 2 that is antisymmetric in time also coexist with solutions where all the modes have symmetric temporal profiles, as (1,2,1) shares the same region of the parameter space than (1,3,1), for instance. The mappings thus suggest that a solution branch can split into different sub-branches, and that it is necessary to conduct a bifurcation analysis to characterize the multistability of the solution, which is beyond the scope of this work.

In addition, this multistability appears for normalized equations. The more extrema a temporal profile has, the larger it is. Hence, the peak power is lower. In practice, the peak power must be high enough to reach the nonlinear regime, yet low enough for higher-order effects to remain negligible. Accounting for this range may remove the apparent coexistence of certain solutions, because some of them fall outside it. This restriction also limits the number of solutions that can be observed.

In conclusion, we highlighted the existence of a large variety of stationary solutions with modal temporal profiles ranging from relatively simple to very complex. The mappings presented in Fig.~\ref{M2_antisym} and~\ref{fig:mappings} illustrate the richness of the possible stationary solutions, but are certainly not exhaustive. A proper bifurcation analysis is required to explore this wide range of solutions. This plethora of multimode solitons must also be confronted with the starting equation, Eq.~\ref{GMMNLSE}, in order to assess their invariance and stability, and to determine how they behave when one of our assumptions is broken. These last questions are addressed in Section~\ref{sec:prop}.

\section{Propagation of multimode solitons}\label{sec:prop}
While the analogous potential method allows us to efficiently find stationary solutions for our reduced model (see Eq.~\ref{Syst_ODE} and Eq.~\ref{Syst_ODE_3modes}), it does not provide any indication about their stability during propagation. It is therefore necessary to simulate the propagation of these solutions in order to check their invariance and to investigate their stability. Hence, the goal of this section is to test our model in its extended version and to understand how higher-order nonlinearities affect the propagation of multimode solitons. We will also qualitatively discuss the stability of the multihumped stationary solutions.

All the multimode solitons presented here were tested by propagating them using an open-source code~\citep{wright_multimode_2018, wiselabaep_wiselabaepgmmnlse-solver-final_2026} that simulates their propagation under the GMMNLSE (see Eq.~\ref{GMMNLSE}). We neglected higher-order effects and imposed the same group velocity on all modes (which makes sense, since we assume that the multimode soliton is already formed), as in the assumptions presented in Section~\ref{sec:theory}. However, we kept the coherent terms and the difference in $\beta_2 $.

First, in Section~\ref{sec:prop_fund}, we discuss the impact of the assumptions exposed in Section~\ref{Chap2} on the propagation of the fundamental multimode soliton. Then, in Section~\ref{sec:prop_mutlihump}, we simulate the noisy propagation of multihumped multimode solitons and qualitatively discuss their stability.
\subsection{Propagation of the fundamental multimode solitons}\label{sec:prop_fund}
\begin{figure}
    \begin{minipage}{0.48\linewidth}
        \begin{overpic}[width=\linewidth]{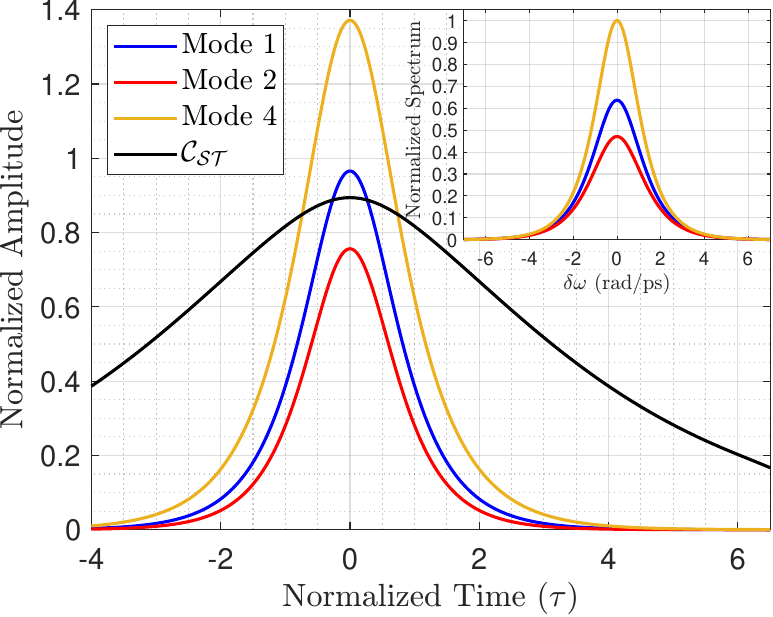}\put(-5,80){(a)}
        \end{overpic}
    \end{minipage}
    \begin{minipage}{0.48\linewidth}
        \begin{overpic}[width=\linewidth]{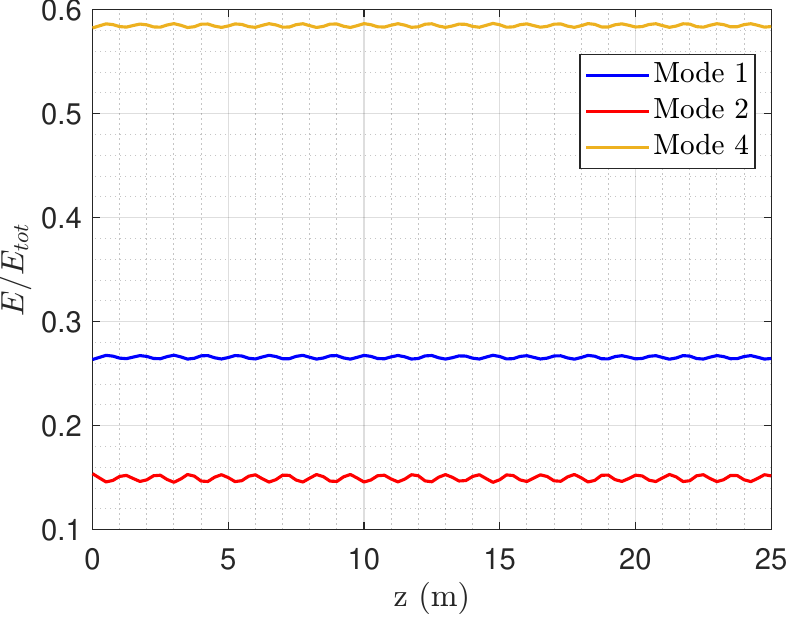}\put(-3,80){(b)}
        \end{overpic}
    \end{minipage}
     \begin{minipage}{0.48\linewidth}
        \begin{overpic}[width=\linewidth]{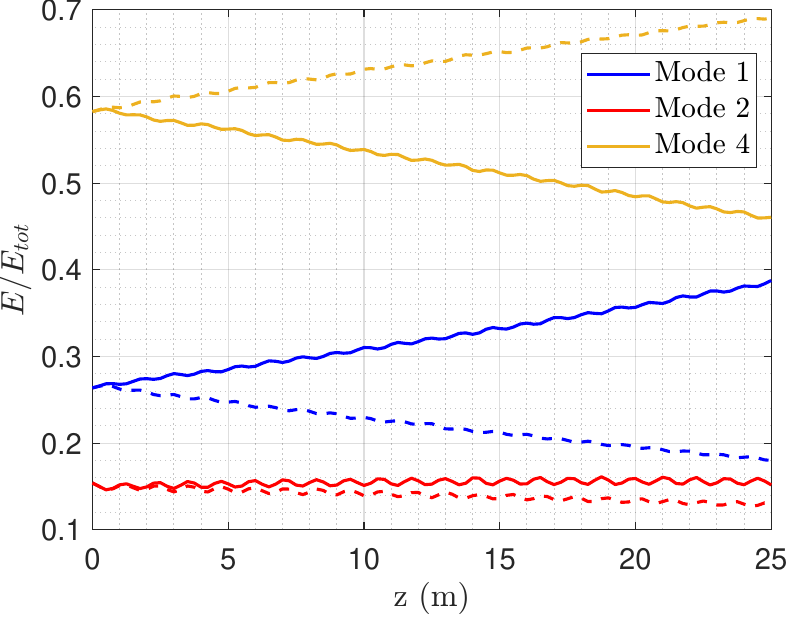}\put(-5,80){(c)}
        \end{overpic}
    \end{minipage}
    \begin{minipage}{0.48\linewidth}
        \begin{overpic}[width=\linewidth]{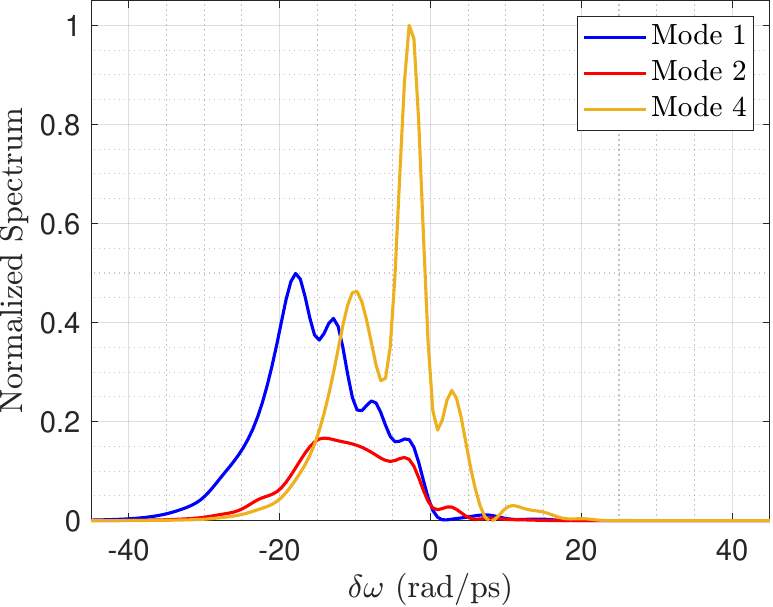}\put(-3,80){(d)}
        \end{overpic}
    \end{minipage}
    \caption{a) Temporal profile and spectrum of a multimode soliton with $K_2/K_1=1.2$ and $K_4/K_1=0.76$, at the entry of the fiber. b) Evolution of the normalized modal energies during the propagation of the multimode soliton in the fiber without Raman effect. c) Evolution of the normalized modal energies during the propagation of the multimode soliton with Raman effect ($f_R=0.18$). The solid lines correspond to the case where the group velocities are standard, and the dotted lines correspond to the case where the signs of the group velocities are reversed. d) Spectra of the modes after 25 m of propagation with Raman effect and (standard) group velocity mismatch. Spectra are interpolated.}
    \label{fig:prop_fund_Raman_velocity}
\end{figure}

Fig.~\ref{fig:prop_fund_Raman_velocity}.a shows a fundamental multimode soliton composed of modes 1 ($LP_{01} $), 2 ($LP_{11a} $), and 4 ($LP_{21a} $). When propagating through the multimode fiber according to Eq.~\ref{GMMNLSE}, \textit{i.e.} considering the extended picture, the temporal profile of each mode remains well invariant. Furthermore, when these profiles are perturbed by, for instance, imposing the same temporal width on all modes, the soliton returns to its equilibrium shape by adjusting the temporal shapes such that the space-time complexity profile is curved again (see also Section~\ref{subsec:2modes.fundamental} and Fig.~\ref{fig:equ_ellipt}). Moreover, the energy exchanges through inter-modal four-wave mixing are periodic, such that the contribution of the coherent terms is indeed negligible. The fundamental solitons are also stable against symmetry breaking and noise, as expected. These confirmations allow us to take our study a step further and examine how our solutions behave when we move beyond the scope of the assumptions defined in Section~\ref{sec:theory}.

The main difference between our constructed soliton and a naturally formed soliton is the shape of the modal spectra. In our case, the spectra are symmetric, as shown in Fig.~\ref{fig:prop_fund_Raman_velocity}.a. For a naturally formed soliton, since the modes propagate at different group velocities and lock to each other through cross-phase modulation, the resulting spectra are asymmetric. We simulate the propagation of the multimode soliton presented in Fig.~\ref{fig:prop_fund_Raman_velocity}.a without canceling the group velocity mismatch. From the point of view of the soliton, it suddenly undergoes a significant perturbation, as the modes tend to separate from each other. Nonetheless, for sufficient peak powers, the soliton adapts itself until it regains its equilibrium shape. In the process, the modal spectra become asymmetric, but this does not affect the periodic energy exchange between modes, as shown in Fig.~\ref{fig:prop_fund_Raman_velocity}.b.

However, when Raman nonlinearity is taken into account ($f_R=0.18 $), the asymmetry of the spectra plays an important role, because it is responsible for a net energy transfer between modes, as shown in Fig.~\ref{fig:prop_fund_Raman_velocity}.c. Indeed, since we are in the anomalous dispersion regime, the spectrum of the fastest mode extends toward lower frequencies, while that of the slowest mode extends toward higher frequencies. Hence, Raman nonlinearity generates Stokes waves in the fastest modes from the energy of the slowest ones. The solid lines in Fig.~\ref{fig:prop_fund_Raman_velocity}.c illustrate this intermodal energy exchange during propagation. The dashed lines correspond to the case where we inverted the sign of the $\beta_1$ coefficients. In this case, mode 4 is the fastest and thus benefits from the energy exchange. 

In addition to this intermodal energy exchange, the soliton also undergoes the well-known Raman self-frequency shift~\citep{mitschke_discovery_1986,gordon_theory_1986,zitelli_characterization_2022}. This is shown in Fig.~\ref{fig:prop_fund_Raman_velocity}.d, where the spectra of the modes after 25 m of propagation are asymmetric but are also all shifted toward lower frequencies. The strength of the intermodal energy exchange depends on the group velocity mismatch, the overlap coefficients, and the peak powers of the modes. For a sufficiently long propagation distance, all the energy will be transferred to the fastest mode, and the soliton will become single-mode~\cite{zitelli_single-mode_2021}.
\begin{figure}
    \begin{minipage}{0.48\linewidth}
        \begin{overpic}[width=\linewidth]{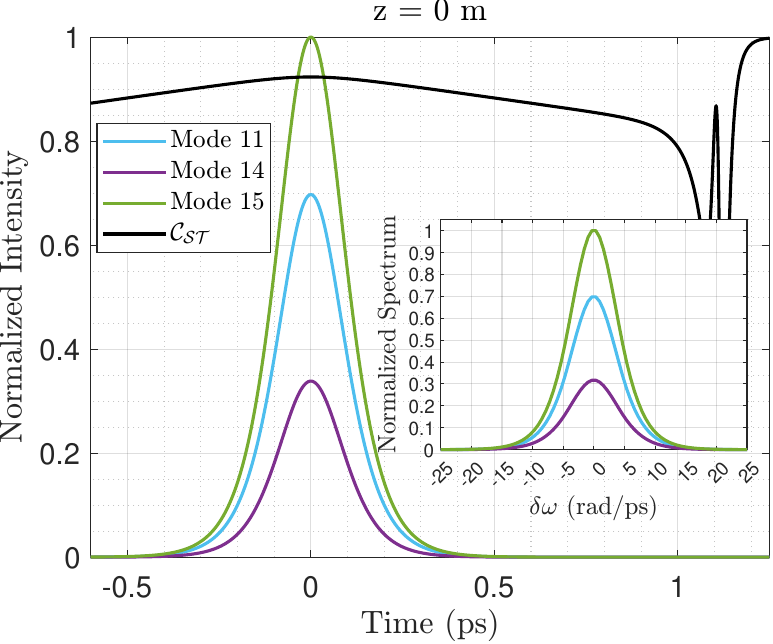}\put(-3,80){(a)}
        \end{overpic}
    \end{minipage}
    \begin{minipage}{0.48\linewidth}
        \begin{overpic}[width=\linewidth]{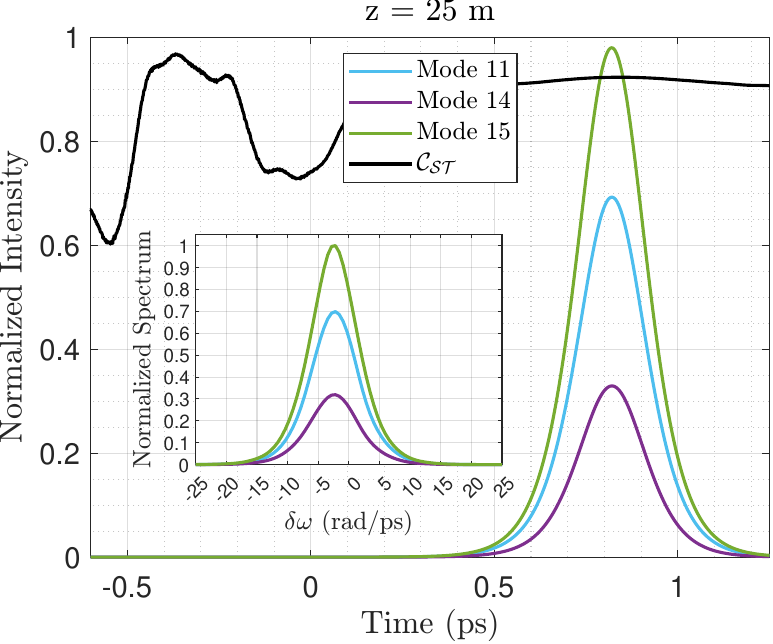}\put(-3,80){(b)}
        \end{overpic}
    \end{minipage}
    \begin{minipage}{0.48\linewidth}
        \begin{overpic}[width=\linewidth]{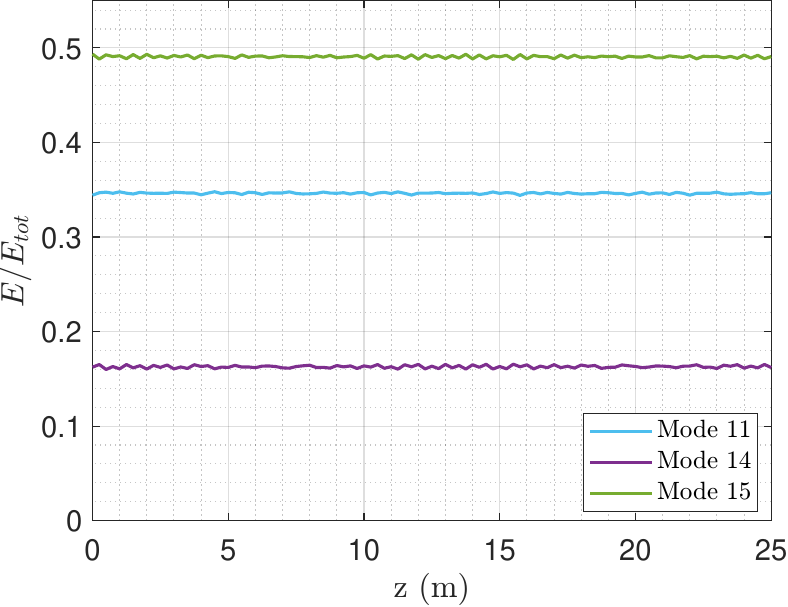}\put(-3,80){(c)}
        \end{overpic}
    \end{minipage}
    \begin{minipage}{0.48\linewidth}
        \begin{overpic}[width=\linewidth]{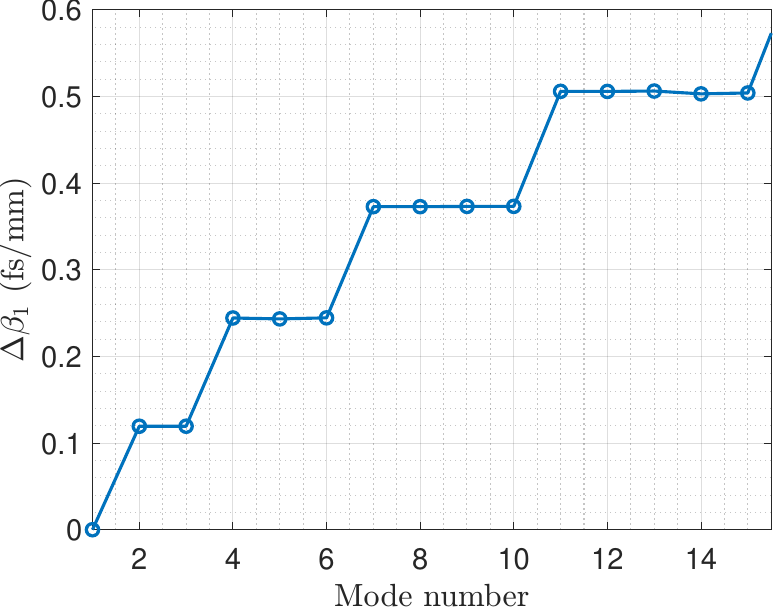}\put(-4,80){(d)}
        \end{overpic}
    \end{minipage}
    \caption{Fundamental multimode soliton composed of modes 11 ($LP_{41a}$), 14 ($LP_{22b}$), and 15 ($LP_{03}$) with the parameters $K_{14}/K_{11}=1.1$ and $K_{15}/K_{11}=1$. a) Normalized intensity profiles and spectra of the modes at the entrance of the fiber; b) Normalized intensity profiles and spectra of the modes after 25 m of propagation; c) Evolution of the normalized modal energies against propagation distance; d) Distribution of $\beta_1$ coefficients against mode number at $\lambda_0=1550$ nm. Spectra are interpolated.}
    \label{fig:SFS}
\end{figure}

It is important to note that Raman-induced energy transfer toward one mode is not inevitable for all solitons, as will be discussed here after. We consider a fundamental multimode soliton made of modes 11 ($LP_{41a}$), 14 ($LP_{22b}$), and 15 ($LP_{03}$), shown in Fig.~\ref{fig:SFS}.a. The space-time complexity varies in time, meaning that the soliton is truly multimode. Note that it is flatter than that of the soliton presented in Fig.~\ref{fig:prop_fund_Raman_velocity}, meaning that the modes have closer durations. We propagate the soliton over 25 m, taking into account coherent terms, group velocity mismatch, and Raman nonlinearity. The temporal profiles of the modes, as well as their spectra after propagation, are shown in Fig.~\ref{fig:SFS}.b. The spectra are still symmetric but shifted from the initial central frequency. There is no net energy exchange between modes, as shown in Fig.~\ref{fig:SFS}.c. This shows that our model remains valid, in this case, even though we are formally outside its range of validity; as mentioned in Section~\ref{fiber}, the beat length between modes 14 and 15 is 20 cm, while the dispersion length of the soliton considered here ($T_0=125$ fs) is approximately 55 cm. The difference from the previous case is that, as shown in Fig.~\ref{fig:SFS}.d, the triplet of modes composing the soliton propagates at very close group velocities — in fact, this is the lowest-order set of three nondegenerate modes with such similar group velocities. Hence, cross-phase modulation does not induce significant asymmetry in the spectra. In this case, the soliton undergoes only the self-frequency shift.

Note that our model does not take fiber imperfections into account. These result in two main effects on the propagation of multimode solitons. The first is that they perturb the pulse, but, as mentioned above, the fundamental multimode soliton is stable and will return to its equilibrium shape. The second is linear random mode coupling, which adds to the Raman-induced energy transfer discussed above and, over sufficiently long propagation distances, makes the fundamental mode act as an attractor, ultimately yielding a single-mode soliton~\cite{zitelli_single-mode_2021}.
\begin{figure}
    \begin{minipage}{0.48\linewidth}
        \begin{overpic}[width=\linewidth]{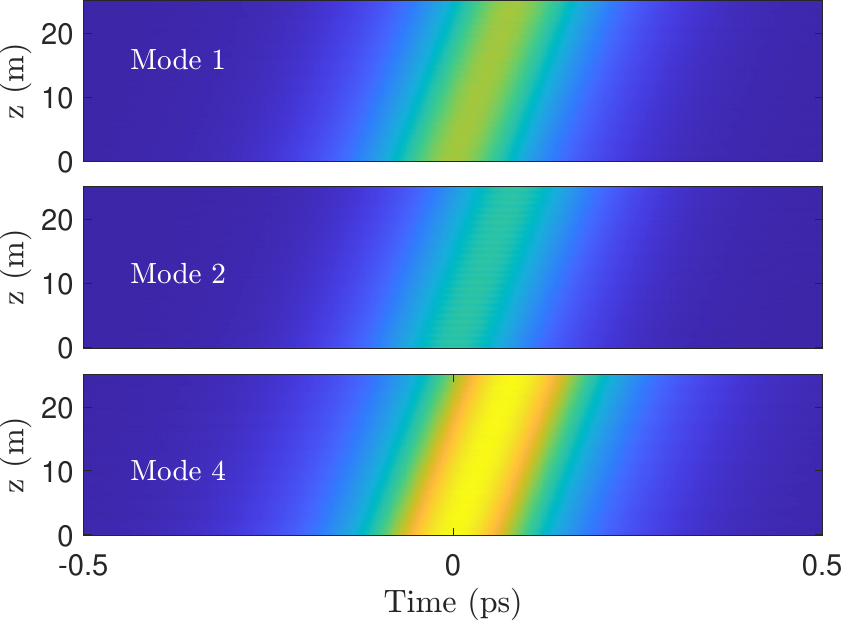}\put(-3,75){(a)}
        \end{overpic}
    \end{minipage}
    \begin{minipage}{0.48\linewidth}
        \begin{overpic}[width=\linewidth]{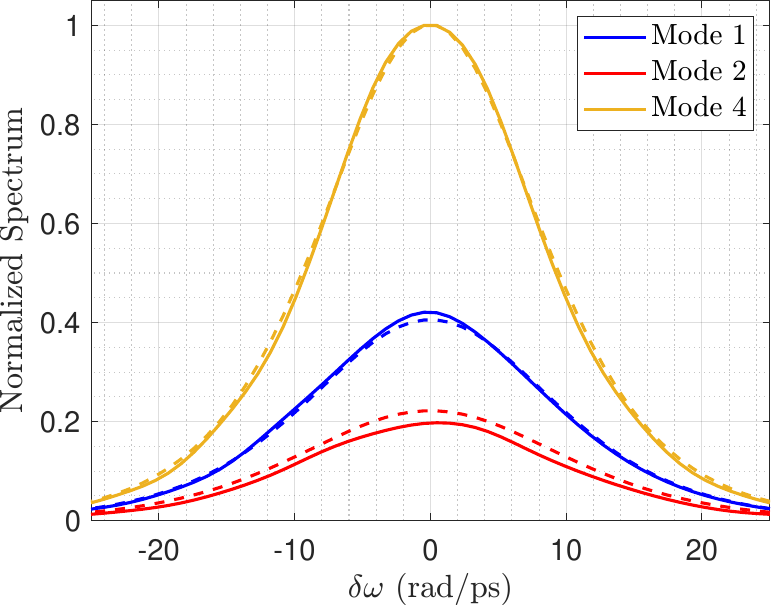}\put(-2,75){(b)}
        \end{overpic}
    \end{minipage}
    \caption{Illustration of the effect of the self-steepening on a fundamental multimode soliton. a) Evolution of the normalized temporal intensity profiles for each mode composing the soliton during the propagation over 25 m in the fiber; b) Interpolated spectra of the modes at the beginning of the propagation (dashed lines) and at the end (solid lines).}
    \label{fig:self-steepening}
\end{figure}

Finally, we address the self-steepening effect on the fundamental multimode soliton. For a pulse short enough that one or several modes exhibit temporal profiles with durations under 100 fs, the self-steepening effect causes a deformation of the temporal profile. Because of the incoherent interaction with the other modes, however, each mode adapts its temporal profile to counteract this deformation. This adaptation, in turn, deforms the modal spectra, which ultimately causes a shift in the group velocity of the pulse. Fig.~\ref{fig:self-steepening}.a shows the evolution of the normalized intensity profiles of the modes 1, 2, and 4 in the $(t,z)$ plane. The reference window propagates at the speed of the soliton would have in absence of self-steepening. The trajectory is tilted, meaning that the soliton propagates more slowly than the reference window. As shown in Fig.~\ref{fig:self-steepening}.a, this shift in group velocity occurs very quickly. This same effect is also visible in the spectral domain. Fig.~\ref{fig:self-steepening}.b compares the spectra at the beginning of the propagation (dashed lines) and at the end (solid lines), i.e., after 25 m of propagation. The spectra are deformed in such a way that those of modes 1 and 2 become slightly asymmetric. The spectrum of mode 4 becomes slightly flatter at its maximum. This velocity shift is a nonlinear effect that depends on the peak power of the multimode soliton: the higher the peak power, the greater the change in group velocity. In conclusion, this higher-order effect does not compromise the stability of the fundamental multimode soliton but changes its velocity.

Note that the stability of the soliton against self-steepening implies that a fundamental soliton can be highly multimode in the femtosecond range. Indeed, the more modes composing the soliton, the greater the difference in duration between the temporal profiles of the modes. Hence, even if the mode with the highest peak power has a duration far longer than 100 fs, the mode with the smallest peak power could still have a duration below that limit. Such a mode could therefore undergo self-steepening without dramatically affecting the overall highly multimode soliton.

\subsection{Propagation of the multihumped solitons}\label{sec:prop_mutlihump}
Finally, we consider the propagation of multihumped multimode solitons in the fiber using our simulation based on Eq.~\ref{GMMNLSE}. As for the fundamental soliton, they are invariant through propagation, and they exchange energy through inter-modal four-wave mixing periodically such that coherent terms do not impact their temporal structure. However, when we introduce noise into the intensity profiles at the beginning of the propagation, or when symmetry breaking is induced manually, many of the multihumped stationary solutions break apart, indicating that they are unstable.

This behavior is also observed in the case of the multihumped elliptical soliton. Symmetry breaking causes all the energy to move into one lobe of the profile randomly. The multihumped elliptical soliton thus collapses into a fundamental elliptical soliton~\citep{kockaert_stability_1999, cambournac_symmetry-breaking_2002}. An interesting outcome of this study is that spatial multimode solitons reveal new regimes that are not accessible with polarization modes. In particular, we can avoid the collapse to the fundamental elliptical soliton, as the spatial version opens new regions of the parameter space that become accessible for experiments. 

For the two-mode solitons presented in Section~\ref{sec:2modes.second_order}, the temporal profiles are mostly unstable, as in the polarization case. Unlike the elliptical soliton, the temporal structure here collapses into several fundamental multimode solitons (one for each lobe), a manifestation of the new dynamical regimes made accessible by nondegenerate spatial multimode solitons. It is worth noting that the stability of such structures, for the elliptical soliton, can be improved depending on the region of the parameter space considered~\cite{ostrovskaya_stability_1999}. The same conclusion can be qualitatively drawn in the case of spatial modes, since, depending on the shape of the trajectory in the potential, the temporal structure can be more or less stable against noise and spontaneous symmetry breaking.

We now turn to the three-mode multihumped case (refer to Section~\ref{subsec: multihumped_3modes}), where the solitons are similarly unstable against noise and/or symmetry breaking. Importantly, as for the two-mode case, some profiles seem to be more stable than others, in the sense that, for the same energy-duration product, certain profiles break up after longer propagation distances than others. The break-up of such sophisticated temporal structures can be quite rich, as they can break into several fundamental three-mode solitons or into other multihumped three-mode solitons. Moreover, the result of the break-up can be picked from a finite set of possible outputs randomly. For instance, the multimode soliton can split into one fundamental multimode soliton and a multihumped multimode soliton (which will break later on), or directly into several fundamental multimode solitons.

\begin{figure}
    \begin{minipage}{0.48\linewidth}
        \begin{overpic}[width=\linewidth]{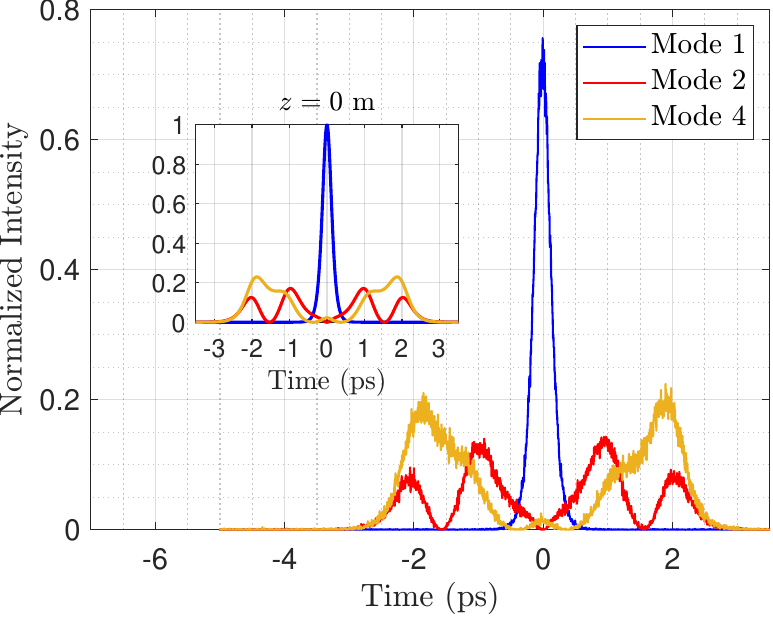}\put(-3,81){(a)}
        \end{overpic}
    \end{minipage}
    \begin{minipage}{0.48\linewidth}
        \begin{overpic}[width=\linewidth]{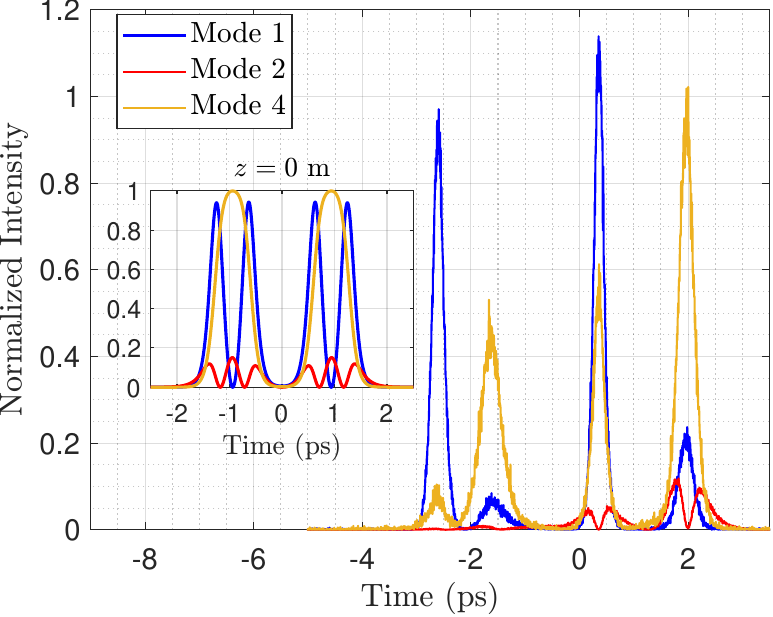}\put(-3,81){(b)}
        \end{overpic}
    \end{minipage}
    \begin{minipage}{0.48\linewidth}
        \begin{overpic}[width=\linewidth]{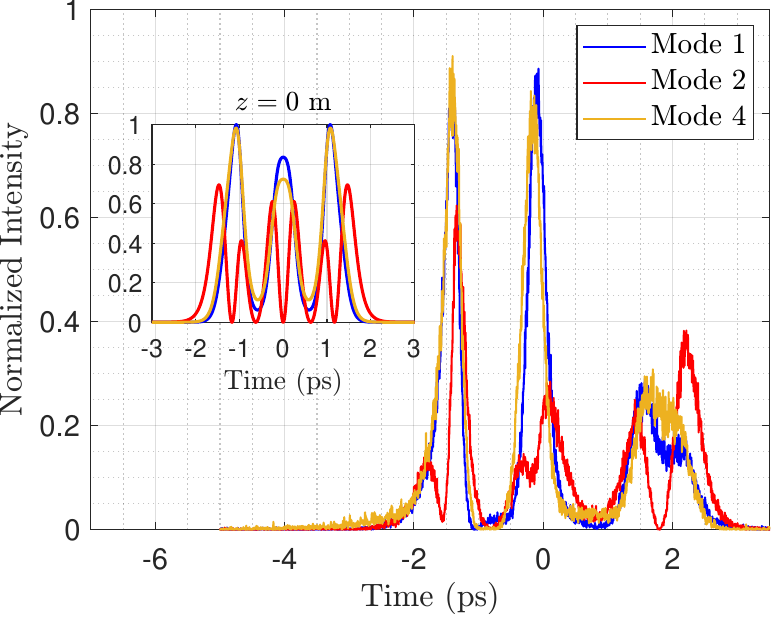}\put(-3,81){(c)}
        \end{overpic}
    \end{minipage}
    \begin{minipage}{0.48\linewidth}
        \begin{overpic}[width=\linewidth]{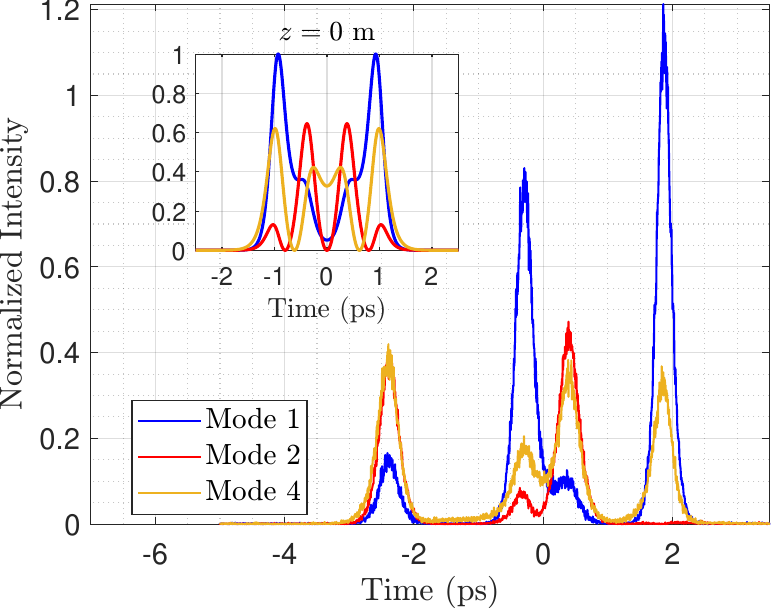}\put(-3,81){(d)}
        \end{overpic}
    \end{minipage}
    \caption{Multihumped three-mode soliton after 50 m of noisy propagation in the multimode fiber. As the fission process is induced by random noise, the figure shows for each soliton one outcome among many. a) (1,4,7); b) (5,6,3); c) (5,6,5); d) (7,4,5). For each case, the normalized intensity profiles (without noise) at the beginning of the propagation are shown.}
    \label{fig:prop_3modesMMS_SecondOrdre}
\end{figure}

As an example, Fig.~\ref{fig:prop_3modesMMS_SecondOrdre} shows the noisy propagation of the multimode solitons presented in Fig.~\ref{fig:exemples_3modesMMS_SecondOrdre} (c-f) after 50 m of propagation in the multimode fiber. Without noise, all the temporal profiles are invariant during propagation. When noise is added, the evolution differs for each soliton. As a reminder, we count the number of extrema in the amplitude profile of each mode to distinguish the different types of solitons (refer to Section~\ref{subsec: multihumped_3modes}).

Fig.~\ref{fig:prop_3modesMMS_SecondOrdre}.a shows that the (1,4,7) soliton is quite robust against noise, since it conserves its shape; nonetheless, when we add symmetry breaking, soliton fission still occurs. In contrast, the other three solitons exhibit far greater sensitivity to noise, and eventually undergo fission on their own. Fig.~\ref{fig:prop_3modesMMS_SecondOrdre}.b shows that the (5,6,3) soliton breaks into two fundamental multimode solitons and two multihumped solitons of type (1,2,1); the latter will break later on into fundamental multimode solitons. A similar but delayed fission is observed for the (5,6,5) soliton in Fig.~\ref{fig:prop_3modesMMS_SecondOrdre}.c : the fission occurs later, or is less dramatic, than for the (5,6,3) soliton, as its lobes are closer to each other. Finally, the (7,4,5) soliton breaks directly into several fundamental multimode solitons, as shown in Fig.~\ref{fig:prop_3modesMMS_SecondOrdre}.d. These results illustrate for each soliton one possible outcome among many as the fission process is due to random noise.

We conclude this section with a remark on the formation of multihumped solitons exhibiting complex structures. Due to the time-reversal invariance of Maxwell's equations, the injection of several fundamental multimode solitons propagating in opposite directions may give rise to a bound state, namely the multihumped soliton. Since this bound state is unstable, it would propagate for a few meters before fissioning, as described previously. This possibility has already been studied for elliptical solitons~\cite{haelterman_bimodal_1993}.
\section{Conclusion}
In conclusion, we have studied fundamental multimode solitons using the analogous potential method and demonstrated the existence of numerous multihumped multimode solitons composed of purely spatial modes. These solitary waves owe their invariance under propagation to the nonlinear incoherent interaction between modes. Starting from the Generalized Multimode Nonlinear Schrödinger Equation, we derived a reduced model whose stationary solutions are the multimode solitons. We then recast this model as the equations of motion of a unit-mass ball and derived the associated potential.

For the two-mode case, we showed that the mathematical formalism developed for the elliptical soliton also applies to multimode solitons made of spatial modes. Unlike in the polarization-mode case, the nonlinear coefficients of the modes can differ, which introduces an asymmetry in the equations. We characterized the temporal shape of the stationary solutions through the trajectories of the ball in the analogous potential, together with the ellipticity, which quantifies the difference in the temporal evolution of the modes. Because cross-phase modulation is partly responsible for the solitonic properties, we proposed a generalization of the soliton order to the multimode case. We also examined how the nonlinear coefficients associated with self- and cross-phase modulation affect the soliton structure, and found that the relative orientation of the transverse profiles can strongly influence the soliton dynamics. Building on this, we established the existence of two-mode multihumped solitons in which one mode exhibits an antisymmetric temporal profile. Owing to the asymmetry in the nonlinear coefficients, and unlike in the elliptical soliton case, these structures are not symmetric under exchange of the roles of the modes composing them.

Going further, we generalized our framework to multimode solitons made of more than two modes, which have no counterpart in the elliptical soliton case. Because ellipticity is defined only for a two-mode basis, we introduced a new metric, the space-time complexity, to quantify the differences between the temporal evolutions of the modes. We also report the existence of new kinds of multihumped multimode solitons, classified by the number of extrema in the temporal profile of each mode. Their temporal profiles range from relatively simple, with a single node in one of the modes, to very complex. Notably, some configurations that cannot exist in the two-mode case appear in the three-mode case. Several solutions coexist in the same region of the parameter space, but this coexistence may be merely apparent, owing to the normalization of the equations. Settling the question of coexistence within this wide range of solutions will require a proper bifurcation analysis.

Finally, we examined the nonlinear propagation of multimode solitons. We found that self-steepening only affects the speed of the fundamental soliton without altering its structure. Raman nonlinearity results in a net modal energy transfer from the slowest modes to the fastest ones. This is due to cross-phase modulation that induces asymmetry in the spectra of the modes. For a multimode soliton made of modes propagating at approximately the same group velocity, no such net energy transfer occurs, and Raman nonlinearity only induces the well-known soliton self-frequency shift. While fundamental solitons are stable against noise and symmetry breaking, this is not necessarily the case for multihumped multimode solitons, although some structures are less unstable than others. The fission of a multihumped multimode soliton generates randomly other, less complex, multihumped solitons as well as fundamental multimode solitons.

In contrast to our recent results showing that space-time excitation of multimode fibers results in splitting into multiple fundamental multimode solitons~\cite{dechanxhe_space-time_2026}, our results in this work show that complicated space-time structures can indeed be solitons. However, their specific space-time shape means they require more arbitrary space-time shaping to properly excite and observe experimentally. In general these results also open the way to a systematic study of the stability and bifurcation structure of multimode solitons in multimode fibers, and illustrate the richness of nonlinear multimode systems.

\appendix
\section{Lagrangian formalism}\label{Anx:Lagrange}
\begin{figure}
    \centering
    \includegraphics[width=\linewidth]{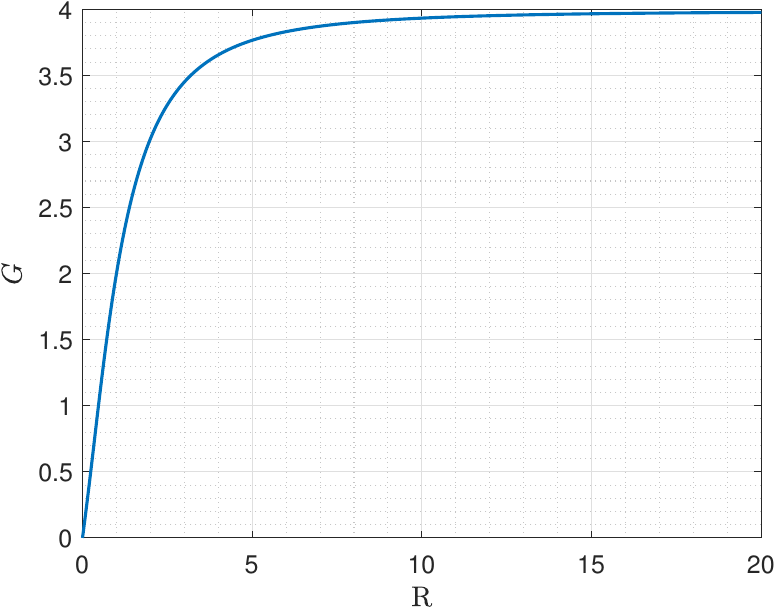}
    \caption{Function $G(R)$ defined by Eq.~\ref{def_G}.}
    \label{fig:G}
\end{figure}

The goal of this appendix is to find the analytical expression given in Eq.~\ref{K} using the Lagrangian approach. By assuming that modes propagate at the same group velocity and that temporal profiles are invariant through propagation, we consider the following ansatz:
\begin{equation}\label{eq:ansatz}
        a_n(\xi,\tau) = \eta_n \sech \left(\frac{\tau}{\tau_n}\right) e^{i\varphi_n}.
\end{equation}
The Lagrangian density of our nonlinear multimode system is given by:
\begin{align}
    \mathcal{L}&=\sum_{n=1}^M\mathcal{L}_n+\mathcal{L}_{int},\\
    \mathcal{L}_n &= i\left(a^{*}_n\pderiv{a_n}{\xi}-a_n\pderiv{a_n^{*}}{\xi}\right)+\sigma_n\left|\pderiv{a_n}{\tau}\right|^2 + N_n^2\left|a_n\right|^4,\\
    \mathcal{L}_{int}&=\sum_{p\neq n}^MC_{np}|a_p|^2|a_n|^2,
\end{align}
where $M$ is the total number of modes considered. By inserting the ansatz of Eq.~\ref{eq:ansatz} into the Lagrangian, and then by integrating against time variable, we get the following reduced Lagrangian density:
\begin{multline}
 L = \sum_{n=1}^M\left[-4\dot{\varphi}_n\eta_n^2\tau_n+\frac{2}{3}\frac{\sigma_n}{\tau_n}\eta_n^2+\frac{4}{3}N_n^2\eta_n^4\tau_n\right]\\
 +\sum_{n=1}^M\sum_{p\neq n}^MC_{np}\eta_p^2\eta_n^2I_{np},
\end{multline}
where $\dot\varphi_n$ is the derivative of the phase of mode $n$ against the longitudinal variable $\zeta$.
Finally, the Euler-Lagrange equation gives:
\begin{equation}
    K_n = \dot\varphi_n = \frac{1}{2}N_n^2\eta_n^2+\frac{1}{4}\sum_{p\neq n}^{M}C_{np}\eta_p^2\left(\frac{I_{np}}{\tau_n}+\pderiv{I_{np}}{\tau_n}\right),
\end{equation}
where $I_{np}=I(\tau_n,\tau_p)$ (the definition of the function is given in Appendix~\ref{Anx:Barnes}). The term $\frac{I_{np}}{\tau_n}+\pderiv{I_{np}}{\tau_n}$ can be written as a function that depends only on the ratio of the temporal widths ($R=T_v/T_u$):
\begin{equation}\label{def_G}
    G(R) = \frac{192}{\pi^2}R^3\sum_{n=0}^{\infty}\sum_{m=0}^{\infty}\frac{(2m+1)}{\left((2n+1)+(2m+1)R\right)^4}.
\end{equation}
This function is continuous, tends to zero when $R\rightarrow 0$, and tends to 4 when $R\rightarrow \infty$, as shown in Fig.~\ref{fig:G}. Notably, $G(1)=2$. Using the equality $\frac{I_{np}}{\tau_n}+\pderiv{I_{np}}{\tau_n}=G(T_p/T_n)$ gives Eq.~\ref{K}.
 
\section{Interaction integral}\label{Anx:Barnes}
The Lagrangian formalism of Section~\ref{Anx:Lagrange} requires computing the following integral:
\begin{equation}
    I(a,b) = \int_{-\infty}^{+\infty}\sech^2(\tau/a)\sech^2(\tau/b)d\tau.
\end{equation}
Using the Parseval-Plancherel theorem with the facts that the functions of the integrand are real and even, we obtain the following equality:
\begin{equation}
     I(a,b) = \frac{8b^2}{\pi^2a}\int_{0}^{+\infty}\frac{u^2}{\sinh(u)\sinh(Ru)}du,
\end{equation}
where $R=b/a$. We compute this integral by expanding the cosech functions in series. This gives:
\begin{subequations}
    \begin{align}
    I(a,b) &= \frac{64a^2b^2}{\pi^2}\sum_{n=0}^{\infty}\sum_{m=0}^{\infty}\frac{1}{\left(a+b+2na+2mb\right)^3},\\
    &=\frac{64a^2b^2}{\pi^2}\;\zeta_2(3,a+b|2a,2b),
        \end{align}
\end{subequations}
where the function $\zeta_2$ is the Barnes double zêta function. The function $I(a,b)$ is homogeneous of degree one, implying that it is the solution of the following differential equation: 
\begin{equation}
    I(a,b)=a\pderiv{I}{a}+b\pderiv{I}{b}.
\end{equation}
\section{Two-mode soliton in the limit case of unilateral optical guiding}\label{Anx:Legendre}
Our goal is to find a solution describing the two-mode soliton (composed of modes $u$ and $v$) in the limit case where $P_v \ll P_u$. In such a case, mode $u$ is not sensitive to mode $v$ while this last is optically guided in time by the former. To start, we consider the set of equations given by Eq.~\ref{Syst_ODE} and we impose that mode $v$ is zero. We take advantage of the invariant property of the normalized set of equations by imposing $K_u=1$. The single-mode solution is given by $a_u(\tau)=u_0(\tau)=\sqrt{\frac{2}{N_u^2}}\sech\left(\sqrt{2}\tau\right)$. We now consider a perturbation of this solution of the type $(u_0+\varepsilon^2x,\varepsilon y)$. This perturbation will allow us to bifurcate to our wanted two-mode solution. Inserting the perturbation in Eq.~\ref{Syst_ODE}, we get: 
\begin{subequations}
    \begin{align}
        \ppderiv{x}{\tau} &= 2\left(1-3N^2_uu_0^2\right)x-2Cu_0y^2, \\
        \ppderiv{y}{\tau} &= 2\left(K_v-Cu_0^2\right) y.
    \end{align}
\end{subequations}
By performing the change of variable $\kappa = \tanh(\sqrt{2}\tau)$ for the second equation, we get a Legendre equation: 
\begin{equation}
    (1-\kappa^2)\frac{d^2y}{d\kappa^2}-2\kappa\frac{dy}{d\kappa}+\left(\frac{2C}{N_u^2}-\frac{K_v}{(1-\kappa^2)}\right)y=0.
\end{equation}
The fundamental solution corresponding to a bright soliton is given by:
\begin{equation}
    y = \sech^{n_{uv}}(\sqrt{2}\tau),
\end{equation}
where the parameter $n_{uv}=\sqrt{K_v}$ depends on the nonlinear coefficients through the relation $n_{uv}(n_{uv}+1)=\frac{2C}{N_u^2}$. By rearranging the last equation, we finally get:
\begin{equation}\label{def_n_i}
    n_{uv} = \frac{1}{2}\left(\sqrt{1+16\frac{\gamma_{uv}}{\gamma_{uu}}}-1\right),
\end{equation}
As mentioned in Section~\ref{sec:2modes.second_order}, other solutions that decay at infinity are given by $P^{\lambda-n_{uv}}_{n_{uv}}(\kappa)$. These solutions exist if $n_{uv}>\lambda$, where $\lambda$ is an integer. In that case, they exhibit $\lambda$ nodes.

\begin{acknowledgments}
Julien Dechanxhe is a FRIA grantee of the Fonds de la Recherche Scientifique (F.R.S.-FNRS). This work is supported by the F.R.S.-FNRS and the French Community of Belgium as part of the funding for his PhD grant.
\end{acknowledgments}

\bibliography{MMS_potentiel}

\end{document}